\documentclass[twocolumn,showpacs,preprintnumbers,amsmath,amssymb,superscriptaddress,longbibliography,prl,aps,10pt]{revtex4-2}
\usepackage{graphicx}% Include figure files
\usepackage{epsfig}% Include figure files
\usepackage{epstopdf}
\usepackage{dcolumn}% Align table columns on decimal point
\usepackage{bm}% bold math
\usepackage{amsmath}
\usepackage{color}

\usepackage{enumitem}
\usepackage{xspace}
\usepackage{lipsum}

\usepackage{orcidlink}
\usepackage{textcomp}

\usepackage[utf8]{inputenc}
\usepackage[T1]{fontenc}
\usepackage{indentfirst}
\usepackage{textgreek}
\usepackage{multirow}
\usepackage{footmisc}

\makeatletter
\def\maketitle{
\@author@finish
\title@column\titleblock@produce
\suppressfloats[t]}
\makeatother

\newcommand{\ph}{\phantom{$-$}}

\newcommand{\mytitle}{Radiative decay and electromagnetic moments \\in $^{229}$Th determined within nuclear DFT}

\newcommand{\myabstract}{Using the nuclear DFT approach with symmetry breaking and restoration, we investigate the electromagnetic properties of the ground and isomeric states in $^{229}$Th. We determine the magnetic dipole transition strength B(M1; $3/2^+_1\rightarrow 5/2^+_1)$ between these two states and discuss the effects of parity breaking, configuration mixing, and time-odd core polarization. We also determine the corresponding spectroscopic magnetic dipole and octupole, and electric quadrupole moments. Because the octupole deformability of the Skyrme functionals used here is not described in sufficient detail, we analyze the results using a set of Skyrme functionals and perform a regression aligned with the measured electric octupole moments of neighboring even-even nuclei. Without parameter adjustment, the results compare favorably with the experimental data but also indicate the need to systematically adjust the octupole degrees of freedom in future functional parametrizations.
}

\begin{document}

\preprint{}

\title{\protect\mytitle}

\author{A. Restrepo-Giraldo\orcidlink{0000-0002-7811-6252}}\affiliation{School of Physics, Engineering and Technology, University of York, Heslington, York YO10 5DD, United Kingdom}

\author{J. Dobaczewski\orcidlink{0000-0002-4158-3770}}
\affiliation{School of Physics, Engineering and Technology, University of York, Heslington, York YO10 5DD, United Kingdom}
\affiliation{Institute of Theoretical Physics, Faculty of Physics, University of Warsaw, ul. Pasteura 5, PL-02-093 Warsaw, Poland}

% \author{P. Becker}
% \affiliation{School of Physics, Engineering and Technology, University of York, Heslington, York YO10 5DD, United Kingdom}

\author{J. Bonnard\orcidlink{0000-0002-5164-3343}}
\affiliation{School of Physics, Engineering and Technology, University of York, Heslington, York YO10 5DD, United Kingdom}

% \author{A. Pastore}
% \affiliation{School of Physics, Engineering and Technology, University of York, Heslington, York YO10 5DD, United Kingdom}

\author{X. Sun\orcidlink{0000-0002-0130-6269}}
\affiliation{School of Physics, Engineering and Technology, University of York, Heslington, York YO10 5DD, United Kingdom}

\date{\today}

\begin{abstract}
\protect\myabstract
\end{abstract}

\maketitle

{\it Introduction}---Since the discovery of the nearly degenerate excited and ground states of $^{229}$Th fifty years ago, many efforts have been undertaken by both experimental and theoretical communities to understand and characterize this unusual emergent feature. To date, it is the lowest excited state in the entire nuclear chart, making it a distinctive isomer of interest for experiments, theory, and potential technological applications~\cite{(He07),PhysRevC.79.064303, Thirolf_2019, l29n-gt5j, PhysRevX.15.021055,(Bee25),HE2008117,arakawa2026probingultralightdarkmatter,Athanasakis-Kaklamanakis2025}. Such an unusual property also makes it a difficult case to study, testing the limits of experimental and theoretical methods.

Its first detection comes from the $\gamma$-spectrum of the $\alpha$-decaying $^{233}$U, where an excited state with energy below 100 eV was proposed to explain its decay scheme~\cite{KROGER197629}. Several subsequent spectroscopic studies helped clarify the excited state and constrained further its energy~\cite{PhysRevC.49.1845, E.Peik_2003, PhysRevC.73.044326, MATINYAN1998199,GULDA200245}. Later, direct detection of internal conversion electrons allowed for a more precise characterization of the isomeric state~\cite{vonderWense2016}. Since then, alternative production channels and measurement techniques have been developed to accurately determine the values of its observables, including electromagnetic moments~\cite{Thielking2018, Moritz2025-eh, PhysRevC.100.024315, Kraemer2023, PhysRevA.111.L050802, xu2025, PhysRevResearch.7.013052}, investigations of internal conversion processes in excited electronic configurations~\cite{PhysRevA.95.032503} and nuclear hyperfine mixing~\cite{8jhh-ktfy, PhysRevA.108.062813}.

Because of the low energy of the radiative transition, the electronic environment strongly influences its properties, making lifetime measurements depend on specific experimental setups and requiring corrections to estimate the corresponding vacuum transition rates. The latest results~\cite{PhysRevLett.132.182501} report a transition energy of $E_\gamma=8.35574(3)$\,eV, a half-life in vacuum of $t_{1/2}=1740(50)$\,s, and a corresponding reduced transition probability of B(M1)=0.0217(6)\,W.u.$_{\text{M1}}=0.0388(12)\,\mu_N^2$~\footnote{\label{footnote_wu}Uncertainties calculated from error propagation of the measurements of $t_{1/2}$ and $E_\gamma$}, marking a significant improvement in precision compared to previous studies~\cite{Thirolf2024}. Note that $1\,\text{W.u.}_{\text{M1}} = 1.7905\,\mu_N^2$.

The electromagnetic moments have been measured for both the 3/2$^+$ isomeric and 5/2$^+$ ground states~\cite{Gerstenkorn1974,PhysRevLett.106.223001, PhysRevA.88.060501, PhysRevLett.127.253001}. The latest measurements of spectroscopic electric quadrupole moments are $Q(3/2^+)=1.77(1)$\,b and $Q(5/2^+)=3.11(2)$\,b~\cite{(Bee25)}. For the magnetic dipole moments, the latest reports are $\mu(3/2^+)= -0.378(8)\,\mu_N$ and $\mu(5/2^+)=0.366(6)\,\mu_N$~\cite{Yamaguchi2024}. The deviation from the Schmidt limits~\cite{Schmidt1937} $\mu_{\text{Sch.}}(5/2)=[1.366,-1.913]\,\mu_N$ and $\mu_{\text{Sch.}}(3/2)=[1.148,-1.913]\,\mu_N$ indicate a strong core polarization effect in this isotope. Magnetic octupole moments are yet to be measured in the actinide region, where
for $\langle r^2\rangle$=33.13(3)\,fm$^2$ in $^{229}$Th~\cite{(Ang13a)}, the Schwartz limits~\cite{PhysRev.97.380} are $\Omega_\text{Sch.}(5/2)= [-0.543,0.181]\,\mu_N$\,b and  $\Omega_\text{Sch.}(3/2)=[-0.380,0.054]\,\mu_N$\,b.

The advances in the experimental field have been motivated by prospective technological applications in areas like chronometry~\cite{PhysRevLett.108.120802,laser_proposal,girvin2025prospectssolidstatenuclearclock,Masuda2019, cw5h-644b}, $\gamma$-ray laser technology~\cite{Ginzburg2004,PhysRevLett.106.162501}, satellite-based navigational and chronological geodesy systems~\cite{Thirolf_2019}. These developments have the potential to enhance a new generation of ultra-precise techniques that will eventually also help advance scientific research~\cite{PhysRevX.15.021055,PhysRevC.79.064303,(Bee25)}. However, such developments depend on precise control of the excitation and de-excitation processes between the $3/2^+$ state and the $5/2^+$ ground state, where M1 multipolarity greatly exceeds the competing E2 by about $10^{13}$ orders of magnitude due to the extremely low energy of the transition~\cite{PhysRevC.92.054324}. Recent advancements have achieved the first optical-controlled laser excitation in doped crystals of Th:CaF$_2$ and Th:LiSrAlF$_6$ along with new proposals being currently developed, demonstrating that these applications are feasible~\cite{wang2026enhancedyieldratetextsuperscript229mth, PhysRevResearch.7.L022036,PhysRevLett.132.182501,Zhang2024, PhysRevLett.133.013201}.

Theoretical efforts to understand and predict the observables of this elusive isomeric transition have been done in the frameworks of phenomenological~\cite{PhysRevC.92.054324}, core-plus-particle~\cite{PhysRevC.73.044326,PhysRevLett.118.212501,Dykhne1998,GULDA200245,PhysRevLett.122.162502,
PhysRevC.103.014313}, and projected shell models~\cite{(Che25)}. In these approaches, reliance on adjusted parameters, effective charges, or effective $g$ factors was crucial for accurately reproducing observables. Previous parameter-free self-consistent approaches have been implemented without parity breaking, that is, neglecting the octupole deformation~\cite{PhysRevC.79.064303}, time-reversal and signature symmetry breaking, which preclude angular-momentum core polarisation~\cite{PhysRevC.79.064303,zhou2025}, parity and rotational symmetry restoration~\cite{PhysRevC.79.064303,PhysRevC.110.034327}, or configuration interaction~\cite{PhysRevC.79.064303,PhysRevC.110.034327,zhou2025} to obtain low-energy structure, multipole moments, and transition strengths.

In this Letter, we present the first parameter-free analysis based on self-consistent nuclear density functional theory (DFT) that incorporates all previously missing elements and employs the methodology of Refs.~\cite{(Sas22c),(Bon23c)} predating the lifetime measurement~\cite{PhysRevLett.132.182501}; see also Refs.~\cite{(Wib25d),(Dob26b),(Dob25g)}. We apply the configuration-interaction multi-reference DFT (MR-DFT) approach with several Skyrme functionals, mixing two sets of paired, octupole-deformed, and symmetry-restored configurations, one for the ground state and the other for the isomer, and estimate the corresponding B(M1) value and electromagnetic moments. A similar configuration-interaction approach has recently been successfully used to calculate two-neutrino double-beta decay matrix elements~\cite{nkh2-3kfl}.

{\it Method\label{sec:method}}---We conducted a multi-step calculation of the ground and isomeric nuclear wave functions of $^{229}$Th to determine the magnetic dipole transition strength B(M1; $3/2^+_1\rightarrow5/2^+_1)$ and the electromagnetic moments $Q(3/2^+)$, $Q(5/2^+)$, $\mu(3/2^+)$, $\mu(5/2^+)$, $\Omega(3/2^+)$ and $\Omega(5/2^+)$ for several Skyrme functionals with pairing interaction, implemented in the code {\sc HFODD}~\cite{(Dob21f),(Dob26a),(Res26)} version 3.33k. We used the pairing strengths, $V_{0,n}$ and $V_{0,p}$, from Ref.~\cite{Athanasakis-Kaklamanakis2025} that reproduce the mass staggering around $^{229}$Th and $^{227}$Ac, respectively, and the Landau parameters $g_0'$~\cite{(Ben02d),(Idi17)} from Refs.~\cite{(Ben02d),(Sas22c)} that constrain the isovector spin-spin coupling constants; see the Supplemental Material~\cite{supp-229Th}.
We present results using seven Skyrme functionals:
SkX$_{\text{c}}$~\cite{(Bro98c)},
SkM*~\cite{(Bar82c)},
UNEDF1~\cite{(Kor10c)},
SIII~\cite{(Bei75b)},
SkO$^\prime$~\cite{(Rei99)},
SLy4~\cite{(Cha98a)}, and
UNEDF0~\cite{(Kor12b)}.

Since none of these Skyrme functionals was adjusted to data on octupole degrees of freedom, we anticipated systematic disagreement among the calculations. Therefore, following Ref.~\cite{(Dob18a)}, we analyzed correlations between the calculated intrinsic octupole moments $Q^3_0$ and the predicted B(M1) and electromagnetic moments, using the experimental values of $Q^3_0$ as reference points. To this end, we computed the ground-state octupole moments of $^{226}$Ra and $^{230}$Th and used the experimental values~\footnote{We use the same convention for $Q^3_0$ as in Ref.~\protect\cite{(Dob18a)}.} for these isotopes, $Q^3_0(^{226}\text{Ra})=1080(30)$\,e\,fm${^3}$~\cite{(Gaf13), WOLLERSHEIM1993261} and $Q^3_0(^{230}\text{Th})=800(40)$\,e\,fm${^3}$~\cite{PhysRevC.10.1146}.

Our computational approach involved blocking specific odd-neutron axial quasiparticle configurations in $^{229}$Th and relaxing parity constraints. This yielded deformed, symmetry-breaking solutions with non-zero intrinsic octupole moments $Q^3_0$. We then projected these solutions onto good parity and angular momentum and formed a mixed wave function.

As discussed in Ref.~\cite{(Dob25g)} and in references cited therein, the DFT magnetic moments critically depend on the so-called time-odd (TO) terms of the mean field~\cite{(Eng75),(Per04c),(Sch19b)}, which are solely responsible for the angular-momentum polarization of the core. To assess their role in determining the M1 transition rates, we performed all calculations in two variants: (i) with the time-even (TE) terms only and (ii) with both TE and TO terms included. In addition, the TE+TO results were obtained for the isovector spin-spin terms of the functionals, with consistent Landau parameters~\cite{(Ben02d),(Idi17)} either adjusted to experimental magnetic moments~\cite{(Sas22c)} or to the Gamow-Teller strengths~\cite{(Ben02d)}; see the Supplemental Material~\cite{supp-229Th}.

{\it Procedure}---Here, we describe the multi-step process used to determine the results of this study.
First, for each Skyrme functional, we determined the axial-parity-breaking paired ground-state wave functions of the neighboring even-even isotope $^{228}$Th, from which we obtained the corresponding single-particle (s.p.)\ wave functions. Then, for each of the angular-momentum projections along the axis of axial symmetry, $\Omega=5/2$ and $\Omega=3/2$~\footnote{We are bound to keep the traditional notation, denoting by the same symbol $\Omega$ the projections of the angular momentum on the axial symmetry axis and of the magnetic octupole moment.}, we selected three of them near the Fermi surface for blocking~\cite {(Rin80b),(Dob09g),(Ber09d)} in the odd-$N$ isotope $^{229}$Th.

In the second step, the $^{228}$Th s.p.\ wave functions served as tags to self-consistently determine the quasiparticle configurations in $^{229}$Th, as explained in Ref.~\cite{(Wib25d)}.
To establish a common naming convention for those configurations, we use the calculated dominant Nilsson labels~\footnote{The dominant Nilsson labels~\protect\cite{(Dob97c)} correspond to the largest component of a given s.p.\ state when it is expanded on the asymptotic Nilsson states $[N_0n_z\Lambda]\Omega$~\protect\cite{(Rin80b)}.} of the $^{228}$Th tag states, which, following Ref.~\cite{(Dob26b)}, we denote with parentheses instead of brackets. They are (642)3/2, (741)3/2, and (631)3/2 for $\Omega=3/2$ and (633)5/2, (752)5/2, and (622)5/2 for $\Omega=5/2$.
%The standard brackets are reserved for the self-consistent solutions in $^{229}$Th. They can differ from the former in some instances.
Note that we also consider parity-broken configurations with Nilsson labels of $N_0=7$, as they may acquire positive-parity components after parity restoration.

The tagging technology enables tracking of blocked quasiparticle states throughout the self-consistent iterations, regardless of their changing energies and deformations. In this way, we obtained the self-consistent symmetry-broken $^{229}$Th wave functions $|\Phi_{n\Omega}\rangle$ for the three lowest configurations, $n=1,2,3$, of $\Omega=5/2$ and $\Omega=3/2$. In some cases, the Nilsson labels of the blocked self-consistent quasiparticle orbitals can differ from those of the tag states, which may occur when the dominance of a given Nilsson state $[N_0n_z\Lambda]\Omega$ is not very strong, as discussed in Ref.~\cite{(Dob25g)}.

Then, in the third step, we restored broken symmetries~\cite{(She21)} by projecting wave functions $|\Phi_{n\Omega}\rangle$ onto good angular momentum $I$ and positive parity $\pi=+$. We denote the resulting band-head wave functions $(I=|\Omega|$) as $|\Phi_{nI^+\Omega}\rangle$, that is, $|\Phi_{n{5}/{2}^+{5}/{2}}\rangle$ and $|\Phi_{n{3}/{2}^+{3}/{2}}\rangle$. Previous tests have shown~\cite{(Wib25d)} that particle-number restoration does not affect magnetic moments and because it increases computation time by two orders of magnitude, we did not perform it in this study.

In the fourth step, we calculated two sets of $3\times3$ matrix elements of the Hamiltonian $H^{I^+\Omega}_{nn'}=\langle\Phi_{nI^+\Omega}|\hat{H}|\Phi_{n'I^+\Omega}\rangle$ and the overlap $N^{I^+\Omega}_{nn'}=\langle\Phi_{nI^+\Omega}|\Phi_{n'I^+\Omega}\rangle$ for $I=\Omega=5/2$ and $I=\Omega=3/2$, along with the set of $6\times6$ matrix elements of the magnetic dipole $\hat{M}$, $M_{nn'}^{II'}=\langle\Phi_{nI^+I}|\hat{M}|\Phi_{n'I^{'+}I^{'}}\rangle$, magnetic octupole $\hat{\Omega}$, $\Omega_{nn'}^{II'}=\langle\Phi_{nI^+I}|\hat{\Omega}|\Phi_{n'I^{'+}I^{'}}\rangle$, and the electric quadrupole $\hat{Q}$, $Q_{nn'}^{II'}=\langle\Phi_{nI^+I}|\hat{Q}|\Phi_{n'I^{'+}I^{'}}\rangle$. For the Skyrme functionals, the matrix elements of the Hamiltonian were determined using the standard methodology defined in the generator coordinate method~\cite{(Rin80b),(Sch19b)}.

Next, in the fifth step, for $I^\pi=5/2^+$ and $I^\pi=3/2^+$, we solved two configuration-interaction equations (the Hill-Wheeler equations~\cite{(Rin80b),(Sch19b)}),
\begin{equation}
\sum_{n'}H^{I^+\Omega}_{nn'}a_{n'{I^+\Omega}}=E\sum_{n'}N^{I^+\Omega}_{nn'}a_{n'{I^+\Omega}},
\end{equation}
where we considered only the solutions with the lowest energies, $E$. This yielded the mixing coefficients $a_{n'{I^+\Omega}}$ and the normalized mixed wave functions $|\Phi_{I^+\Omega}\rangle=\sum_{n'}a_{n'{I^+\Omega}}|\Phi_{n'I^+\Omega}\rangle$. We note that the states $|\Phi_{n'I^+\Omega}\rangle$ are generally non-orthogonal ($N^{I^+\Omega}_{nn'}\neq\delta_{nn'}$); therefore, the mixing coefficients $a_{n'{I^+\Omega}}$ cannot be interpreted as probability amplitudes.

In the final sixth step, we could then determine the MR-DFT matrix element of the M1 transition between the $I^\pi=5/2^+$ and $I^\pi=3/2^+$ states,
\begin{equation}
    \langle\Phi_{5/2^+5/2}|\hat{M}|\Phi_{3/2^+3/2}\rangle =
    \sum_{nn'}a_{n5/2^+5/2}^* M_{nn'}^{\frac{5}{2}\frac{3}{2}} a_{n'3/2^+3/2},
\end{equation}
and the corresponding reduced transition probability,
\begin{equation}
    \mbox{B(M1}:3/2^+_1\rightarrow 5/2^+_1)=\tfrac{1}{4}|\langle\Phi_{5/2^+5/2}|\hat{M}|\Phi_{3/2^+3/2}\rangle|^2.
    \label{mixed_BM1}
\end{equation}
Similarly, we determine the spectroscopic electromagnetic moments of the $I^\pi=5/2^+$ and $I^\pi=3/2^+$ states as follows,
\begin{eqnarray}
    \langle\Phi_{I^+I}|\hat{\mathcal{O}}|\Phi_{I^+I}\rangle &=&
    \sum_{nn'}a_{nI^+I}^* \mathcal{O}_{nn'}^{II} a_{n'I^+I},
    % \langle\Phi_{I^+I}|\hat{\Omega}|\Phi_{I^+I}\rangle &=&
    % \sum_{nn'}a_{nI^+I}^* \Omega_{nn'}^{II} a_{n'I^+I}, \\
    % \langle\Phi_{I^+I}|\hat{Q}|\Phi_{I^+I}\rangle &=&
    % \sum_{nn'}a_{nI^+I}^* Q_{nn'}^{II} a_{n'I^+I}.
\end{eqnarray}
where $\mathcal{O} = M,  \Omega$ or $Q$.

\begin{figure}[h]
\centering\includegraphics[width=0.48\textwidth]{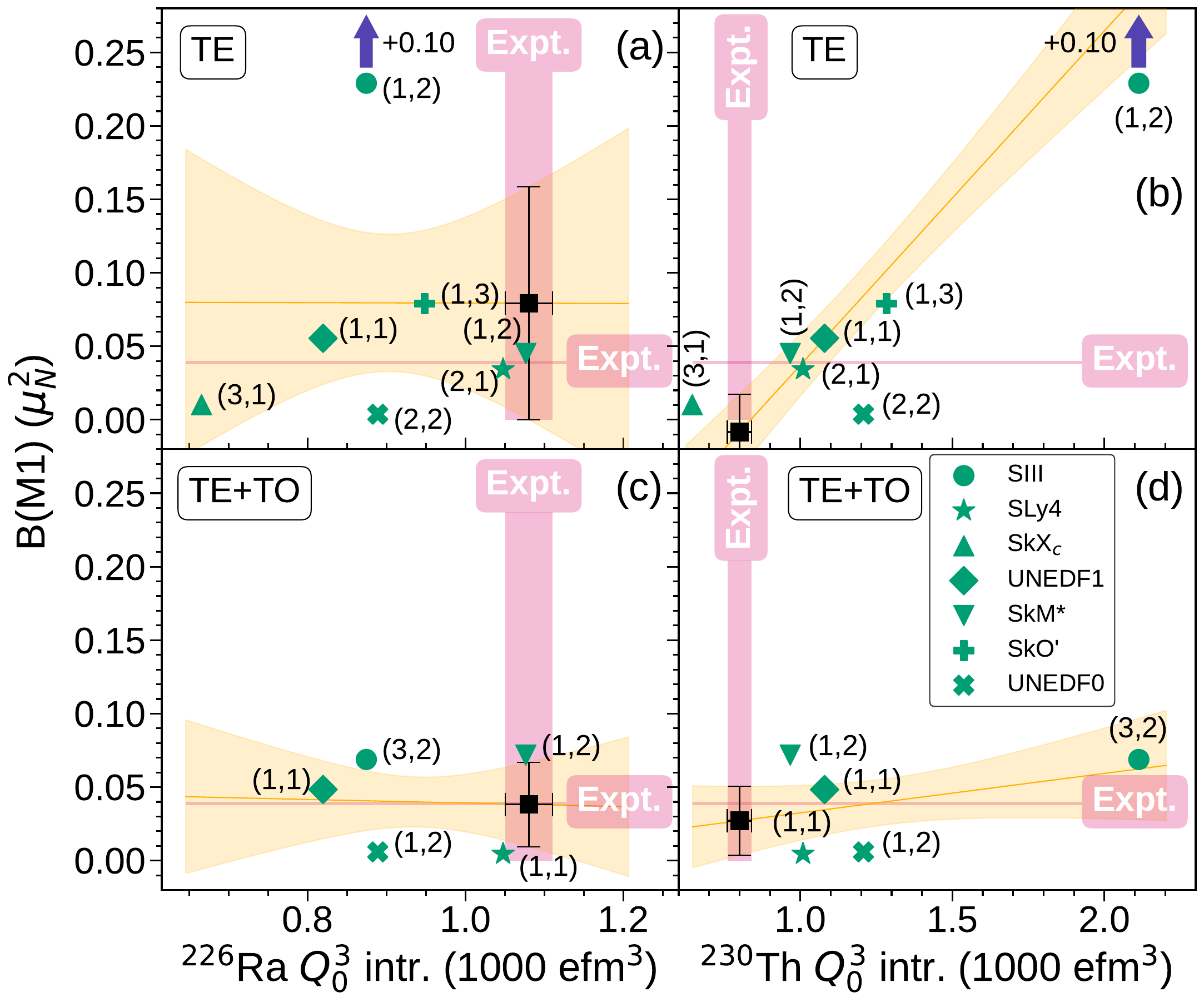}
\caption{B(M1) values with respect to the intrinsic octupole moments $Q^3_0$ of $^{226}$Ra (left panels) and $^{230}$Th (right panels), determined for functionals reduced to the time-even mean fields (TE, upper panels) and with complete time-even and time-odd mean fields (TE+TO, lower panels). The numbers in parentheses indicate the number of successfully mixed wave functions for the two configurations, namely $(5/2^+,3/2^+)$. Vertical and horizontal stripes labelled Expt.\ are the experimental values of $Q^3_0$ and B(M1), respectively. Thick lines and shaded bands denote the regression results and their uncertainties, respectively~\protect\cite{(Dob18a)}. Arrows denote shifted points outside the scale of the figure.\label{Corr1a}}
\end{figure}
{\it Results}---Figure~\ref{Corr1a} presents our study's principal findings. It compares the TE B(M1) values (upper panels) with the TE+TO values (lower panels) derived from the measured intrinsic octupole moments in $^{226}$Ra (left panels) and $^{230}$Th (right panels). In each case, results for different Skyrme functionals were analyzed using the linear regression method of Ref.~\cite{(Dob18a)} and extrapolated, with uncertainties, to the measured points.

Encouragingly, the TE+TO results extrapolated to the experimental data for $^{226}$Ra and $^{230}$Th are consistent with each other, yielding B(M1)=0.04(3) and 0.03(2)\,$\mu_N^2$, respectively. It is also gratifying that both agree with the experimental value of 0.0388(12)\,$\mu_N^2$ ~\cite{PhysRevLett.132.182501} (uncertainties calculated from error propagation). However, the large spread in results obtained with different Skyrme functionals leads to a theoretical uncertainty that is much larger than the experimental uncertainty.

The TE results, characterized by large outlier SIII values, read B(M1)=0.08(8) and $-$0.01(3)\,$\mu_N^2$, respectively. They are less compatible with experiment, less consistent with one another, and may exhibit greater uncertainty. The current results thus provide some indication of the importance of time-odd mean fields in describing the B(M1) value in $^{229}$Th.

The main challenge in obtaining the results shown in Fig.~\ref{Corr1a} is that the Hamiltonian mixing matrix elements $H^{I^+\Omega}_{nn'}$ are often singular due to non-zero self-interaction and self-pairing energies that characterize Skyrme functionals (see discussion in Ref.~\cite{(She21)}). This problem has not been satisfactorily resolved yet, although the search for a solution continues~\cite{(Sad13),(Sad13b),(Ben17)}. In this work, we removed all singular cases from the analysis (see the Supplemental Material~\cite{supp-229Th}), and the number retained is shown in parentheses in Fig.~\ref{Corr1a}.

As shown in Fig.~\ref{Corr2a}, configuration interaction has a limited effect on the B(M1) values. The figure shows values calculated for only one 3/2$^+$ configuration [631]3/2 and one 5/2$^+$ configuration [633]5/2. For the TE+TO option, the results are nearly identical to those obtained with the mixed configurations. However, we had to discard all points that did not converge properly; see the Supplemental Material~\cite{supp-229Th}.

\begin{figure}[t]
     \centering
     \includegraphics[width=0.48\textwidth]{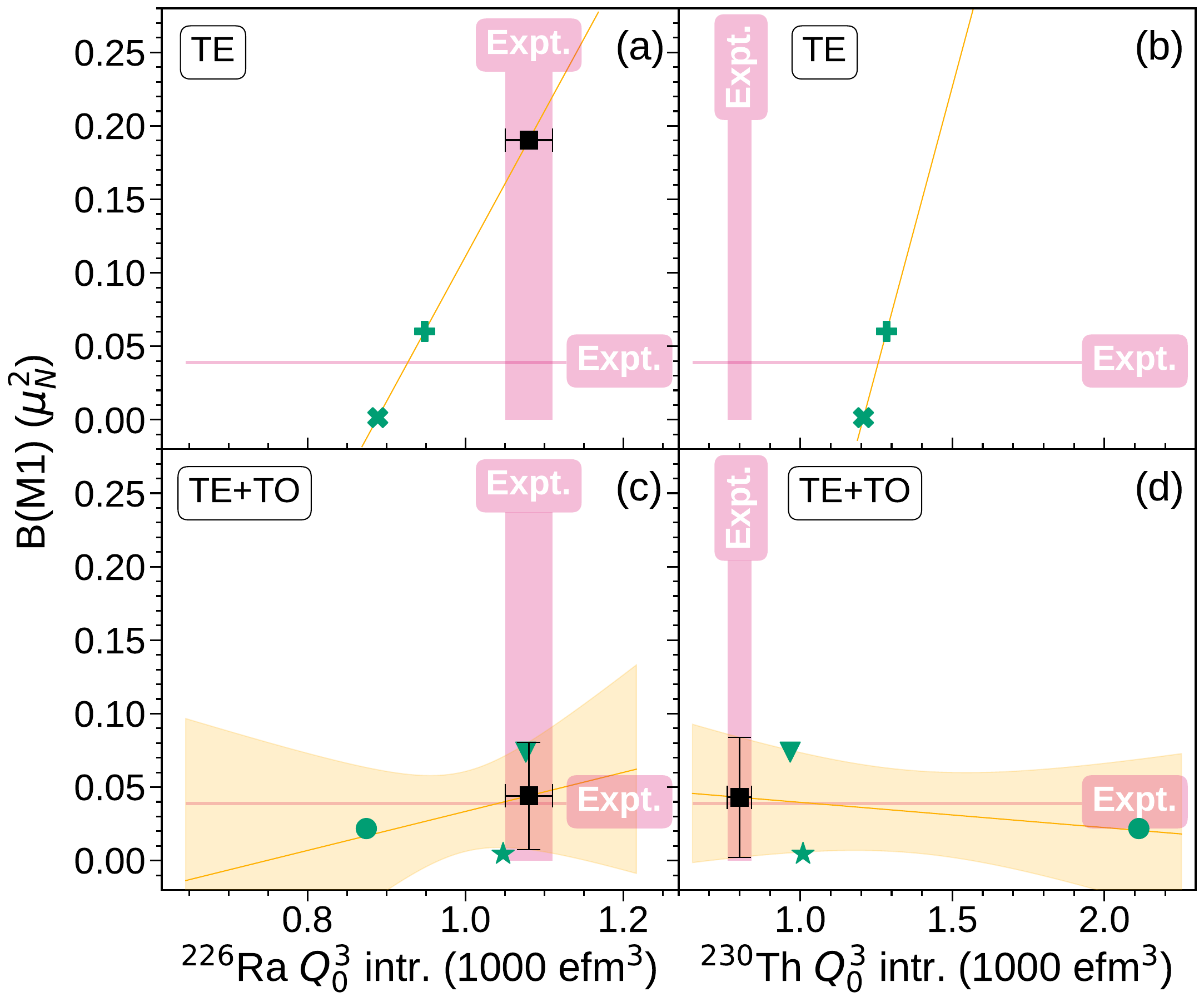}
\caption{Same as in Fig.~\protect\ref{Corr1a} but for the B(M1) values calculated between converged individual (not mixed) blocked self-consistent states [631]3/2 and [633]5/2.\label{Corr2a}}
\end{figure}

\begin{figure}[t]
     \centering
     \includegraphics[width=0.48\textwidth]{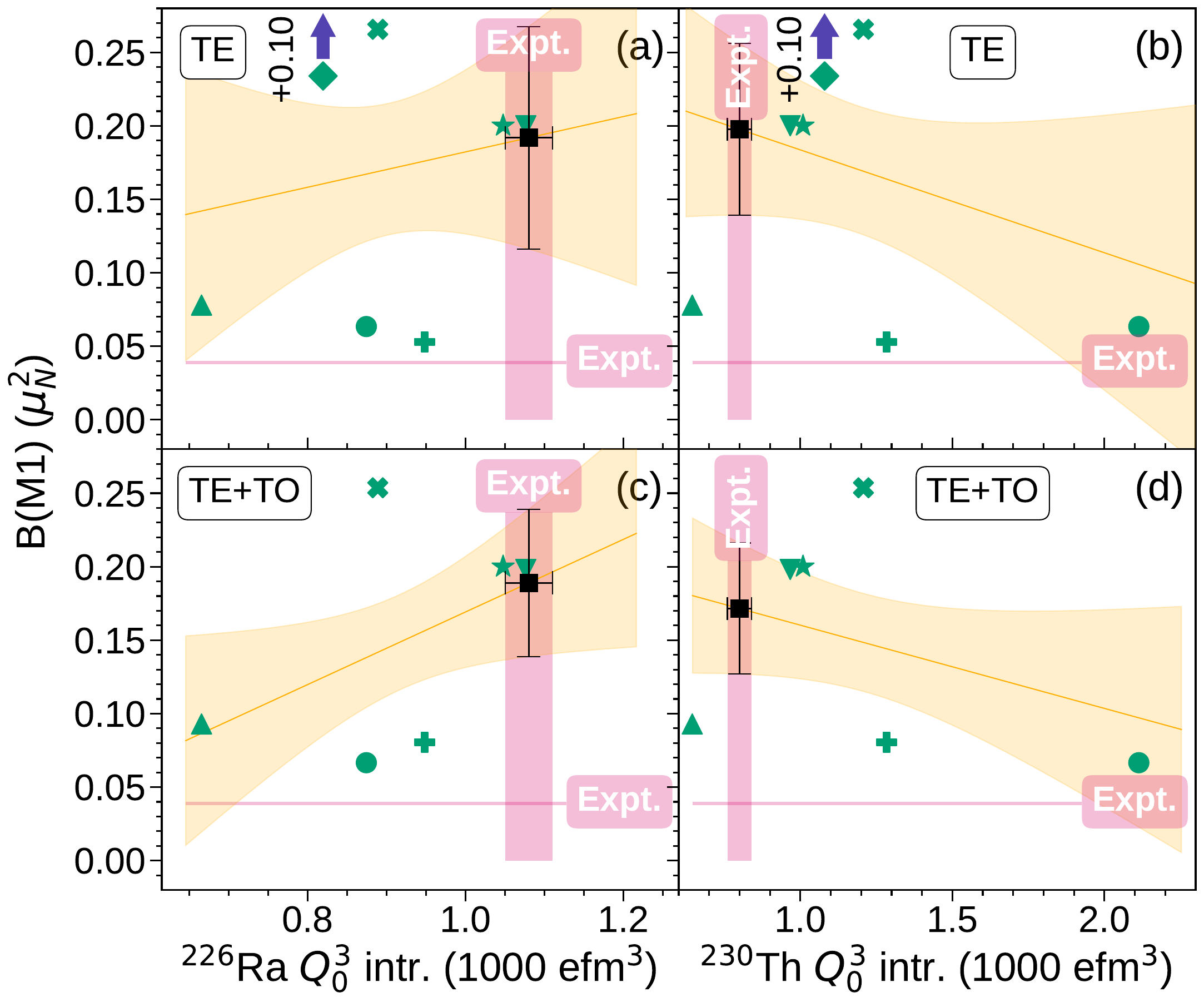}
\caption{Same as in Fig.~\protect\ref{Corr2a} but for the B(M1) values calculated between parity-conserving states.\label{Corr3a}}
\end{figure}

Finally, in Fig.~\ref{Corr3a}, we present results obtained by preserving the parity of the [631]3/2 and [633]5/2 states in $^{229}$Th. Neglecting the octupole deformation leads to a significant discrepancy with the experimental B(M1) value and is therefore unacceptable.
\begin{table*}[t!]
\caption{Same as in Fig.~\protect\ref{Summary} but for the weighted averages of the results determined for the $^{226}$Ra and $^{230}$Th mixed and TE+TO regressions. Columns $\Delta$  and $\Delta$ (\%) show differences and percentage differences, respectively, between DFT and experiment.}
\label{table:weighted}
%\vspace*{2mm}
\centering
\begin{tabular}{l c l l l l}
\hline
$^{229}$Th                                         & $I^\pi_1$                    & \hspace{4mm}DFT & ~~\hspace{2mm}Expt.         & $\hspace{6mm}\Delta$  & $\hspace{2mm}\Delta$ (\%)   \\
\hline B(M1) ($\mu_N^2$)                           & $3/2^+_1\rightarrow 5/2^+_1$ &  \ph0.03(2)   &  \ph0.0388(12)  &   $-$0.006(17)  &    $-$20(40)    \\
\hline \multirow{2}{*}{$\mu$  ($\mu_N$)}         & $5/2^+_1$                    &  \ph0.3(3)   &  \ph0.366(6)    &    $-$0.10(17)   &    $-$30(50)            \\
                                                   & $3/2^+_1$                    &  $-$0.18(7)   &  $-$0.378(8)    &   \ph0.19(7)    &    $-$50(20)            \\
\hline \multirow{2}{*}{$Q$ (b)}                    & $5/2^+_1$                    &  \ph2.98(5)   &  \ph3.11(2)     &   $-$0.13(5)    &    $-$4(2)             \\
                                                   & $3/2^+_1$                    &  \ph1.728(12)   &  \ph1.77(1)   &   $-$0.042(15)  &    $-$2.4(9)             \\
\hline \multirow{2}{*}{$\Omega$ ($\mu_N$\,b)}      & $5/2^+_1$                    &  $-$0.054(10)   &   ~~~~---     &    ~~~~---     &    ~~~~---           \\
                                                   & $3/2^+_1$                    &  \ph0.129(6)   &    ~~~~---     &    ~~~~---     &    ~~~~---           \\
\hline
\end{tabular}
\end{table*}

Figure~\ref{Summary}(a) compares all the B(M1) results discussed above with the experimental value. To benchmark our methods against other experimentally known electromagnetic observables in $^{229}$Th, panels (b)--(e) of Fig.~\ref{Summary} summarize results for the magnetic dipole moments, plotted on an absolute scale in panels (b) and (c), and the electric quadrupole moments, plotted on a percentage-deviation scale, $\Delta{Q}=(Q_{\text{DFT}}-Q_{\text{Expt}})/Q_{\text{Expt}}$, in panels (d) and (e). In panels (f) and (g), we show the DFT predictions for yet-unmeasured magnetic octupole moments.

\begin{figure}[t]
     \centering
     \includegraphics[width=0.48\textwidth]{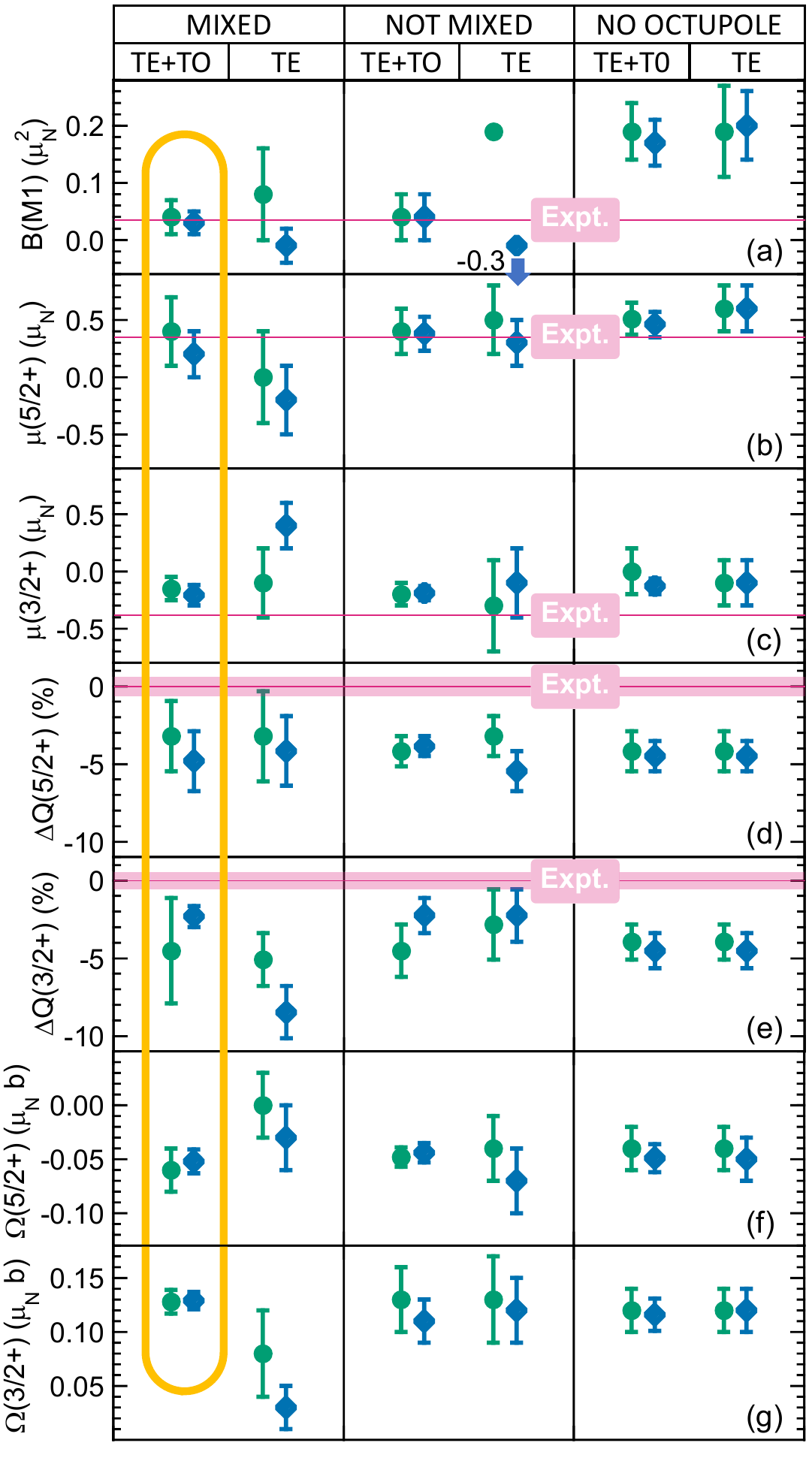}
\caption{Summary results obtained in this Letter, compared with the experimental data~\cite{PhysRevLett.132.182501,(Bee25),Yamaguchi2024}. The shown values are: reduced transition probabilities B(M1) (a), magnetic dipole moments $\mu$ of the 5/2$^+$ (b) and 3/2$^+$ (c) states, percentage deviations from the data $\Delta{}Q$ of the electric quadrupole moments of the of the 5/2$^+$ (d) and 3/2$^+$ (e) states, and magnetic octupole moments $\Omega$ of the 5/2$^+$ (f) and 3/2$^+$ (g) states. Circles and diamonds correspond to the regressions aligned with the measured octupole moments of $^{226}$Ra and $^{230}$Th, respectively.
\label{Summary}}
\end{figure}

We observe that the TE+TO moments of both the 5/2$^+$ and 3/2$^+$ states do not strongly depend on the version of our calculation; however, this is not the case for the mixed magnetic moments obtained by omitting the TO terms. Additionally, results obtained for individual functionals may depend significantly on the inclusion of TO terms; see the Supplemental Material~\cite{supp-229Th}. All results are closest to the data for the most advanced version of calculation, marked by the oval in Fig.~\ref{Summary}. In Table~\ref{table:weighted}, we list the weighted averages of the results determined for the $^{226}$Ra and $^{230}$Th regressions.

The magnetic moment of the 5/2$^+$ ground state is well reproduced, whereas that of the isomer 3/2$^+$ is underestimated by about a factor of two. Nonetheless, this discrepancy does not hinder the accurate replication of the B(M1) value in panel (a). The quadrupole moments of both states are reproduced with high accuracy; the calculations agree with the data to within a few percent. The magnetic octupole moments are predicted with relatively small uncertainties. They appear to be weakly affected by differences in the electric octupole moments. Numerical results and plots showing the regression analyses for the electromagnetic moments are included in the Supplemental Material~\cite{supp-229Th}.

{\it Conclusions}---We presented the first application of nuclear DFT to the challenging problem of describing the electromagnetic properties of $^{229}$Th, including the decay rate of its low-energy isomer. We addressed three fundamental questions in describing these properties: the role of configuration interaction, time-odd core polarization, and octupole correlations within a fully symmetry-conserving approach. Since the configurations of the isomer and ground state can be confidently based on two distinct deformed neutron orbitals, the main physics aspects of the problem concern details of their structure. Our results indicate that the role of configuration interaction is probably small, that of time-odd polarization is significant, and that including the effects of octupole correlations is fundamental. We reached these conclusions within a theory that allows symmetry breaking while fully respecting symmetry restoration.

The consistent description of the data obtained without parameter adjustments is gratifying. Still, the crucial need to properly account for octupole correlations points to the most essential way to improve the approach's predictive power: fine-tuning the functional’s octupole polarizability within global adjustments to experimental data. In practice, there is also a fundamental drawback to the existing functionals, which all struggle with self-interactions and self-pairing, rendering many applications erratic and unacceptable. An intensified effort in these two directions is strongly warranted.

\bigskip
\begin{acknowledgments}
We thank Herlik Wibowo for fruitful discussions and Pierre Becker and Alessandro Pastore for their early involvement in the project.
COLFUTURO  financially supported ARG.
This work was partially supported by the STFC Grant Nos.~ST/V001035/1 and~ST/Y000285/1, and by a Leverhulme Trust Research Project Grant.
We thank the CSC-IT Center for Science Ltd., Finland, and the IFT Computer Center at the University of Warsaw, Poland, for the allocation of computational resources. This project was partly undertaken on the Viking Cluster, a high-performance compute facility provided by the University of York. We are grateful for computational support from the University of York High Performance Computing service, Viking, and the Research Computing team.
We thank Grammarly for its support with English writing.
\end{acknowledgments}

{\it Data availability}---The data that support the findings of this article are openly available~\cite{rep-229Th}.

%\bibliography{229Th,jd.251119,jacwit42}

\begin{thebibliography}{95}%
\makeatletter
\providecommand \@ifxundefined [1]{%
 \@ifx{#1\undefined}
}%
\providecommand \@ifnum [1]{%
 \ifnum #1\expandafter \@firstoftwo
 \else \expandafter \@secondoftwo
 \fi
}%
\providecommand \@ifx [1]{%
 \ifx #1\expandafter \@firstoftwo
 \else \expandafter \@secondoftwo
 \fi
}%
\providecommand \natexlab [1]{#1}%
\providecommand \enquote  [1]{``#1''}%
\providecommand \bibnamefont  [1]{#1}%
\providecommand \bibfnamefont [1]{#1}%
\providecommand \citenamefont [1]{#1}%
\providecommand \href@noop [0]{\@secondoftwo}%
\providecommand \href [0]{\begingroup \@sanitize@url \@href}%
\providecommand \@href[1]{\@@startlink{#1}\@@href}%
\providecommand \@@href[1]{\endgroup#1\@@endlink}%
\providecommand \@sanitize@url [0]{\catcode `\\12\catcode `\$12\catcode `\&12\catcode `\#12\catcode `\^12\catcode `\_12\catcode `\%12\relax}%
\providecommand \@@startlink[1]{}%
\providecommand \@@endlink[0]{}%
\providecommand \url  [0]{\begingroup\@sanitize@url \@url }%
\providecommand \@url [1]{\endgroup\@href {#1}{\urlprefix }}%
\providecommand \urlprefix  [0]{URL }%
\providecommand \Eprint [0]{\href }%
\providecommand \doibase [0]{https://doi.org/}%
\providecommand \selectlanguage [0]{\@gobble}%
\providecommand \bibinfo  [0]{\@secondoftwo}%
\providecommand \bibfield  [0]{\@secondoftwo}%
\providecommand \translation [1]{[#1]}%
\providecommand \BibitemOpen [0]{}%
\providecommand \bibitemStop [0]{}%
\providecommand \bibitemNoStop [0]{.\EOS\space}%
\providecommand \EOS [0]{\spacefactor3000\relax}%
\providecommand \BibitemShut  [1]{\csname bibitem#1\endcsname}%
\let\auto@bib@innerbib\@empty
%</preamble>
\bibitem [{\citenamefont {He}\ and\ \citenamefont {Ren}(2007)}]{(He07)}%
  \BibitemOpen
  \bibfield  {author} {\bibinfo {author} {\bibfnamefont {X.-t.}\ \bibnamefont {He}}\ and\ \bibinfo {author} {\bibfnamefont {Z.-z.}\ \bibnamefont {Ren}},\ }\bibfield  {title} {\bibinfo {title} {Enhanced sensitivity to variation of fundamental constants in the transitions of {$^{229}$Th} and {$^{249}$Bk}},\ }\href {https://doi.org/10.1088/0954-3899/34/7/003} {\bibfield  {journal} {\bibinfo  {journal} {Journal of Physics G: Nuclear and Particle Physics}\ }\textbf {\bibinfo {volume} {34}},\ \bibinfo {pages} {1611} (\bibinfo {year} {2007})}\BibitemShut {NoStop}%
\bibitem [{\citenamefont {Litvinova}\ \emph {et~al.}(2009)\citenamefont {Litvinova}, \citenamefont {Feldmeier}, \citenamefont {Dobaczewski},\ and\ \citenamefont {Flambaum}}]{PhysRevC.79.064303}%
  \BibitemOpen
  \bibfield  {author} {\bibinfo {author} {\bibfnamefont {E.}~\bibnamefont {Litvinova}}, \bibinfo {author} {\bibfnamefont {H.}~\bibnamefont {Feldmeier}}, \bibinfo {author} {\bibfnamefont {J.}~\bibnamefont {Dobaczewski}},\ and\ \bibinfo {author} {\bibfnamefont {V.}~\bibnamefont {Flambaum}},\ }\bibfield  {title} {\bibinfo {title} {Nuclear structure of lowest $^{229}\mathrm{Th}$ states and time-dependent fundamental constants},\ }\href {https://doi.org/10.1103/PhysRevC.79.064303} {\bibfield  {journal} {\bibinfo  {journal} {Phys. Rev. C}\ }\textbf {\bibinfo {volume} {79}},\ \bibinfo {pages} {064303} (\bibinfo {year} {2009})}\BibitemShut {NoStop}%
\bibitem [{\citenamefont {Thirolf}\ \emph {et~al.}(2019)\citenamefont {Thirolf}, \citenamefont {Seiferle},\ and\ \citenamefont {von~der Wense}}]{Thirolf_2019}%
  \BibitemOpen
  \bibfield  {author} {\bibinfo {author} {\bibfnamefont {P.~G.}\ \bibnamefont {Thirolf}}, \bibinfo {author} {\bibfnamefont {B.}~\bibnamefont {Seiferle}},\ and\ \bibinfo {author} {\bibfnamefont {L.}~\bibnamefont {von~der Wense}},\ }\bibfield  {title} {\bibinfo {title} {The 229-thorium isomer: doorway to the road from the atomic clock to the nuclear clock},\ }\href {https://doi.org/10.1088/1361-6455/ab29b8} {\bibfield  {journal} {\bibinfo  {journal} {Journal of Physics B: Atomic, Molecular and Optical Physics}\ }\textbf {\bibinfo {volume} {52}},\ \bibinfo {pages} {203001} (\bibinfo {year} {2019})}\BibitemShut {NoStop}%
\bibitem [{\citenamefont {Caputo}\ \emph {et~al.}(2025)\citenamefont {Caputo}, \citenamefont {Gazit}, \citenamefont {Hammer}, \citenamefont {Kopp}, \citenamefont {Paz}, \citenamefont {Perez},\ and\ \citenamefont {Springmann}}]{l29n-gt5j}%
  \BibitemOpen
  \bibfield  {author} {\bibinfo {author} {\bibfnamefont {A.}~\bibnamefont {Caputo}}, \bibinfo {author} {\bibfnamefont {D.}~\bibnamefont {Gazit}}, \bibinfo {author} {\bibfnamefont {H.-W.}\ \bibnamefont {Hammer}}, \bibinfo {author} {\bibfnamefont {J.}~\bibnamefont {Kopp}}, \bibinfo {author} {\bibfnamefont {G.}~\bibnamefont {Paz}}, \bibinfo {author} {\bibfnamefont {G.}~\bibnamefont {Perez}},\ and\ \bibinfo {author} {\bibfnamefont {K.}~\bibnamefont {Springmann}},\ }\bibfield  {title} {\bibinfo {title} {Sensitivity of nuclear clocks to new physics},\ }\href {https://doi.org/10.1103/l29n-gt5j} {\bibfield  {journal} {\bibinfo  {journal} {Phys. Rev. C}\ }\textbf {\bibinfo {volume} {112}},\ \bibinfo {pages} {L031302} (\bibinfo {year} {2025})}\BibitemShut {NoStop}%
\bibitem [{\citenamefont {Fuchs}\ \emph {et~al.}(2025)\citenamefont {Fuchs}, \citenamefont {Kirk}, \citenamefont {Madge}, \citenamefont {Paranjape}, \citenamefont {Peik}, \citenamefont {Perez}, \citenamefont {Ratzinger},\ and\ \citenamefont {Tiedau}}]{PhysRevX.15.021055}%
  \BibitemOpen
  \bibfield  {author} {\bibinfo {author} {\bibfnamefont {E.}~\bibnamefont {Fuchs}}, \bibinfo {author} {\bibfnamefont {F.}~\bibnamefont {Kirk}}, \bibinfo {author} {\bibfnamefont {E.}~\bibnamefont {Madge}}, \bibinfo {author} {\bibfnamefont {C.}~\bibnamefont {Paranjape}}, \bibinfo {author} {\bibfnamefont {E.}~\bibnamefont {Peik}}, \bibinfo {author} {\bibfnamefont {G.}~\bibnamefont {Perez}}, \bibinfo {author} {\bibfnamefont {W.}~\bibnamefont {Ratzinger}},\ and\ \bibinfo {author} {\bibfnamefont {J.}~\bibnamefont {Tiedau}},\ }\bibfield  {title} {\bibinfo {title} {Searching for dark matter with the $^{229}\mathrm{Th}$ nuclear lineshape from laser spectroscopy},\ }\href {https://doi.org/10.1103/PhysRevX.15.021055} {\bibfield  {journal} {\bibinfo  {journal} {Phys. Rev. X}\ }\textbf {\bibinfo {volume} {15}},\ \bibinfo {pages} {021055} (\bibinfo {year} {2025})}\BibitemShut {NoStop}%
\bibitem [{\citenamefont {Beeks}\ \emph {et~al.}(2025)\citenamefont {Beeks}, \citenamefont {Kazakov}, \citenamefont {Schaden}, \citenamefont {Morawetz}, \citenamefont {Toscani De~Col}, \citenamefont {Riebner}, \citenamefont {Bartokos}, \citenamefont {Sikorsky}, \citenamefont {Schumm}, \citenamefont {Zhang}, \citenamefont {Ooi}, \citenamefont {Higgins}, \citenamefont {Doyle}, \citenamefont {Ye},\ and\ \citenamefont {Safronova}}]{(Bee25)}%
  \BibitemOpen
  \bibfield  {author} {\bibinfo {author} {\bibfnamefont {K.}~\bibnamefont {Beeks}}, \bibinfo {author} {\bibfnamefont {G.~A.}\ \bibnamefont {Kazakov}}, \bibinfo {author} {\bibfnamefont {F.}~\bibnamefont {Schaden}}, \bibinfo {author} {\bibfnamefont {I.}~\bibnamefont {Morawetz}}, \bibinfo {author} {\bibfnamefont {L.}~\bibnamefont {Toscani De~Col}}, \bibinfo {author} {\bibfnamefont {T.}~\bibnamefont {Riebner}}, \bibinfo {author} {\bibfnamefont {M.}~\bibnamefont {Bartokos}}, \bibinfo {author} {\bibfnamefont {T.}~\bibnamefont {Sikorsky}}, \bibinfo {author} {\bibfnamefont {T.}~\bibnamefont {Schumm}}, \bibinfo {author} {\bibfnamefont {C.}~\bibnamefont {Zhang}}, \bibinfo {author} {\bibfnamefont {T.}~\bibnamefont {Ooi}}, \bibinfo {author} {\bibfnamefont {J.~S.}\ \bibnamefont {Higgins}}, \bibinfo {author} {\bibfnamefont {J.~F.}\ \bibnamefont {Doyle}}, \bibinfo {author} {\bibfnamefont {J.}~\bibnamefont {Ye}},\ and\ \bibinfo {author} {\bibfnamefont {M.~S.}\ \bibnamefont {Safronova}},\ }\bibfield  {title} {\bibinfo {title}
  {Fine-structure constant sensitivity of the {Th-229} nuclear clock transition},\ }\href {https://doi.org/10.1038/s41467-025-64191-7} {\bibfield  {journal} {\bibinfo  {journal} {Nature Communications}\ }\textbf {\bibinfo {volume} {16}},\ \bibinfo {pages} {9147} (\bibinfo {year} {2025})}\BibitemShut {NoStop}%
\bibitem [{\citenamefont {tao He}\ and\ \citenamefont {zhou Ren}(2008)}]{HE2008117}%
  \BibitemOpen
  \bibfield  {author} {\bibinfo {author} {\bibfnamefont {X.}~\bibnamefont {tao He}}\ and\ \bibinfo {author} {\bibfnamefont {Z.}~\bibnamefont {zhou Ren}},\ }\bibfield  {title} {\bibinfo {title} {Temporal variation of the fine structure constant and the strong interaction parameter in the {$^{229}$Th} transition},\ }\href {https://doi.org/https://doi.org/10.1016/j.nuclphysa.2008.03.002} {\bibfield  {journal} {\bibinfo  {journal} {Nuclear Physics A}\ }\textbf {\bibinfo {volume} {806}},\ \bibinfo {pages} {117} (\bibinfo {year} {2008})}\BibitemShut {NoStop}%
\bibitem [{\citenamefont {Arakawa}\ \emph {et~al.}(2026)\citenamefont {Arakawa}, \citenamefont {Doyle}, \citenamefont {Fuchs}, \citenamefont {Higgins}, \citenamefont {Kirk}, \citenamefont {Li}, \citenamefont {Ooi}, \citenamefont {Perez}, \citenamefont {Ratzinger}, \citenamefont {Safronova}, \citenamefont {Schumm}, \citenamefont {Ye},\ and\ \citenamefont {Zhang}}]{arakawa2026probingultralightdarkmatter}%
  \BibitemOpen
  \bibfield  {author} {\bibinfo {author} {\bibfnamefont {J.}~\bibnamefont {Arakawa}}, \bibinfo {author} {\bibfnamefont {J.~F.}\ \bibnamefont {Doyle}}, \bibinfo {author} {\bibfnamefont {E.}~\bibnamefont {Fuchs}}, \bibinfo {author} {\bibfnamefont {J.~S.}\ \bibnamefont {Higgins}}, \bibinfo {author} {\bibfnamefont {F.}~\bibnamefont {Kirk}}, \bibinfo {author} {\bibfnamefont {K.}~\bibnamefont {Li}}, \bibinfo {author} {\bibfnamefont {T.}~\bibnamefont {Ooi}}, \bibinfo {author} {\bibfnamefont {G.}~\bibnamefont {Perez}}, \bibinfo {author} {\bibfnamefont {W.}~\bibnamefont {Ratzinger}}, \bibinfo {author} {\bibfnamefont {M.~S.}\ \bibnamefont {Safronova}}, \bibinfo {author} {\bibfnamefont {T.}~\bibnamefont {Schumm}}, \bibinfo {author} {\bibfnamefont {J.}~\bibnamefont {Ye}},\ and\ \bibinfo {author} {\bibfnamefont {C.}~\bibnamefont {Zhang}},\ }\href {https://arxiv.org/abs/2602.16804} {\bibinfo {title} {Probing ultralight dark matter at the {Mega-Planck} scale with the thorium nuclear clock}} (\bibinfo {year} {2026}),\
  \Eprint {https://arxiv.org/abs/2602.16804} {arXiv:2602.16804 [hep-ph]} \BibitemShut {NoStop}%
\bibitem [{\citenamefont {Athanasakis-Kaklamanakis}\ \emph {et~al.}(2025)\citenamefont {Athanasakis-Kaklamanakis}, \citenamefont {Au}, \citenamefont {Kyuberis}, \citenamefont {Z{\"u}lch}, \citenamefont {Gaul}, \citenamefont {Wibowo}, \citenamefont {Skripnikov}, \citenamefont {Lalanne}, \citenamefont {Reilly}, \citenamefont {Koszor{\'u}s}, \citenamefont {Bara}, \citenamefont {Ballof}, \citenamefont {Berger}, \citenamefont {Bernerd}, \citenamefont {Borschevsky}, \citenamefont {Breier}, \citenamefont {Chrysalidis}, \citenamefont {Cocolios}, \citenamefont {de~Groote}, \citenamefont {Dorne}, \citenamefont {Dobaczewski}, \citenamefont {Fajardo~Zambrano}, \citenamefont {Flanagan}, \citenamefont {Franchoo}, \citenamefont {Johnson}, \citenamefont {Garcia~Ruiz}, \citenamefont {Hanstorp}, \citenamefont {Kujanp{\"a}{\"a}}, \citenamefont {Liu}, \citenamefont {Lynch}, \citenamefont {McGlone}, \citenamefont {Mosyagin}, \citenamefont {Neyens}, \citenamefont {Nichols}, \citenamefont {Nies}, \citenamefont {Pastrana},
  \citenamefont {Rothe}, \citenamefont {Ryssens}, \citenamefont {van~den Borne}, \citenamefont {Wessolek}, \citenamefont {Wilkins},\ and\ \citenamefont {Yang}}]{Athanasakis-Kaklamanakis2025}%
  \BibitemOpen
  \bibfield  {author} {\bibinfo {author} {\bibfnamefont {M.}~\bibnamefont {Athanasakis-Kaklamanakis}}, \bibinfo {author} {\bibfnamefont {M.}~\bibnamefont {Au}}, \bibinfo {author} {\bibfnamefont {A.}~\bibnamefont {Kyuberis}}, \bibinfo {author} {\bibfnamefont {C.}~\bibnamefont {Z{\"u}lch}}, \bibinfo {author} {\bibfnamefont {K.}~\bibnamefont {Gaul}}, \bibinfo {author} {\bibfnamefont {H.}~\bibnamefont {Wibowo}}, \bibinfo {author} {\bibfnamefont {L.}~\bibnamefont {Skripnikov}}, \bibinfo {author} {\bibfnamefont {L.}~\bibnamefont {Lalanne}}, \bibinfo {author} {\bibfnamefont {J.~R.}\ \bibnamefont {Reilly}}, \bibinfo {author} {\bibfnamefont {{\'A}.}~\bibnamefont {Koszor{\'u}s}}, \bibinfo {author} {\bibfnamefont {S.}~\bibnamefont {Bara}}, \bibinfo {author} {\bibfnamefont {J.}~\bibnamefont {Ballof}}, \bibinfo {author} {\bibfnamefont {R.}~\bibnamefont {Berger}}, \bibinfo {author} {\bibfnamefont {C.}~\bibnamefont {Bernerd}}, \bibinfo {author} {\bibfnamefont {A.}~\bibnamefont {Borschevsky}}, \bibinfo {author}
  {\bibfnamefont {A.~A.}\ \bibnamefont {Breier}}, \bibinfo {author} {\bibfnamefont {K.}~\bibnamefont {Chrysalidis}}, \bibinfo {author} {\bibfnamefont {T.~E.}\ \bibnamefont {Cocolios}}, \bibinfo {author} {\bibfnamefont {R.~P.}\ \bibnamefont {de~Groote}}, \bibinfo {author} {\bibfnamefont {A.}~\bibnamefont {Dorne}}, \bibinfo {author} {\bibfnamefont {J.}~\bibnamefont {Dobaczewski}}, \bibinfo {author} {\bibfnamefont {C.~M.}\ \bibnamefont {Fajardo~Zambrano}}, \bibinfo {author} {\bibfnamefont {K.~T.}\ \bibnamefont {Flanagan}}, \bibinfo {author} {\bibfnamefont {S.}~\bibnamefont {Franchoo}}, \bibinfo {author} {\bibfnamefont {J.~D.}\ \bibnamefont {Johnson}}, \bibinfo {author} {\bibfnamefont {R.~F.}\ \bibnamefont {Garcia~Ruiz}}, \bibinfo {author} {\bibfnamefont {D.}~\bibnamefont {Hanstorp}}, \bibinfo {author} {\bibfnamefont {S.}~\bibnamefont {Kujanp{\"a}{\"a}}}, \bibinfo {author} {\bibfnamefont {Y.~C.}\ \bibnamefont {Liu}}, \bibinfo {author} {\bibfnamefont {K.~M.}\ \bibnamefont {Lynch}}, \bibinfo {author} {\bibfnamefont
  {A.}~\bibnamefont {McGlone}}, \bibinfo {author} {\bibfnamefont {N.~S.}\ \bibnamefont {Mosyagin}}, \bibinfo {author} {\bibfnamefont {G.}~\bibnamefont {Neyens}}, \bibinfo {author} {\bibfnamefont {M.}~\bibnamefont {Nichols}}, \bibinfo {author} {\bibfnamefont {L.}~\bibnamefont {Nies}}, \bibinfo {author} {\bibfnamefont {F.}~\bibnamefont {Pastrana}}, \bibinfo {author} {\bibfnamefont {S.}~\bibnamefont {Rothe}}, \bibinfo {author} {\bibfnamefont {W.}~\bibnamefont {Ryssens}}, \bibinfo {author} {\bibfnamefont {B.}~\bibnamefont {van~den Borne}}, \bibinfo {author} {\bibfnamefont {J.}~\bibnamefont {Wessolek}}, \bibinfo {author} {\bibfnamefont {S.~G.}\ \bibnamefont {Wilkins}},\ and\ \bibinfo {author} {\bibfnamefont {X.~F.}\ \bibnamefont {Yang}},\ }\bibfield  {title} {\bibinfo {title} {Laser spectroscopy and {CP}-violation sensitivity of actinium monofluoride},\ }\href {https://doi.org/10.1038/s41586-025-09814-1} {\bibfield  {journal} {\bibinfo  {journal} {Nature}\ }\textbf {\bibinfo {volume} {648}},\ \bibinfo {pages}
  {562} (\bibinfo {year} {2025})}\BibitemShut {NoStop}%
\bibitem [{\citenamefont {Kroger}\ and\ \citenamefont {Reich}(1976)}]{KROGER197629}%
  \BibitemOpen
  \bibfield  {author} {\bibinfo {author} {\bibfnamefont {L.}~\bibnamefont {Kroger}}\ and\ \bibinfo {author} {\bibfnamefont {C.}~\bibnamefont {Reich}},\ }\bibfield  {title} {\bibinfo {title} {Features of the low-energy level scheme of {$^{229}$Th} as observed in the $\alpha$-decay of {$^{233}$U}},\ }\href {https://doi.org/https://doi.org/10.1016/0375-9474(76)90494-2} {\bibfield  {journal} {\bibinfo  {journal} {Nuclear Physics A}\ }\textbf {\bibinfo {volume} {259}},\ \bibinfo {pages} {29} (\bibinfo {year} {1976})}\BibitemShut {NoStop}%
\bibitem [{\citenamefont {Helmer}\ and\ \citenamefont {Reich}(1994)}]{PhysRevC.49.1845}%
  \BibitemOpen
  \bibfield  {author} {\bibinfo {author} {\bibfnamefont {R.~G.}\ \bibnamefont {Helmer}}\ and\ \bibinfo {author} {\bibfnamefont {C.~W.}\ \bibnamefont {Reich}},\ }\bibfield  {title} {\bibinfo {title} {An excited state of $^{229}\mathrm{Th}$ at {3.5 eV}},\ }\href {https://doi.org/10.1103/PhysRevC.49.1845} {\bibfield  {journal} {\bibinfo  {journal} {Phys. Rev. C}\ }\textbf {\bibinfo {volume} {49}},\ \bibinfo {pages} {1845} (\bibinfo {year} {1994})}\BibitemShut {NoStop}%
\bibitem [{\citenamefont {Peik}\ and\ \citenamefont {Tamm}(2003)}]{E.Peik_2003}%
  \BibitemOpen
  \bibfield  {author} {\bibinfo {author} {\bibfnamefont {E.}~\bibnamefont {Peik}}\ and\ \bibinfo {author} {\bibfnamefont {C.}~\bibnamefont {Tamm}},\ }\bibfield  {title} {\bibinfo {title} {Nuclear laser spectroscopy of the {3.5 eV} transition in {Th-229}},\ }\href {https://doi.org/10.1209/epl/i2003-00210-x} {\bibfield  {journal} {\bibinfo  {journal} {Europhysics Letters}\ }\textbf {\bibinfo {volume} {61}},\ \bibinfo {pages} {181} (\bibinfo {year} {2003})}\BibitemShut {NoStop}%
\bibitem [{\citenamefont {Ruchowska}\ \emph {et~al.}(2006)\citenamefont {Ruchowska}, \citenamefont {P\l{}\'ociennik}, \citenamefont {\ifmmode~\dot{Z}\else \.{Z}\fi{}ylicz}, \citenamefont {Mach}, \citenamefont {Kvasil}, \citenamefont {Algora}, \citenamefont {Amzal}, \citenamefont {B\"ack}, \citenamefont {Borge}, \citenamefont {Boutami}, \citenamefont {Butler}, \citenamefont {Cederk\"all}, \citenamefont {Cederwall}, \citenamefont {Fogelberg}, \citenamefont {Fraile}, \citenamefont {Fynbo}, \citenamefont {Hageb\o{}}, \citenamefont {Hoff}, \citenamefont {Gausemel}, \citenamefont {Jungclaus}, \citenamefont {Kaczarowski}, \citenamefont {Kerek}, \citenamefont {Kurcewicz}, \citenamefont {Lagergren}, \citenamefont {Nacher}, \citenamefont {Rubio}, \citenamefont {Syntfeld}, \citenamefont {Tengblad}, \citenamefont {Wasilewski},\ and\ \citenamefont {Weissman}}]{PhysRevC.73.044326}%
  \BibitemOpen
  \bibfield  {author} {\bibinfo {author} {\bibfnamefont {E.}~\bibnamefont {Ruchowska}}, \bibinfo {author} {\bibfnamefont {W.~A.}\ \bibnamefont {P\l{}\'ociennik}}, \bibinfo {author} {\bibfnamefont {J.}~\bibnamefont {\ifmmode~\dot{Z}\else \.{Z}\fi{}ylicz}}, \bibinfo {author} {\bibfnamefont {H.}~\bibnamefont {Mach}}, \bibinfo {author} {\bibfnamefont {J.}~\bibnamefont {Kvasil}}, \bibinfo {author} {\bibfnamefont {A.}~\bibnamefont {Algora}}, \bibinfo {author} {\bibfnamefont {N.}~\bibnamefont {Amzal}}, \bibinfo {author} {\bibfnamefont {T.}~\bibnamefont {B\"ack}}, \bibinfo {author} {\bibfnamefont {M.~G.}\ \bibnamefont {Borge}}, \bibinfo {author} {\bibfnamefont {R.}~\bibnamefont {Boutami}}, \bibinfo {author} {\bibfnamefont {P.~A.}\ \bibnamefont {Butler}}, \bibinfo {author} {\bibfnamefont {J.}~\bibnamefont {Cederk\"all}}, \bibinfo {author} {\bibfnamefont {B.}~\bibnamefont {Cederwall}}, \bibinfo {author} {\bibfnamefont {B.}~\bibnamefont {Fogelberg}}, \bibinfo {author} {\bibfnamefont {L.~M.}\ \bibnamefont {Fraile}},
  \bibinfo {author} {\bibfnamefont {H.~O.~U.}\ \bibnamefont {Fynbo}}, \bibinfo {author} {\bibfnamefont {E.}~\bibnamefont {Hageb\o{}}}, \bibinfo {author} {\bibfnamefont {P.}~\bibnamefont {Hoff}}, \bibinfo {author} {\bibfnamefont {H.}~\bibnamefont {Gausemel}}, \bibinfo {author} {\bibfnamefont {A.}~\bibnamefont {Jungclaus}}, \bibinfo {author} {\bibfnamefont {R.}~\bibnamefont {Kaczarowski}}, \bibinfo {author} {\bibfnamefont {A.}~\bibnamefont {Kerek}}, \bibinfo {author} {\bibfnamefont {W.}~\bibnamefont {Kurcewicz}}, \bibinfo {author} {\bibfnamefont {K.}~\bibnamefont {Lagergren}}, \bibinfo {author} {\bibfnamefont {E.}~\bibnamefont {Nacher}}, \bibinfo {author} {\bibfnamefont {B.}~\bibnamefont {Rubio}}, \bibinfo {author} {\bibfnamefont {A.}~\bibnamefont {Syntfeld}}, \bibinfo {author} {\bibfnamefont {O.}~\bibnamefont {Tengblad}}, \bibinfo {author} {\bibfnamefont {A.~A.}\ \bibnamefont {Wasilewski}},\ and\ \bibinfo {author} {\bibfnamefont {L.}~\bibnamefont {Weissman}},\ }\bibfield  {title} {\bibinfo {title} {Nuclear
  structure of $^{229}\mathrm{Th}$},\ }\href {https://doi.org/10.1103/PhysRevC.73.044326} {\bibfield  {journal} {\bibinfo  {journal} {Phys. Rev. C}\ }\textbf {\bibinfo {volume} {73}},\ \bibinfo {pages} {044326} (\bibinfo {year} {2006})}\BibitemShut {NoStop}%
\bibitem [{\citenamefont {Matinyan}(1998)}]{MATINYAN1998199}%
  \BibitemOpen
  \bibfield  {author} {\bibinfo {author} {\bibfnamefont {S.}~\bibnamefont {Matinyan}},\ }\bibfield  {title} {\bibinfo {title} {Lasers as a bridge between atomic and nuclear physics},\ }\href {https://doi.org/https://doi.org/10.1016/S0370-1573(97)00084-7} {\bibfield  {journal} {\bibinfo  {journal} {Physics Reports}\ }\textbf {\bibinfo {volume} {298}},\ \bibinfo {pages} {199} (\bibinfo {year} {1998})}\BibitemShut {NoStop}%
\bibitem [{\citenamefont {Gulda}\ \emph {et~al.}(2002)\citenamefont {Gulda}, \citenamefont {Kurcewicz}, \citenamefont {Aas}, \citenamefont {Borge}, \citenamefont {Burke}, \citenamefont {Fogelberg}, \citenamefont {Grant}, \citenamefont {Hagebø}, \citenamefont {Kaffrell}, \citenamefont {Kvasil}, \citenamefont {Løvhøiden}, \citenamefont {Mach}, \citenamefont {Mackova}, \citenamefont {Martinez}, \citenamefont {Nyman}, \citenamefont {Rubio}, \citenamefont {Tain}, \citenamefont {Tengblad},\ and\ \citenamefont {Thorsteinsen}}]{GULDA200245}%
  \BibitemOpen
  \bibfield  {author} {\bibinfo {author} {\bibfnamefont {K.}~\bibnamefont {Gulda}}, \bibinfo {author} {\bibfnamefont {W.}~\bibnamefont {Kurcewicz}}, \bibinfo {author} {\bibfnamefont {A.}~\bibnamefont {Aas}}, \bibinfo {author} {\bibfnamefont {M.}~\bibnamefont {Borge}}, \bibinfo {author} {\bibfnamefont {D.}~\bibnamefont {Burke}}, \bibinfo {author} {\bibfnamefont {B.}~\bibnamefont {Fogelberg}}, \bibinfo {author} {\bibfnamefont {I.}~\bibnamefont {Grant}}, \bibinfo {author} {\bibfnamefont {E.}~\bibnamefont {Hagebø}}, \bibinfo {author} {\bibfnamefont {N.}~\bibnamefont {Kaffrell}}, \bibinfo {author} {\bibfnamefont {J.}~\bibnamefont {Kvasil}}, \bibinfo {author} {\bibfnamefont {G.}~\bibnamefont {Løvhøiden}}, \bibinfo {author} {\bibfnamefont {H.}~\bibnamefont {Mach}}, \bibinfo {author} {\bibfnamefont {A.}~\bibnamefont {Mackova}}, \bibinfo {author} {\bibfnamefont {T.}~\bibnamefont {Martinez}}, \bibinfo {author} {\bibfnamefont {G.}~\bibnamefont {Nyman}}, \bibinfo {author} {\bibfnamefont {B.}~\bibnamefont {Rubio}},
  \bibinfo {author} {\bibfnamefont {J.}~\bibnamefont {Tain}}, \bibinfo {author} {\bibfnamefont {O.}~\bibnamefont {Tengblad}},\ and\ \bibinfo {author} {\bibfnamefont {T.}~\bibnamefont {Thorsteinsen}},\ }\bibfield  {title} {\bibinfo {title} {The nuclear structure of {$^{229}$Th}},\ }\href {https://doi.org/https://doi.org/10.1016/S0375-9474(01)01456-7} {\bibfield  {journal} {\bibinfo  {journal} {Nuclear Physics A}\ }\textbf {\bibinfo {volume} {703}},\ \bibinfo {pages} {45} (\bibinfo {year} {2002})}\BibitemShut {NoStop}%
\bibitem [{\citenamefont {von~der Wense}\ \emph {et~al.}(2016)\citenamefont {von~der Wense}, \citenamefont {Seiferle}, \citenamefont {Laatiaoui}, \citenamefont {Neumayr}, \citenamefont {Maier}, \citenamefont {Wirth}, \citenamefont {Mokry}, \citenamefont {Runke}, \citenamefont {Eberhardt}, \citenamefont {D{\"u}llmann}, \citenamefont {Trautmann},\ and\ \citenamefont {Thirolf}}]{vonderWense2016}%
  \BibitemOpen
  \bibfield  {author} {\bibinfo {author} {\bibfnamefont {L.}~\bibnamefont {von~der Wense}}, \bibinfo {author} {\bibfnamefont {B.}~\bibnamefont {Seiferle}}, \bibinfo {author} {\bibfnamefont {M.}~\bibnamefont {Laatiaoui}}, \bibinfo {author} {\bibfnamefont {J.~B.}\ \bibnamefont {Neumayr}}, \bibinfo {author} {\bibfnamefont {H.-J.}\ \bibnamefont {Maier}}, \bibinfo {author} {\bibfnamefont {H.-F.}\ \bibnamefont {Wirth}}, \bibinfo {author} {\bibfnamefont {C.}~\bibnamefont {Mokry}}, \bibinfo {author} {\bibfnamefont {J.}~\bibnamefont {Runke}}, \bibinfo {author} {\bibfnamefont {K.}~\bibnamefont {Eberhardt}}, \bibinfo {author} {\bibfnamefont {C.~E.}\ \bibnamefont {D{\"u}llmann}}, \bibinfo {author} {\bibfnamefont {N.~G.}\ \bibnamefont {Trautmann}},\ and\ \bibinfo {author} {\bibfnamefont {P.~G.}\ \bibnamefont {Thirolf}},\ }\bibfield  {title} {\bibinfo {title} {Direct detection of the {$^{229}$Th} nuclear clock transition},\ }\href {https://doi.org/10.1038/nature17669} {\bibfield  {journal} {\bibinfo  {journal} {Nature}\
  }\textbf {\bibinfo {volume} {533}},\ \bibinfo {pages} {47} (\bibinfo {year} {2016})}\BibitemShut {NoStop}%
\bibitem [{\citenamefont {Thielking}\ \emph {et~al.}(2018)\citenamefont {Thielking}, \citenamefont {Okhapkin}, \citenamefont {G{\l}owacki}, \citenamefont {Meier}, \citenamefont {von~der Wense}, \citenamefont {Seiferle}, \citenamefont {D{\"u}llmann}, \citenamefont {Thirolf},\ and\ \citenamefont {Peik}}]{Thielking2018}%
  \BibitemOpen
  \bibfield  {author} {\bibinfo {author} {\bibfnamefont {J.}~\bibnamefont {Thielking}}, \bibinfo {author} {\bibfnamefont {M.~V.}\ \bibnamefont {Okhapkin}}, \bibinfo {author} {\bibfnamefont {P.}~\bibnamefont {G{\l}owacki}}, \bibinfo {author} {\bibfnamefont {D.~M.}\ \bibnamefont {Meier}}, \bibinfo {author} {\bibfnamefont {L.}~\bibnamefont {von~der Wense}}, \bibinfo {author} {\bibfnamefont {B.}~\bibnamefont {Seiferle}}, \bibinfo {author} {\bibfnamefont {C.~E.}\ \bibnamefont {D{\"u}llmann}}, \bibinfo {author} {\bibfnamefont {P.~G.}\ \bibnamefont {Thirolf}},\ and\ \bibinfo {author} {\bibfnamefont {E.}~\bibnamefont {Peik}},\ }\bibfield  {title} {\bibinfo {title} {Laser spectroscopic characterization of the nuclear-clock isomer {$^{229m}$Th}},\ }\href {https://doi.org/10.1038/s41586-018-0011-8} {\bibfield  {journal} {\bibinfo  {journal} {Nature}\ }\textbf {\bibinfo {volume} {556}},\ \bibinfo {pages} {321} (\bibinfo {year} {2018})}\BibitemShut {NoStop}%
\bibitem [{\citenamefont {Moritz}\ \emph {et~al.}(2025)\citenamefont {Moritz}, \citenamefont {Scharl}, \citenamefont {Wiesinger}, \citenamefont {Holthoff}, \citenamefont {Teschler}, \citenamefont {Hussain}, \citenamefont {Crespo L{\'o}pez-Urrutia}, \citenamefont {Dickel}, \citenamefont {Ding}, \citenamefont {D{\"u}llmann}, \citenamefont {Hudson}, \citenamefont {Kraemer}, \citenamefont {L{\"o}bell}, \citenamefont {Mokry}, \citenamefont {Runke}, \citenamefont {Seiferle}, \citenamefont {von~der Wense}, \citenamefont {Zacherl},\ and\ \citenamefont {Thirolf}}]{Moritz2025-eh}%
  \BibitemOpen
  \bibfield  {author} {\bibinfo {author} {\bibfnamefont {D.}~\bibnamefont {Moritz}}, \bibinfo {author} {\bibfnamefont {K.}~\bibnamefont {Scharl}}, \bibinfo {author} {\bibfnamefont {M.}~\bibnamefont {Wiesinger}}, \bibinfo {author} {\bibfnamefont {G.}~\bibnamefont {Holthoff}}, \bibinfo {author} {\bibfnamefont {T.}~\bibnamefont {Teschler}}, \bibinfo {author} {\bibfnamefont {M.~I.}\ \bibnamefont {Hussain}}, \bibinfo {author} {\bibfnamefont {J.~R.}\ \bibnamefont {Crespo L{\'o}pez-Urrutia}}, \bibinfo {author} {\bibfnamefont {T.}~\bibnamefont {Dickel}}, \bibinfo {author} {\bibfnamefont {S.}~\bibnamefont {Ding}}, \bibinfo {author} {\bibfnamefont {C.~E.}\ \bibnamefont {D{\"u}llmann}}, \bibinfo {author} {\bibfnamefont {E.~R.}\ \bibnamefont {Hudson}}, \bibinfo {author} {\bibfnamefont {S.}~\bibnamefont {Kraemer}}, \bibinfo {author} {\bibfnamefont {L.}~\bibnamefont {L{\"o}bell}}, \bibinfo {author} {\bibfnamefont {C.}~\bibnamefont {Mokry}}, \bibinfo {author} {\bibfnamefont {J.}~\bibnamefont {Runke}}, \bibinfo {author}
  {\bibfnamefont {B.}~\bibnamefont {Seiferle}}, \bibinfo {author} {\bibfnamefont {L.}~\bibnamefont {von~der Wense}}, \bibinfo {author} {\bibfnamefont {F.}~\bibnamefont {Zacherl}},\ and\ \bibinfo {author} {\bibfnamefont {P.~G.}\ \bibnamefont {Thirolf}},\ }\bibfield  {title} {\bibinfo {title} {A cryogenic paul trap for probing the nuclear isomeric excited state {$^{m}$Th$^{3+}$}},\ }\href@noop {} {\bibfield  {journal} {\bibinfo  {journal} {The European Physical Journal D}\ }\textbf {\bibinfo {volume} {79}},\ \bibinfo {pages} {127} (\bibinfo {year} {2025})}\BibitemShut {NoStop}%
\bibitem [{\citenamefont {Verlinde}\ \emph {et~al.}(2019)\citenamefont {Verlinde}, \citenamefont {Kraemer}, \citenamefont {Moens}, \citenamefont {Chrysalidis}, \citenamefont {Correia}, \citenamefont {Cottenier}, \citenamefont {De~Witte}, \citenamefont {Fedorov}, \citenamefont {Fedosseev}, \citenamefont {Ferrer}, \citenamefont {Fraile}, \citenamefont {Geldhof}, \citenamefont {Granados}, \citenamefont {Laatiaoui}, \citenamefont {Lima}, \citenamefont {Lin}, \citenamefont {Manea}, \citenamefont {Marsh}, \citenamefont {Moore}, \citenamefont {Pereira}, \citenamefont {Raeder}, \citenamefont {Van~den Bergh}, \citenamefont {Van~Duppen}, \citenamefont {Vantomme}, \citenamefont {Verstraelen}, \citenamefont {Wahl},\ and\ \citenamefont {Wilkins}}]{PhysRevC.100.024315}%
  \BibitemOpen
  \bibfield  {author} {\bibinfo {author} {\bibfnamefont {M.}~\bibnamefont {Verlinde}}, \bibinfo {author} {\bibfnamefont {S.}~\bibnamefont {Kraemer}}, \bibinfo {author} {\bibfnamefont {J.}~\bibnamefont {Moens}}, \bibinfo {author} {\bibfnamefont {K.}~\bibnamefont {Chrysalidis}}, \bibinfo {author} {\bibfnamefont {J.~G.}\ \bibnamefont {Correia}}, \bibinfo {author} {\bibfnamefont {S.}~\bibnamefont {Cottenier}}, \bibinfo {author} {\bibfnamefont {H.}~\bibnamefont {De~Witte}}, \bibinfo {author} {\bibfnamefont {D.~V.}\ \bibnamefont {Fedorov}}, \bibinfo {author} {\bibfnamefont {V.~N.}\ \bibnamefont {Fedosseev}}, \bibinfo {author} {\bibfnamefont {R.}~\bibnamefont {Ferrer}}, \bibinfo {author} {\bibfnamefont {L.~M.}\ \bibnamefont {Fraile}}, \bibinfo {author} {\bibfnamefont {S.}~\bibnamefont {Geldhof}}, \bibinfo {author} {\bibfnamefont {C.~A.}\ \bibnamefont {Granados}}, \bibinfo {author} {\bibfnamefont {M.}~\bibnamefont {Laatiaoui}}, \bibinfo {author} {\bibfnamefont {T.~A.~L.}\ \bibnamefont {Lima}}, \bibinfo {author}
  {\bibfnamefont {P.-C.}\ \bibnamefont {Lin}}, \bibinfo {author} {\bibfnamefont {V.}~\bibnamefont {Manea}}, \bibinfo {author} {\bibfnamefont {B.~A.}\ \bibnamefont {Marsh}}, \bibinfo {author} {\bibfnamefont {I.}~\bibnamefont {Moore}}, \bibinfo {author} {\bibfnamefont {L.~M.~C.}\ \bibnamefont {Pereira}}, \bibinfo {author} {\bibfnamefont {S.}~\bibnamefont {Raeder}}, \bibinfo {author} {\bibfnamefont {P.}~\bibnamefont {Van~den Bergh}}, \bibinfo {author} {\bibfnamefont {P.}~\bibnamefont {Van~Duppen}}, \bibinfo {author} {\bibfnamefont {A.}~\bibnamefont {Vantomme}}, \bibinfo {author} {\bibfnamefont {E.}~\bibnamefont {Verstraelen}}, \bibinfo {author} {\bibfnamefont {U.}~\bibnamefont {Wahl}},\ and\ \bibinfo {author} {\bibfnamefont {S.~G.}\ \bibnamefont {Wilkins}},\ }\bibfield  {title} {\bibinfo {title} {Alternative approach to populate and study the $^{229}\mathrm{Th}$ nuclear clock isomer},\ }\href {https://doi.org/10.1103/PhysRevC.100.024315} {\bibfield  {journal} {\bibinfo  {journal} {Phys. Rev. C}\ }\textbf
  {\bibinfo {volume} {100}},\ \bibinfo {pages} {024315} (\bibinfo {year} {2019})}\BibitemShut {NoStop}%
\bibitem [{\citenamefont {Kraemer}\ \emph {et~al.}(2023)\citenamefont {Kraemer}, \citenamefont {Moens}, \citenamefont {Athanasakis-Kaklamanakis}, \citenamefont {Bara}, \citenamefont {Beeks}, \citenamefont {Chhetri}, \citenamefont {Chrysalidis}, \citenamefont {Claessens}, \citenamefont {Cocolios}, \citenamefont {Correia}, \citenamefont {Witte}, \citenamefont {Ferrer}, \citenamefont {Geldhof}, \citenamefont {Heinke}, \citenamefont {Hosseini}, \citenamefont {Huyse}, \citenamefont {K{\"o}ster}, \citenamefont {Kudryavtsev}, \citenamefont {Laatiaoui}, \citenamefont {Lica}, \citenamefont {Magchiels}, \citenamefont {Manea}, \citenamefont {Merckling}, \citenamefont {Pereira}, \citenamefont {Raeder}, \citenamefont {Schumm}, \citenamefont {Sels}, \citenamefont {Thirolf}, \citenamefont {Tunhuma}, \citenamefont {Van Den~Bergh}, \citenamefont {Van~Duppen}, \citenamefont {Vantomme}, \citenamefont {Verlinde}, \citenamefont {Villarreal},\ and\ \citenamefont {Wahl}}]{Kraemer2023}%
  \BibitemOpen
  \bibfield  {author} {\bibinfo {author} {\bibfnamefont {S.}~\bibnamefont {Kraemer}}, \bibinfo {author} {\bibfnamefont {J.}~\bibnamefont {Moens}}, \bibinfo {author} {\bibfnamefont {M.}~\bibnamefont {Athanasakis-Kaklamanakis}}, \bibinfo {author} {\bibfnamefont {S.}~\bibnamefont {Bara}}, \bibinfo {author} {\bibfnamefont {K.}~\bibnamefont {Beeks}}, \bibinfo {author} {\bibfnamefont {P.}~\bibnamefont {Chhetri}}, \bibinfo {author} {\bibfnamefont {K.}~\bibnamefont {Chrysalidis}}, \bibinfo {author} {\bibfnamefont {A.}~\bibnamefont {Claessens}}, \bibinfo {author} {\bibfnamefont {T.~E.}\ \bibnamefont {Cocolios}}, \bibinfo {author} {\bibfnamefont {J.~G.~M.}\ \bibnamefont {Correia}}, \bibinfo {author} {\bibfnamefont {H.~D.}\ \bibnamefont {Witte}}, \bibinfo {author} {\bibfnamefont {R.}~\bibnamefont {Ferrer}}, \bibinfo {author} {\bibfnamefont {S.}~\bibnamefont {Geldhof}}, \bibinfo {author} {\bibfnamefont {R.}~\bibnamefont {Heinke}}, \bibinfo {author} {\bibfnamefont {N.}~\bibnamefont {Hosseini}}, \bibinfo {author}
  {\bibfnamefont {M.}~\bibnamefont {Huyse}}, \bibinfo {author} {\bibfnamefont {U.}~\bibnamefont {K{\"o}ster}}, \bibinfo {author} {\bibfnamefont {Y.}~\bibnamefont {Kudryavtsev}}, \bibinfo {author} {\bibfnamefont {M.}~\bibnamefont {Laatiaoui}}, \bibinfo {author} {\bibfnamefont {R.}~\bibnamefont {Lica}}, \bibinfo {author} {\bibfnamefont {G.}~\bibnamefont {Magchiels}}, \bibinfo {author} {\bibfnamefont {V.}~\bibnamefont {Manea}}, \bibinfo {author} {\bibfnamefont {C.}~\bibnamefont {Merckling}}, \bibinfo {author} {\bibfnamefont {L.~M.~C.}\ \bibnamefont {Pereira}}, \bibinfo {author} {\bibfnamefont {S.}~\bibnamefont {Raeder}}, \bibinfo {author} {\bibfnamefont {T.}~\bibnamefont {Schumm}}, \bibinfo {author} {\bibfnamefont {S.}~\bibnamefont {Sels}}, \bibinfo {author} {\bibfnamefont {P.~G.}\ \bibnamefont {Thirolf}}, \bibinfo {author} {\bibfnamefont {S.~M.}\ \bibnamefont {Tunhuma}}, \bibinfo {author} {\bibfnamefont {P.}~\bibnamefont {Van Den~Bergh}}, \bibinfo {author} {\bibfnamefont {P.}~\bibnamefont {Van~Duppen}},
  \bibinfo {author} {\bibfnamefont {A.}~\bibnamefont {Vantomme}}, \bibinfo {author} {\bibfnamefont {M.}~\bibnamefont {Verlinde}}, \bibinfo {author} {\bibfnamefont {R.}~\bibnamefont {Villarreal}},\ and\ \bibinfo {author} {\bibfnamefont {U.}~\bibnamefont {Wahl}},\ }\bibfield  {title} {\bibinfo {title} {Observation of the radiative decay of the {$^{229}$Th} nuclear clock isomer},\ }\href {https://doi.org/10.1038/s41586-023-05894-z} {\bibfield  {journal} {\bibinfo  {journal} {Nature}\ }\textbf {\bibinfo {volume} {617}},\ \bibinfo {pages} {706} (\bibinfo {year} {2023})}\BibitemShut {NoStop}%
\bibitem [{\citenamefont {Zitzer}\ \emph {et~al.}(2025)\citenamefont {Zitzer}, \citenamefont {Tiedau}, \citenamefont {D\"ullmann}, \citenamefont {Okhapkin},\ and\ \citenamefont {Peik}}]{PhysRevA.111.L050802}%
  \BibitemOpen
  \bibfield  {author} {\bibinfo {author} {\bibfnamefont {G.}~\bibnamefont {Zitzer}}, \bibinfo {author} {\bibfnamefont {J.}~\bibnamefont {Tiedau}}, \bibinfo {author} {\bibfnamefont {C.~E.}\ \bibnamefont {D\"ullmann}}, \bibinfo {author} {\bibfnamefont {M.~V.}\ \bibnamefont {Okhapkin}},\ and\ \bibinfo {author} {\bibfnamefont {E.}~\bibnamefont {Peik}},\ }\bibfield  {title} {\bibinfo {title} {Laser spectroscopy on the hyperfine structure and isotope shift of sympathetically cooled $^{229}\mathrm{Th}^{3+}$ ions},\ }\href {https://doi.org/10.1103/PhysRevA.111.L050802} {\bibfield  {journal} {\bibinfo  {journal} {Phys. Rev. A}\ }\textbf {\bibinfo {volume} {111}},\ \bibinfo {pages} {L050802} (\bibinfo {year} {2025})}\BibitemShut {NoStop}%
\bibitem [{\citenamefont {Xu}\ \emph {et~al.}(2025)\citenamefont {Xu}, \citenamefont {Xiao}, \citenamefont {Cheng}, \citenamefont {Zhang},\ and\ \citenamefont {Yu}}]{xu2025}%
  \BibitemOpen
  \bibfield  {author} {\bibinfo {author} {\bibfnamefont {Y.-Y.}\ \bibnamefont {Xu}}, \bibinfo {author} {\bibfnamefont {Q.}~\bibnamefont {Xiao}}, \bibinfo {author} {\bibfnamefont {J.-H.}\ \bibnamefont {Cheng}}, \bibinfo {author} {\bibfnamefont {W.-Y.}\ \bibnamefont {Zhang}},\ and\ \bibinfo {author} {\bibfnamefont {T.-P.}\ \bibnamefont {Yu}},\ }\href {https://arxiv.org/abs/2510.08212} {\bibinfo {title} {Charge state regulation of nuclear excitation by electron capture in {$^{229}$Th} ions}} (\bibinfo {year} {2025}),\ \Eprint {https://arxiv.org/abs/2510.08212} {arXiv:2510.08212 [nucl-th]} \BibitemShut {NoStop}%
\bibitem [{\citenamefont {Pineda}\ \emph {et~al.}(2025)\citenamefont {Pineda}, \citenamefont {Chhetri}, \citenamefont {Bara}, \citenamefont {Elskens}, \citenamefont {Casci}, \citenamefont {Alexandrova}, \citenamefont {Au}, \citenamefont {Athanasakis-Kaklamanakis}, \citenamefont {Bartokos}, \citenamefont {Beeks}, \citenamefont {Bernerd}, \citenamefont {Claessens}, \citenamefont {Chrysalidis}, \citenamefont {Cocolios}, \citenamefont {Correia}, \citenamefont {De~Witte}, \citenamefont {Elwell}, \citenamefont {Ferrer}, \citenamefont {Heinke}, \citenamefont {Hudson}, \citenamefont {Ivandikov}, \citenamefont {Kudryavtsev}, \citenamefont {K\"oster}, \citenamefont {Kraemer}, \citenamefont {Laatiaoui}, \citenamefont {Lica}, \citenamefont {Merckling}, \citenamefont {Morawetz}, \citenamefont {Morgan}, \citenamefont {Moritz}, \citenamefont {Pereira}, \citenamefont {Raeder}, \citenamefont {Rothe}, \citenamefont {Schaden}, \citenamefont {Scharl}, \citenamefont {Schumm}, \citenamefont {Stegemann}, \citenamefont {Terhune},
  \citenamefont {Thirolf}, \citenamefont {Tunhuma}, \citenamefont {Van Den~Bergh}, \citenamefont {Van~Duppen}, \citenamefont {Vantomme}, \citenamefont {Wahl},\ and\ \citenamefont {Yue}}]{PhysRevResearch.7.013052}%
  \BibitemOpen
  \bibfield  {author} {\bibinfo {author} {\bibfnamefont {S.~V.}\ \bibnamefont {Pineda}}, \bibinfo {author} {\bibfnamefont {P.}~\bibnamefont {Chhetri}}, \bibinfo {author} {\bibfnamefont {S.}~\bibnamefont {Bara}}, \bibinfo {author} {\bibfnamefont {Y.}~\bibnamefont {Elskens}}, \bibinfo {author} {\bibfnamefont {S.}~\bibnamefont {Casci}}, \bibinfo {author} {\bibfnamefont {A.~N.}\ \bibnamefont {Alexandrova}}, \bibinfo {author} {\bibfnamefont {M.}~\bibnamefont {Au}}, \bibinfo {author} {\bibfnamefont {M.}~\bibnamefont {Athanasakis-Kaklamanakis}}, \bibinfo {author} {\bibfnamefont {M.}~\bibnamefont {Bartokos}}, \bibinfo {author} {\bibfnamefont {K.}~\bibnamefont {Beeks}}, \bibinfo {author} {\bibfnamefont {C.}~\bibnamefont {Bernerd}}, \bibinfo {author} {\bibfnamefont {A.}~\bibnamefont {Claessens}}, \bibinfo {author} {\bibfnamefont {K.}~\bibnamefont {Chrysalidis}}, \bibinfo {author} {\bibfnamefont {T.~E.}\ \bibnamefont {Cocolios}}, \bibinfo {author} {\bibfnamefont {J.~G.}\ \bibnamefont {Correia}}, \bibinfo {author}
  {\bibfnamefont {H.}~\bibnamefont {De~Witte}}, \bibinfo {author} {\bibfnamefont {R.}~\bibnamefont {Elwell}}, \bibinfo {author} {\bibfnamefont {R.}~\bibnamefont {Ferrer}}, \bibinfo {author} {\bibfnamefont {R.}~\bibnamefont {Heinke}}, \bibinfo {author} {\bibfnamefont {E.~R.}\ \bibnamefont {Hudson}}, \bibinfo {author} {\bibfnamefont {F.}~\bibnamefont {Ivandikov}}, \bibinfo {author} {\bibfnamefont {Y.}~\bibnamefont {Kudryavtsev}}, \bibinfo {author} {\bibfnamefont {U.}~\bibnamefont {K\"oster}}, \bibinfo {author} {\bibfnamefont {S.}~\bibnamefont {Kraemer}}, \bibinfo {author} {\bibfnamefont {M.}~\bibnamefont {Laatiaoui}}, \bibinfo {author} {\bibfnamefont {R.}~\bibnamefont {Lica}}, \bibinfo {author} {\bibfnamefont {C.}~\bibnamefont {Merckling}}, \bibinfo {author} {\bibfnamefont {I.}~\bibnamefont {Morawetz}}, \bibinfo {author} {\bibfnamefont {H.~W.~T.}\ \bibnamefont {Morgan}}, \bibinfo {author} {\bibfnamefont {D.}~\bibnamefont {Moritz}}, \bibinfo {author} {\bibfnamefont {L.~M.~C.}\ \bibnamefont {Pereira}}, \bibinfo
  {author} {\bibfnamefont {S.}~\bibnamefont {Raeder}}, \bibinfo {author} {\bibfnamefont {S.}~\bibnamefont {Rothe}}, \bibinfo {author} {\bibfnamefont {F.}~\bibnamefont {Schaden}}, \bibinfo {author} {\bibfnamefont {K.}~\bibnamefont {Scharl}}, \bibinfo {author} {\bibfnamefont {T.}~\bibnamefont {Schumm}}, \bibinfo {author} {\bibfnamefont {S.}~\bibnamefont {Stegemann}}, \bibinfo {author} {\bibfnamefont {J.}~\bibnamefont {Terhune}}, \bibinfo {author} {\bibfnamefont {P.~G.}\ \bibnamefont {Thirolf}}, \bibinfo {author} {\bibfnamefont {S.~M.}\ \bibnamefont {Tunhuma}}, \bibinfo {author} {\bibfnamefont {P.}~\bibnamefont {Van Den~Bergh}}, \bibinfo {author} {\bibfnamefont {P.}~\bibnamefont {Van~Duppen}}, \bibinfo {author} {\bibfnamefont {A.}~\bibnamefont {Vantomme}}, \bibinfo {author} {\bibfnamefont {U.}~\bibnamefont {Wahl}},\ and\ \bibinfo {author} {\bibfnamefont {Z.}~\bibnamefont {Yue}},\ }\bibfield  {title} {\bibinfo {title} {Radiative decay of the $^{229m}\mathrm{Th}$ nuclear clock isomer in different host materials},\
  }\href {https://doi.org/10.1103/PhysRevResearch.7.013052} {\bibfield  {journal} {\bibinfo  {journal} {Phys. Rev. Res.}\ }\textbf {\bibinfo {volume} {7}},\ \bibinfo {pages} {013052} (\bibinfo {year} {2025})}\BibitemShut {NoStop}%
\bibitem [{\citenamefont {Bilous}\ \emph {et~al.}(2017)\citenamefont {Bilous}, \citenamefont {Kazakov}, \citenamefont {Moore}, \citenamefont {Schumm},\ and\ \citenamefont {P\'alffy}}]{PhysRevA.95.032503}%
  \BibitemOpen
  \bibfield  {author} {\bibinfo {author} {\bibfnamefont {P.~V.}\ \bibnamefont {Bilous}}, \bibinfo {author} {\bibfnamefont {G.~A.}\ \bibnamefont {Kazakov}}, \bibinfo {author} {\bibfnamefont {I.~D.}\ \bibnamefont {Moore}}, \bibinfo {author} {\bibfnamefont {T.}~\bibnamefont {Schumm}},\ and\ \bibinfo {author} {\bibfnamefont {A.}~\bibnamefont {P\'alffy}},\ }\bibfield  {title} {\bibinfo {title} {Internal conversion from excited electronic states of $^{229}\mathrm{Th}$ ions},\ }\href {https://doi.org/10.1103/PhysRevA.95.032503} {\bibfield  {journal} {\bibinfo  {journal} {Phys. Rev. A}\ }\textbf {\bibinfo {volume} {95}},\ \bibinfo {pages} {032503} (\bibinfo {year} {2017})}\BibitemShut {NoStop}%
\bibitem [{\citenamefont {Zhou}\ and\ \citenamefont {Wang}(2026)}]{8jhh-ktfy}%
  \BibitemOpen
  \bibfield  {author} {\bibinfo {author} {\bibfnamefont {J.}~\bibnamefont {Zhou}}\ and\ \bibinfo {author} {\bibfnamefont {X.}~\bibnamefont {Wang}},\ }\bibfield  {title} {\bibinfo {title} {Effects of nuclear hyperfine mixing on ${^{229}\mathrm{Th}}^{3+}$ ions},\ }\href {https://doi.org/10.1103/8jhh-ktfy} {\bibfield  {journal} {\bibinfo  {journal} {Phys. Rev. Lett.}\ }\textbf {\bibinfo {volume} {136}},\ \bibinfo {pages} {092503} (\bibinfo {year} {2026})}\BibitemShut {NoStop}%
\bibitem [{\citenamefont {Dzuba}\ and\ \citenamefont {Flambaum}(2023)}]{PhysRevA.108.062813}%
  \BibitemOpen
  \bibfield  {author} {\bibinfo {author} {\bibfnamefont {V.~A.}\ \bibnamefont {Dzuba}}\ and\ \bibinfo {author} {\bibfnamefont {V.~V.}\ \bibnamefont {Flambaum}},\ }\bibfield  {title} {\bibinfo {title} {Shift of nuclear clock transition frequency in $^{229}\mathrm{Th}$ ions due to hyperfine interaction},\ }\href {https://doi.org/10.1103/PhysRevA.108.062813} {\bibfield  {journal} {\bibinfo  {journal} {Phys. Rev. A}\ }\textbf {\bibinfo {volume} {108}},\ \bibinfo {pages} {062813} (\bibinfo {year} {2023})}\BibitemShut {NoStop}%
\bibitem [{\citenamefont {Tiedau}\ \emph {et~al.}(2024)\citenamefont {Tiedau}, \citenamefont {Okhapkin}, \citenamefont {Zhang}, \citenamefont {Thielking}, \citenamefont {Zitzer}, \citenamefont {Peik}, \citenamefont {Schaden}, \citenamefont {Pronebner}, \citenamefont {Morawetz}, \citenamefont {De~Col}, \citenamefont {Schneider}, \citenamefont {Leitner}, \citenamefont {Pressler}, \citenamefont {Kazakov}, \citenamefont {Beeks}, \citenamefont {Sikorsky},\ and\ \citenamefont {Schumm}}]{PhysRevLett.132.182501}%
  \BibitemOpen
  \bibfield  {author} {\bibinfo {author} {\bibfnamefont {J.}~\bibnamefont {Tiedau}}, \bibinfo {author} {\bibfnamefont {M.~V.}\ \bibnamefont {Okhapkin}}, \bibinfo {author} {\bibfnamefont {K.}~\bibnamefont {Zhang}}, \bibinfo {author} {\bibfnamefont {J.}~\bibnamefont {Thielking}}, \bibinfo {author} {\bibfnamefont {G.}~\bibnamefont {Zitzer}}, \bibinfo {author} {\bibfnamefont {E.}~\bibnamefont {Peik}}, \bibinfo {author} {\bibfnamefont {F.}~\bibnamefont {Schaden}}, \bibinfo {author} {\bibfnamefont {T.}~\bibnamefont {Pronebner}}, \bibinfo {author} {\bibfnamefont {I.}~\bibnamefont {Morawetz}}, \bibinfo {author} {\bibfnamefont {L.~T.}\ \bibnamefont {De~Col}}, \bibinfo {author} {\bibfnamefont {F.}~\bibnamefont {Schneider}}, \bibinfo {author} {\bibfnamefont {A.}~\bibnamefont {Leitner}}, \bibinfo {author} {\bibfnamefont {M.}~\bibnamefont {Pressler}}, \bibinfo {author} {\bibfnamefont {G.~A.}\ \bibnamefont {Kazakov}}, \bibinfo {author} {\bibfnamefont {K.}~\bibnamefont {Beeks}}, \bibinfo {author} {\bibfnamefont
  {T.}~\bibnamefont {Sikorsky}},\ and\ \bibinfo {author} {\bibfnamefont {T.}~\bibnamefont {Schumm}},\ }\bibfield  {title} {\bibinfo {title} {Laser excitation of the {Th-229} nucleus},\ }\href {https://doi.org/10.1103/PhysRevLett.132.182501} {\bibfield  {journal} {\bibinfo  {journal} {Phys. Rev. Lett.}\ }\textbf {\bibinfo {volume} {132}},\ \bibinfo {pages} {182501} (\bibinfo {year} {2024})}\BibitemShut {NoStop}%
\bibitem [{Note1()}]{Note1}%
  \BibitemOpen
  \bibinfo {note} {\label {footnote_wu}Uncertainties calculated from error propagation of the measurements of $t_{1/2}$ and $E_\gamma $}\BibitemShut {NoStop}%
\bibitem [{\citenamefont {Thirolf}(2024)}]{Thirolf2024}%
  \BibitemOpen
  \bibfield  {author} {\bibinfo {author} {\bibfnamefont {P.}~\bibnamefont {Thirolf}},\ }\bibfield  {title} {\bibinfo {title} {Shedding light on the {Thorium-229} nuclear clock isomer},\ }\href {https://doi.org/10.1103/Physics.17.71} {\bibfield  {journal} {\bibinfo  {journal} {Physics}\ }\textbf {\bibinfo {volume} {17}},\ \bibinfo {pages} {71} (\bibinfo {year} {2024})}\BibitemShut {NoStop}%
\bibitem [{\citenamefont {Gerstenkorn}\ \emph {et~al.}(1974)\citenamefont {Gerstenkorn}, \citenamefont {Luc}, \citenamefont {Verges}, \citenamefont {Englekemeir}, \citenamefont {Gindler},\ and\ \citenamefont {Tomkins}}]{Gerstenkorn1974}%
  \BibitemOpen
  \bibfield  {author} {\bibinfo {author} {\bibfnamefont {S.}~\bibnamefont {Gerstenkorn}}, \bibinfo {author} {\bibfnamefont {P.}~\bibnamefont {Luc}}, \bibinfo {author} {\bibfnamefont {J.}~\bibnamefont {Verges}}, \bibinfo {author} {\bibfnamefont {D.~W.}\ \bibnamefont {Englekemeir}}, \bibinfo {author} {\bibfnamefont {J.~E.}\ \bibnamefont {Gindler}},\ and\ \bibinfo {author} {\bibfnamefont {F.~S.}\ \bibnamefont {Tomkins}},\ }\bibfield  {title} {\bibinfo {title} {Structures hyperfines du spectre d'étincelle, moment magnétique et quadrupolaire de l'isotope {$^{229}$Th}},\ }\href {https://doi.org/10.1051/jphys:01974003506048300} {\bibfield  {journal} {\bibinfo  {journal} {J. Phys. (Paris)}\ }\textbf {\bibinfo {volume} {35}},\ \bibinfo {pages} {483} (\bibinfo {year} {1974})}\BibitemShut {NoStop}%
\bibitem [{\citenamefont {Campbell}\ \emph {et~al.}(2011)\citenamefont {Campbell}, \citenamefont {Radnaev},\ and\ \citenamefont {Kuzmich}}]{PhysRevLett.106.223001}%
  \BibitemOpen
  \bibfield  {author} {\bibinfo {author} {\bibfnamefont {C.~J.}\ \bibnamefont {Campbell}}, \bibinfo {author} {\bibfnamefont {A.~G.}\ \bibnamefont {Radnaev}},\ and\ \bibinfo {author} {\bibfnamefont {A.}~\bibnamefont {Kuzmich}},\ }\bibfield  {title} {\bibinfo {title} {Wigner crystals of $^{229}\mathrm{Th}$ for optical excitation of the nuclear isomer},\ }\href {https://doi.org/10.1103/PhysRevLett.106.223001} {\bibfield  {journal} {\bibinfo  {journal} {Phys. Rev. Lett.}\ }\textbf {\bibinfo {volume} {106}},\ \bibinfo {pages} {223001} (\bibinfo {year} {2011})}\BibitemShut {NoStop}%
\bibitem [{\citenamefont {Safronova}\ \emph {et~al.}(2013)\citenamefont {Safronova}, \citenamefont {Safronova}, \citenamefont {Radnaev}, \citenamefont {Campbell},\ and\ \citenamefont {Kuzmich}}]{PhysRevA.88.060501}%
  \BibitemOpen
  \bibfield  {author} {\bibinfo {author} {\bibfnamefont {M.~S.}\ \bibnamefont {Safronova}}, \bibinfo {author} {\bibfnamefont {U.~I.}\ \bibnamefont {Safronova}}, \bibinfo {author} {\bibfnamefont {A.~G.}\ \bibnamefont {Radnaev}}, \bibinfo {author} {\bibfnamefont {C.~J.}\ \bibnamefont {Campbell}},\ and\ \bibinfo {author} {\bibfnamefont {A.}~\bibnamefont {Kuzmich}},\ }\bibfield  {title} {\bibinfo {title} {Magnetic dipole and electric quadrupole moments of the {$^{229}$Th} nucleus},\ }\href {https://doi.org/10.1103/PhysRevA.88.060501} {\bibfield  {journal} {\bibinfo  {journal} {Phys. Rev. A}\ }\textbf {\bibinfo {volume} {88}},\ \bibinfo {pages} {060501} (\bibinfo {year} {2013})}\BibitemShut {NoStop}%
\bibitem [{\citenamefont {Porsev}\ \emph {et~al.}(2021)\citenamefont {Porsev}, \citenamefont {Safronova},\ and\ \citenamefont {Kozlov}}]{PhysRevLett.127.253001}%
  \BibitemOpen
  \bibfield  {author} {\bibinfo {author} {\bibfnamefont {S.~G.}\ \bibnamefont {Porsev}}, \bibinfo {author} {\bibfnamefont {M.~S.}\ \bibnamefont {Safronova}},\ and\ \bibinfo {author} {\bibfnamefont {M.~G.}\ \bibnamefont {Kozlov}},\ }\bibfield  {title} {\bibinfo {title} {Precision calculation of hyperfine constants for extracting nuclear moments of $^{229}\mathrm{Th}$},\ }\href {https://doi.org/10.1103/PhysRevLett.127.253001} {\bibfield  {journal} {\bibinfo  {journal} {Phys. Rev. Lett.}\ }\textbf {\bibinfo {volume} {127}},\ \bibinfo {pages} {253001} (\bibinfo {year} {2021})}\BibitemShut {NoStop}%
\bibitem [{\citenamefont {Yamaguchi}\ \emph {et~al.}(2024)\citenamefont {Yamaguchi}, \citenamefont {Shigekawa}, \citenamefont {Haba}, \citenamefont {Kikunaga}, \citenamefont {Shirasaki}, \citenamefont {Wada},\ and\ \citenamefont {Katori}}]{Yamaguchi2024}%
  \BibitemOpen
  \bibfield  {author} {\bibinfo {author} {\bibfnamefont {A.}~\bibnamefont {Yamaguchi}}, \bibinfo {author} {\bibfnamefont {Y.}~\bibnamefont {Shigekawa}}, \bibinfo {author} {\bibfnamefont {H.}~\bibnamefont {Haba}}, \bibinfo {author} {\bibfnamefont {H.}~\bibnamefont {Kikunaga}}, \bibinfo {author} {\bibfnamefont {K.}~\bibnamefont {Shirasaki}}, \bibinfo {author} {\bibfnamefont {M.}~\bibnamefont {Wada}},\ and\ \bibinfo {author} {\bibfnamefont {H.}~\bibnamefont {Katori}},\ }\bibfield  {title} {\bibinfo {title} {Laser spectroscopy of triply charged {$^{229}$Th} isomer for a nuclear clock},\ }\href {https://doi.org/10.1038/s41586-024-07296-1} {\bibfield  {journal} {\bibinfo  {journal} {Nature}\ }\textbf {\bibinfo {volume} {629}},\ \bibinfo {pages} {62} (\bibinfo {year} {2024})}\BibitemShut {NoStop}%
\bibitem [{\citenamefont {Schmidt}(1937)}]{Schmidt1937}%
  \BibitemOpen
  \bibfield  {author} {\bibinfo {author} {\bibfnamefont {T.}~\bibnamefont {Schmidt}},\ }\bibfield  {title} {\bibinfo {title} {Über die magnetischen momente der atomkerne},\ }\href {https://doi.org/10.1007/BF01338728} {\bibfield  {journal} {\bibinfo  {journal} {Zeitschrift für Physik}\ }\textbf {\bibinfo {volume} {106}},\ \bibinfo {pages} {358} (\bibinfo {year} {1937})}\BibitemShut {NoStop}%
\bibitem [{\citenamefont {Angeli}\ and\ \citenamefont {Marinova}(2013)}]{(Ang13a)}%
  \BibitemOpen
  \bibfield  {author} {\bibinfo {author} {\bibfnamefont {I.}~\bibnamefont {Angeli}}\ and\ \bibinfo {author} {\bibfnamefont {K.}~\bibnamefont {Marinova}},\ }\bibfield  {title} {\bibinfo {title} {Table of experimental nuclear ground state charge radii: An update},\ }\href {https://doi.org/10.1016/j.adt.2011.12.006} {\bibfield  {journal} {\bibinfo  {journal} {At. Data Nucl. Data Tables}\ }\textbf {\bibinfo {volume} {99}},\ \bibinfo {pages} {69} (\bibinfo {year} {2013})}\BibitemShut {NoStop}%
\bibitem [{\citenamefont {Schwartz}(1955)}]{PhysRev.97.380}%
  \BibitemOpen
  \bibfield  {author} {\bibinfo {author} {\bibfnamefont {C.}~\bibnamefont {Schwartz}},\ }\bibfield  {title} {\bibinfo {title} {Theory of hyperfine structure},\ }\href {https://doi.org/10.1103/PhysRev.97.380} {\bibfield  {journal} {\bibinfo  {journal} {Phys. Rev.}\ }\textbf {\bibinfo {volume} {97}},\ \bibinfo {pages} {380} (\bibinfo {year} {1955})}\BibitemShut {NoStop}%
\bibitem [{\citenamefont {Campbell}\ \emph {et~al.}(2012)\citenamefont {Campbell}, \citenamefont {Radnaev}, \citenamefont {Kuzmich}, \citenamefont {Dzuba}, \citenamefont {Flambaum},\ and\ \citenamefont {Derevianko}}]{PhysRevLett.108.120802}%
  \BibitemOpen
  \bibfield  {author} {\bibinfo {author} {\bibfnamefont {C.~J.}\ \bibnamefont {Campbell}}, \bibinfo {author} {\bibfnamefont {A.~G.}\ \bibnamefont {Radnaev}}, \bibinfo {author} {\bibfnamefont {A.}~\bibnamefont {Kuzmich}}, \bibinfo {author} {\bibfnamefont {V.~A.}\ \bibnamefont {Dzuba}}, \bibinfo {author} {\bibfnamefont {V.~V.}\ \bibnamefont {Flambaum}},\ and\ \bibinfo {author} {\bibfnamefont {A.}~\bibnamefont {Derevianko}},\ }\bibfield  {title} {\bibinfo {title} {Single-ion nuclear clock for metrology at the 19th decimal place},\ }\href {https://doi.org/10.1103/PhysRevLett.108.120802} {\bibfield  {journal} {\bibinfo  {journal} {Phys. Rev. Lett.}\ }\textbf {\bibinfo {volume} {108}},\ \bibinfo {pages} {120802} (\bibinfo {year} {2012})}\BibitemShut {NoStop}%
\bibitem [{\citenamefont {Morgan}\ \emph {et~al.}(2025)\citenamefont {Morgan}, \citenamefont {Elwell}, \citenamefont {Terhune}, \citenamefont {Tran~Tan}, \citenamefont {Perera}, \citenamefont {Derevianko}, \citenamefont {Alexandrova},\ and\ \citenamefont {Hudson}}]{laser_proposal}%
  \BibitemOpen
  \bibfield  {author} {\bibinfo {author} {\bibfnamefont {H.~W.~T.}\ \bibnamefont {Morgan}}, \bibinfo {author} {\bibfnamefont {R.}~\bibnamefont {Elwell}}, \bibinfo {author} {\bibfnamefont {J.~E.~S.}\ \bibnamefont {Terhune}}, \bibinfo {author} {\bibfnamefont {H.~B.}\ \bibnamefont {Tran~Tan}}, \bibinfo {author} {\bibfnamefont {U.~C.}\ \bibnamefont {Perera}}, \bibinfo {author} {\bibfnamefont {A.}~\bibnamefont {Derevianko}}, \bibinfo {author} {\bibfnamefont {A.~N.}\ \bibnamefont {Alexandrova}},\ and\ \bibinfo {author} {\bibfnamefont {E.~R.}\ \bibnamefont {Hudson}},\ }\bibfield  {title} {\bibinfo {title} {Proposal and theoretical investigation of {$^{229}$Th}-doped nonlinear optical crystals for compact solid-state clocks},\ }\href {https://doi.org/10.1063/5.0247867} {\bibfield  {journal} {\bibinfo  {journal} {Applied Physics Letters}\ }\textbf {\bibinfo {volume} {126}},\ \bibinfo {pages} {111101} (\bibinfo {year} {2025})}\BibitemShut {NoStop}%
\bibitem [{\citenamefont {Girvin}\ and\ \citenamefont {Radzihovsky}(2025)}]{girvin2025prospectssolidstatenuclearclock}%
  \BibitemOpen
  \bibfield  {author} {\bibinfo {author} {\bibfnamefont {S.~M.}\ \bibnamefont {Girvin}}\ and\ \bibinfo {author} {\bibfnamefont {L.}~\bibnamefont {Radzihovsky}},\ }\href {https://arxiv.org/abs/2511.13017} {\bibinfo {title} {Prospects for a solid-state nuclear clock}} (\bibinfo {year} {2025}),\ \Eprint {https://arxiv.org/abs/2511.13017} {arXiv:2511.13017 [physics.atom-ph]} \BibitemShut {NoStop}%
\bibitem [{\citenamefont {Masuda}\ \emph {et~al.}(2019)\citenamefont {Masuda}, \citenamefont {Yoshimi}, \citenamefont {Fujieda}, \citenamefont {Fujimoto}, \citenamefont {Haba}, \citenamefont {Hara}, \citenamefont {Hiraki}, \citenamefont {Kaino}, \citenamefont {Kasamatsu}, \citenamefont {Kitao}, \citenamefont {Konashi}, \citenamefont {Miyamoto}, \citenamefont {Okai}, \citenamefont {Okubo}, \citenamefont {Sasao}, \citenamefont {Seto}, \citenamefont {Schumm}, \citenamefont {Shigekawa}, \citenamefont {Suzuki}, \citenamefont {Stellmer}, \citenamefont {Tamasaku}, \citenamefont {Uetake}, \citenamefont {Watanabe}, \citenamefont {Watanabe}, \citenamefont {Yasuda}, \citenamefont {Yamaguchi}, \citenamefont {Yoda}, \citenamefont {Yokokita}, \citenamefont {Yoshimura},\ and\ \citenamefont {Yoshimura}}]{Masuda2019}%
  \BibitemOpen
  \bibfield  {author} {\bibinfo {author} {\bibfnamefont {T.}~\bibnamefont {Masuda}}, \bibinfo {author} {\bibfnamefont {A.}~\bibnamefont {Yoshimi}}, \bibinfo {author} {\bibfnamefont {A.}~\bibnamefont {Fujieda}}, \bibinfo {author} {\bibfnamefont {H.}~\bibnamefont {Fujimoto}}, \bibinfo {author} {\bibfnamefont {H.}~\bibnamefont {Haba}}, \bibinfo {author} {\bibfnamefont {H.}~\bibnamefont {Hara}}, \bibinfo {author} {\bibfnamefont {T.}~\bibnamefont {Hiraki}}, \bibinfo {author} {\bibfnamefont {H.}~\bibnamefont {Kaino}}, \bibinfo {author} {\bibfnamefont {Y.}~\bibnamefont {Kasamatsu}}, \bibinfo {author} {\bibfnamefont {S.}~\bibnamefont {Kitao}}, \bibinfo {author} {\bibfnamefont {K.}~\bibnamefont {Konashi}}, \bibinfo {author} {\bibfnamefont {Y.}~\bibnamefont {Miyamoto}}, \bibinfo {author} {\bibfnamefont {K.}~\bibnamefont {Okai}}, \bibinfo {author} {\bibfnamefont {S.}~\bibnamefont {Okubo}}, \bibinfo {author} {\bibfnamefont {N.}~\bibnamefont {Sasao}}, \bibinfo {author} {\bibfnamefont {M.}~\bibnamefont {Seto}}, \bibinfo
  {author} {\bibfnamefont {T.}~\bibnamefont {Schumm}}, \bibinfo {author} {\bibfnamefont {Y.}~\bibnamefont {Shigekawa}}, \bibinfo {author} {\bibfnamefont {K.}~\bibnamefont {Suzuki}}, \bibinfo {author} {\bibfnamefont {S.}~\bibnamefont {Stellmer}}, \bibinfo {author} {\bibfnamefont {K.}~\bibnamefont {Tamasaku}}, \bibinfo {author} {\bibfnamefont {S.}~\bibnamefont {Uetake}}, \bibinfo {author} {\bibfnamefont {M.}~\bibnamefont {Watanabe}}, \bibinfo {author} {\bibfnamefont {T.}~\bibnamefont {Watanabe}}, \bibinfo {author} {\bibfnamefont {Y.}~\bibnamefont {Yasuda}}, \bibinfo {author} {\bibfnamefont {A.}~\bibnamefont {Yamaguchi}}, \bibinfo {author} {\bibfnamefont {Y.}~\bibnamefont {Yoda}}, \bibinfo {author} {\bibfnamefont {T.}~\bibnamefont {Yokokita}}, \bibinfo {author} {\bibfnamefont {M.}~\bibnamefont {Yoshimura}},\ and\ \bibinfo {author} {\bibfnamefont {K.}~\bibnamefont {Yoshimura}},\ }\bibfield  {title} {\bibinfo {title} {X-ray pumping of the {$^{229}$Th} nuclear clock isomer},\ }\href
  {https://doi.org/10.1038/s41586-019-1542-3} {\bibfield  {journal} {\bibinfo  {journal} {Nature}\ }\textbf {\bibinfo {volume} {573}},\ \bibinfo {pages} {238} (\bibinfo {year} {2019})}\BibitemShut {NoStop}%
\bibitem [{\citenamefont {Xiao}\ \emph {et~al.}(2026)\citenamefont {Xiao}, \citenamefont {Penyazkov}, \citenamefont {Yu}, \citenamefont {Huang}, \citenamefont {Li}, \citenamefont {Shi}, \citenamefont {Yu}, \citenamefont {Mo},\ and\ \citenamefont {Ding}}]{cw5h-644b}%
  \BibitemOpen
  \bibfield  {author} {\bibinfo {author} {\bibfnamefont {Q.}~\bibnamefont {Xiao}}, \bibinfo {author} {\bibfnamefont {G.}~\bibnamefont {Penyazkov}}, \bibinfo {author} {\bibfnamefont {R.}~\bibnamefont {Yu}}, \bibinfo {author} {\bibfnamefont {B.}~\bibnamefont {Huang}}, \bibinfo {author} {\bibfnamefont {J.}~\bibnamefont {Li}}, \bibinfo {author} {\bibfnamefont {J.}~\bibnamefont {Shi}}, \bibinfo {author} {\bibfnamefont {Y.}~\bibnamefont {Yu}}, \bibinfo {author} {\bibfnamefont {Y.}~\bibnamefont {Mo}},\ and\ \bibinfo {author} {\bibfnamefont {S.}~\bibnamefont {Ding}},\ }\bibfield  {title} {\bibinfo {title} {Proposal for the generation of continuous-wave vacuum-ultraviolet laser light for {Th}-229 isomer precision spectroscopy},\ }\href {https://doi.org/10.1103/cw5h-644b} {\bibfield  {journal} {\bibinfo  {journal} {Phys. Rev. Appl.}\ }\textbf {\bibinfo {volume} {25}},\ \bibinfo {pages} {024034} (\bibinfo {year} {2026})}\BibitemShut {NoStop}%
\bibitem [{\citenamefont {Ginzburg}(2004)}]{Ginzburg2004}%
  \BibitemOpen
  \bibfield  {author} {\bibinfo {author} {\bibfnamefont {V.~L.}\ \bibnamefont {Ginzburg}},\ }\bibfield  {title} {\bibinfo {title} {On superconductivity and superfluidity},\ }\href {https://doi.org/10.1070/PU2004v047n11ABEH001825} {\bibfield  {journal} {\bibinfo  {journal} {Physics–Uspekhi}\ }\textbf {\bibinfo {volume} {47}},\ \bibinfo {pages} {1155} (\bibinfo {year} {2004})},\ \bibinfo {note} {{Nobel Lecture, 8 December 2003}}\BibitemShut {NoStop}%
\bibitem [{\citenamefont {Tkalya}(2011)}]{PhysRevLett.106.162501}%
  \BibitemOpen
  \bibfield  {author} {\bibinfo {author} {\bibfnamefont {E.~V.}\ \bibnamefont {Tkalya}},\ }\bibfield  {title} {\bibinfo {title} {Proposal for a nuclear gamma-ray laser of optical range},\ }\href {https://doi.org/10.1103/PhysRevLett.106.162501} {\bibfield  {journal} {\bibinfo  {journal} {Phys. Rev. Lett.}\ }\textbf {\bibinfo {volume} {106}},\ \bibinfo {pages} {162501} (\bibinfo {year} {2011})}\BibitemShut {NoStop}%
\bibitem [{\citenamefont {Tkalya}\ \emph {et~al.}(2015)\citenamefont {Tkalya}, \citenamefont {Schneider}, \citenamefont {Jeet},\ and\ \citenamefont {Hudson}}]{PhysRevC.92.054324}%
  \BibitemOpen
  \bibfield  {author} {\bibinfo {author} {\bibfnamefont {E.~V.}\ \bibnamefont {Tkalya}}, \bibinfo {author} {\bibfnamefont {C.}~\bibnamefont {Schneider}}, \bibinfo {author} {\bibfnamefont {J.}~\bibnamefont {Jeet}},\ and\ \bibinfo {author} {\bibfnamefont {E.~R.}\ \bibnamefont {Hudson}},\ }\bibfield  {title} {\bibinfo {title} {Radiative lifetime and energy of the low-energy isomeric level in $^{229}\mathrm{Th}$},\ }\href {https://doi.org/10.1103/PhysRevC.92.054324} {\bibfield  {journal} {\bibinfo  {journal} {Phys. Rev. C}\ }\textbf {\bibinfo {volume} {92}},\ \bibinfo {pages} {054324} (\bibinfo {year} {2015})}\BibitemShut {NoStop}%
\bibitem [{\citenamefont {Wang}\ \emph {et~al.}(2026)\citenamefont {Wang}, \citenamefont {Yang}, \citenamefont {Li}, \citenamefont {Yue}, \citenamefont {Zhao}, \citenamefont {Wang}, \citenamefont {Fu},\ and\ \citenamefont {Ma}}]{wang2026enhancedyieldratetextsuperscript229mth}%
  \BibitemOpen
  \bibfield  {author} {\bibinfo {author} {\bibfnamefont {Y.}~\bibnamefont {Wang}}, \bibinfo {author} {\bibfnamefont {Y.}~\bibnamefont {Yang}}, \bibinfo {author} {\bibfnamefont {Y.}~\bibnamefont {Li}}, \bibinfo {author} {\bibfnamefont {D.}~\bibnamefont {Yue}}, \bibinfo {author} {\bibfnamefont {K.}~\bibnamefont {Zhao}}, \bibinfo {author} {\bibfnamefont {Y.}~\bibnamefont {Wang}}, \bibinfo {author} {\bibfnamefont {C.}~\bibnamefont {Fu}},\ and\ \bibinfo {author} {\bibfnamefont {Y.}~\bibnamefont {Ma}},\ }\href {https://arxiv.org/abs/2601.22417} {\bibinfo {title} {Enhanced yield rate of {$^{229m}$Th} via cascade decay in storage rings and electron beam ion traps}} (\bibinfo {year} {2026}),\ \Eprint {https://arxiv.org/abs/2601.22417} {arXiv:2601.22417 [nucl-th]} \BibitemShut {NoStop}%
\bibitem [{\citenamefont {Schaden}\ \emph {et~al.}(2025)\citenamefont {Schaden}, \citenamefont {Riebner}, \citenamefont {Morawetz}, \citenamefont {De~Col}, \citenamefont {Kazakov}, \citenamefont {Beeks}, \citenamefont {Sikorsky}, \citenamefont {Schumm}, \citenamefont {Zhang}, \citenamefont {Lal}, \citenamefont {Zitzer}, \citenamefont {Tiedau}, \citenamefont {Okhapkin},\ and\ \citenamefont {Peik}}]{PhysRevResearch.7.L022036}%
  \BibitemOpen
  \bibfield  {author} {\bibinfo {author} {\bibfnamefont {F.}~\bibnamefont {Schaden}}, \bibinfo {author} {\bibfnamefont {T.}~\bibnamefont {Riebner}}, \bibinfo {author} {\bibfnamefont {I.}~\bibnamefont {Morawetz}}, \bibinfo {author} {\bibfnamefont {L.~T.}\ \bibnamefont {De~Col}}, \bibinfo {author} {\bibfnamefont {G.~A.}\ \bibnamefont {Kazakov}}, \bibinfo {author} {\bibfnamefont {K.}~\bibnamefont {Beeks}}, \bibinfo {author} {\bibfnamefont {T.}~\bibnamefont {Sikorsky}}, \bibinfo {author} {\bibfnamefont {T.}~\bibnamefont {Schumm}}, \bibinfo {author} {\bibfnamefont {K.}~\bibnamefont {Zhang}}, \bibinfo {author} {\bibfnamefont {V.}~\bibnamefont {Lal}}, \bibinfo {author} {\bibfnamefont {G.}~\bibnamefont {Zitzer}}, \bibinfo {author} {\bibfnamefont {J.}~\bibnamefont {Tiedau}}, \bibinfo {author} {\bibfnamefont {M.~V.}\ \bibnamefont {Okhapkin}},\ and\ \bibinfo {author} {\bibfnamefont {E.}~\bibnamefont {Peik}},\ }\bibfield  {title} {\bibinfo {title} {Laser-induced quenching of the {Th-229} nuclear clock isomer in calcium
  fluoride},\ }\href {https://doi.org/10.1103/PhysRevResearch.7.L022036} {\bibfield  {journal} {\bibinfo  {journal} {Phys. Rev. Res.}\ }\textbf {\bibinfo {volume} {7}},\ \bibinfo {pages} {L022036} (\bibinfo {year} {2025})}\BibitemShut {NoStop}%
\bibitem [{\citenamefont {Zhang}\ \emph {et~al.}(2024)\citenamefont {Zhang}, \citenamefont {Ooi}, \citenamefont {Higgins}, \citenamefont {Doyle}, \citenamefont {von~der Wense}, \citenamefont {Beeks}, \citenamefont {Leitner}, \citenamefont {Kazakov}, \citenamefont {Li}, \citenamefont {Thirolf}, \citenamefont {Schumm},\ and\ \citenamefont {Ye}}]{Zhang2024}%
  \BibitemOpen
  \bibfield  {author} {\bibinfo {author} {\bibfnamefont {C.}~\bibnamefont {Zhang}}, \bibinfo {author} {\bibfnamefont {T.}~\bibnamefont {Ooi}}, \bibinfo {author} {\bibfnamefont {J.~S.}\ \bibnamefont {Higgins}}, \bibinfo {author} {\bibfnamefont {J.~F.}\ \bibnamefont {Doyle}}, \bibinfo {author} {\bibfnamefont {L.}~\bibnamefont {von~der Wense}}, \bibinfo {author} {\bibfnamefont {K.}~\bibnamefont {Beeks}}, \bibinfo {author} {\bibfnamefont {A.}~\bibnamefont {Leitner}}, \bibinfo {author} {\bibfnamefont {G.~A.}\ \bibnamefont {Kazakov}}, \bibinfo {author} {\bibfnamefont {P.}~\bibnamefont {Li}}, \bibinfo {author} {\bibfnamefont {P.~G.}\ \bibnamefont {Thirolf}}, \bibinfo {author} {\bibfnamefont {T.}~\bibnamefont {Schumm}},\ and\ \bibinfo {author} {\bibfnamefont {J.}~\bibnamefont {Ye}},\ }\bibfield  {title} {\bibinfo {title} {Frequency ratio of the {$^{229m}$Th} nuclear isomeric transition and the {$^{87}$Sr} atomic clock},\ }\href {https://doi.org/10.1038/s41586-024-07839-6} {\bibfield  {journal} {\bibinfo  {journal}
  {Nature}\ }\textbf {\bibinfo {volume} {633}},\ \bibinfo {pages} {63} (\bibinfo {year} {2024})}\BibitemShut {NoStop}%
\bibitem [{\citenamefont {Elwell}\ \emph {et~al.}(2024)\citenamefont {Elwell}, \citenamefont {Schneider}, \citenamefont {Jeet}, \citenamefont {Terhune}, \citenamefont {Morgan}, \citenamefont {Alexandrova}, \citenamefont {Tran~Tan}, \citenamefont {Derevianko},\ and\ \citenamefont {Hudson}}]{PhysRevLett.133.013201}%
  \BibitemOpen
  \bibfield  {author} {\bibinfo {author} {\bibfnamefont {R.}~\bibnamefont {Elwell}}, \bibinfo {author} {\bibfnamefont {C.}~\bibnamefont {Schneider}}, \bibinfo {author} {\bibfnamefont {J.}~\bibnamefont {Jeet}}, \bibinfo {author} {\bibfnamefont {J.~E.~S.}\ \bibnamefont {Terhune}}, \bibinfo {author} {\bibfnamefont {H.~W.~T.}\ \bibnamefont {Morgan}}, \bibinfo {author} {\bibfnamefont {A.~N.}\ \bibnamefont {Alexandrova}}, \bibinfo {author} {\bibfnamefont {H.~B.}\ \bibnamefont {Tran~Tan}}, \bibinfo {author} {\bibfnamefont {A.}~\bibnamefont {Derevianko}},\ and\ \bibinfo {author} {\bibfnamefont {E.~R.}\ \bibnamefont {Hudson}},\ }\bibfield  {title} {\bibinfo {title} {Laser excitation of the $^{229}\mathrm{Th}$ nuclear isomeric transition in a solid-state host},\ }\href {https://doi.org/10.1103/PhysRevLett.133.013201} {\bibfield  {journal} {\bibinfo  {journal} {Phys. Rev. Lett.}\ }\textbf {\bibinfo {volume} {133}},\ \bibinfo {pages} {013201} (\bibinfo {year} {2024})}\BibitemShut {NoStop}%
\bibitem [{\citenamefont {Minkov}\ and\ \citenamefont {P\'alffy}(2017)}]{PhysRevLett.118.212501}%
  \BibitemOpen
  \bibfield  {author} {\bibinfo {author} {\bibfnamefont {N.}~\bibnamefont {Minkov}}\ and\ \bibinfo {author} {\bibfnamefont {A.}~\bibnamefont {P\'alffy}},\ }\bibfield  {title} {\bibinfo {title} {Reduced transition probabilities for the gamma decay of the {7.8 eV} isomer in $^{229}\mathrm{Th}$},\ }\href {https://doi.org/10.1103/PhysRevLett.118.212501} {\bibfield  {journal} {\bibinfo  {journal} {Phys. Rev. Lett.}\ }\textbf {\bibinfo {volume} {118}},\ \bibinfo {pages} {212501} (\bibinfo {year} {2017})}\BibitemShut {NoStop}%
\bibitem [{\citenamefont {Dykhne}\ and\ \citenamefont {Tkalya}(1998)}]{Dykhne1998}%
  \BibitemOpen
  \bibfield  {author} {\bibinfo {author} {\bibfnamefont {A.~M.}\ \bibnamefont {Dykhne}}\ and\ \bibinfo {author} {\bibfnamefont {E.~V.}\ \bibnamefont {Tkalya}},\ }\bibfield  {title} {\bibinfo {title} {Matrix element of the anomalously low-energy {(3.5{\textpm}0.5 eV)} transition in {$^{229}$Th} and the isomer lifetime},\ }\href {https://doi.org/10.1134/1.567659} {\bibfield  {journal} {\bibinfo  {journal} {Journal of Experimental and Theoretical Physics Letters}\ }\textbf {\bibinfo {volume} {67}},\ \bibinfo {pages} {251} (\bibinfo {year} {1998})}\BibitemShut {NoStop}%
\bibitem [{\citenamefont {Minkov}\ and\ \citenamefont {P\'alffy}(2019)}]{PhysRevLett.122.162502}%
  \BibitemOpen
  \bibfield  {author} {\bibinfo {author} {\bibfnamefont {N.}~\bibnamefont {Minkov}}\ and\ \bibinfo {author} {\bibfnamefont {A.}~\bibnamefont {P\'alffy}},\ }\bibfield  {title} {\bibinfo {title} {Theoretical predictions for the magnetic dipole moment of $^{229m}\mathrm{Th}$},\ }\href {https://doi.org/10.1103/PhysRevLett.122.162502} {\bibfield  {journal} {\bibinfo  {journal} {Phys. Rev. Lett.}\ }\textbf {\bibinfo {volume} {122}},\ \bibinfo {pages} {162502} (\bibinfo {year} {2019})}\BibitemShut {NoStop}%
\bibitem [{\citenamefont {Minkov}\ and\ \citenamefont {P\'alffy}(2021)}]{PhysRevC.103.014313}%
  \BibitemOpen
  \bibfield  {author} {\bibinfo {author} {\bibfnamefont {N.}~\bibnamefont {Minkov}}\ and\ \bibinfo {author} {\bibfnamefont {A.}~\bibnamefont {P\'alffy}},\ }\bibfield  {title} {\bibinfo {title} {$^{229m}\mathrm{Th}$ isomer from a nuclear model perspective},\ }\href {https://doi.org/10.1103/PhysRevC.103.014313} {\bibfield  {journal} {\bibinfo  {journal} {Phys. Rev. C}\ }\textbf {\bibinfo {volume} {103}},\ \bibinfo {pages} {014313} (\bibinfo {year} {2021})}\BibitemShut {NoStop}%
\bibitem [{\citenamefont {Chen}\ \emph {et~al.}(2025)\citenamefont {Chen}, \citenamefont {Wang},\ and\ \citenamefont {Wu}}]{(Che25)}%
  \BibitemOpen
  \bibfield  {author} {\bibinfo {author} {\bibfnamefont {Z.-R.}\ \bibnamefont {Chen}}, \bibinfo {author} {\bibfnamefont {L.-J.}\ \bibnamefont {Wang}},\ and\ \bibinfo {author} {\bibfnamefont {Y.}~\bibnamefont {Wu}},\ }\bibfield  {title} {\bibinfo {title} {Microscopic nuclear structure study of {$^{229}$Th} by projected shell model},\ }\href {https://doi.org/https://doi.org/10.1016/j.physletb.2025.139858} {\bibfield  {journal} {\bibinfo  {journal} {Physics Letters B}\ }\textbf {\bibinfo {volume} {869}},\ \bibinfo {pages} {139858} (\bibinfo {year} {2025})}\BibitemShut {NoStop}%
\bibitem [{\citenamefont {Zhou}\ and\ \citenamefont {Yao}(2025)}]{zhou2025}%
  \BibitemOpen
  \bibfield  {author} {\bibinfo {author} {\bibfnamefont {E.~F.}\ \bibnamefont {Zhou}}\ and\ \bibinfo {author} {\bibfnamefont {J.~M.}\ \bibnamefont {Yao}},\ }\href {https://arxiv.org/abs/2511.05984} {\bibinfo {title} {Microscopic study of low-lying states in odd-mass nuclei for atomic electric dipole moment searches}} (\bibinfo {year} {2025}),\ \Eprint {https://arxiv.org/abs/2511.05984} {arXiv:2511.05984 [nucl-th]} \BibitemShut {NoStop}%
\bibitem [{\citenamefont {Minkov}\ \emph {et~al.}(2024)\citenamefont {Minkov}, \citenamefont {P\'alffy}, \citenamefont {Quentin},\ and\ \citenamefont {Bonneau}}]{PhysRevC.110.034327}%
  \BibitemOpen
  \bibfield  {author} {\bibinfo {author} {\bibfnamefont {N.}~\bibnamefont {Minkov}}, \bibinfo {author} {\bibfnamefont {A.}~\bibnamefont {P\'alffy}}, \bibinfo {author} {\bibfnamefont {P.}~\bibnamefont {Quentin}},\ and\ \bibinfo {author} {\bibfnamefont {L.}~\bibnamefont {Bonneau}},\ }\bibfield  {title} {\bibinfo {title} {{Skyrme-Hartree-Fock-BCS} approach to {$^{229m}\mathrm{Th}$} and neighboring nuclei},\ }\href {https://doi.org/10.1103/PhysRevC.110.034327} {\bibfield  {journal} {\bibinfo  {journal} {Phys. Rev. C}\ }\textbf {\bibinfo {volume} {110}},\ \bibinfo {pages} {034327} (\bibinfo {year} {2024})}\BibitemShut {NoStop}%
\bibitem [{\citenamefont {Sassarini}\ \emph {et~al.}(2022)\citenamefont {Sassarini}, \citenamefont {Dobaczewski}, \citenamefont {Bonnard},\ and\ \citenamefont {{Garcia Ruiz}}}]{(Sas22c)}%
  \BibitemOpen
  \bibfield  {author} {\bibinfo {author} {\bibfnamefont {P.~L.}\ \bibnamefont {Sassarini}}, \bibinfo {author} {\bibfnamefont {J.}~\bibnamefont {Dobaczewski}}, \bibinfo {author} {\bibfnamefont {J.}~\bibnamefont {Bonnard}},\ and\ \bibinfo {author} {\bibfnamefont {R.~F.}\ \bibnamefont {{Garcia Ruiz}}},\ }\bibfield  {title} {\bibinfo {title} {Nuclear {DFT} analysis of electromagnetic moments in odd near doubly magic nuclei},\ }\href {https://doi.org/10.1088/1361-6471/ac900a} {\bibfield  {journal} {\bibinfo  {journal} {Journal of Physics G: Nuclear and Particle Physics}\ }\textbf {\bibinfo {volume} {49}},\ \bibinfo {pages} {11LT01} (\bibinfo {year} {2022})}\BibitemShut {NoStop}%
\bibitem [{\citenamefont {Bonnard}\ \emph {et~al.}(2023)\citenamefont {Bonnard}, \citenamefont {Dobaczewski}, \citenamefont {Danneaux},\ and\ \citenamefont {Kortelainen}}]{(Bon23c)}%
  \BibitemOpen
  \bibfield  {author} {\bibinfo {author} {\bibfnamefont {J.}~\bibnamefont {Bonnard}}, \bibinfo {author} {\bibfnamefont {J.}~\bibnamefont {Dobaczewski}}, \bibinfo {author} {\bibfnamefont {G.}~\bibnamefont {Danneaux}},\ and\ \bibinfo {author} {\bibfnamefont {M.}~\bibnamefont {Kortelainen}},\ }\bibfield  {title} {\bibinfo {title} {Nuclear {DFT} electromagnetic moments in heavy deformed open-shell odd nuclei},\ }\href {https://doi.org/10.1016/j.physletb.2023.138014} {\bibfield  {journal} {\bibinfo  {journal} {Physics Letters B}\ }\textbf {\bibinfo {volume} {843}},\ \bibinfo {pages} {138014} (\bibinfo {year} {2023})}\BibitemShut {NoStop}%
\bibitem [{\citenamefont {Wibowo}\ \emph {et~al.}(2025)\citenamefont {Wibowo}, \citenamefont {Backes}, \citenamefont {Dobaczewski}, \citenamefont {de~Groote}, \citenamefont {Nagpal}, \citenamefont {S\'anchez-Fern\'andez}, \citenamefont {Sun},\ and\ \citenamefont {Wood}}]{(Wib25d)}%
  \BibitemOpen
  \bibfield  {author} {\bibinfo {author} {\bibfnamefont {H.}~\bibnamefont {Wibowo}}, \bibinfo {author} {\bibfnamefont {B.~C.}\ \bibnamefont {Backes}}, \bibinfo {author} {\bibfnamefont {J.}~\bibnamefont {Dobaczewski}}, \bibinfo {author} {\bibfnamefont {R.~P.}\ \bibnamefont {de~Groote}}, \bibinfo {author} {\bibfnamefont {A.}~\bibnamefont {Nagpal}}, \bibinfo {author} {\bibfnamefont {A.}~\bibnamefont {S\'anchez-Fern\'andez}}, \bibinfo {author} {\bibfnamefont {X.}~\bibnamefont {Sun}},\ and\ \bibinfo {author} {\bibfnamefont {J.~L.}\ \bibnamefont {Wood}},\ }\bibfield  {title} {\bibinfo {title} {Electromagnetic moments in the {Sn-Gd} region determined within nuclear {DFT}},\ }\href {https://doi.org/10.1088/1361-6471/ade0dd} {\bibfield  {journal} {\bibinfo  {journal} {Journal of Physics G: Nuclear and Particle Physics}\ }\textbf {\bibinfo {volume} {52}},\ \bibinfo {pages} {065104} (\bibinfo {year} {2025})}\BibitemShut {NoStop}%
\bibitem [{\citenamefont {Dobaczewski}\ \emph {et~al.}(2026)\citenamefont {Dobaczewski}, \citenamefont {Stuchbery}, \citenamefont {Danneaux}, \citenamefont {Nagpal}, \citenamefont {Sassarini},\ and\ \citenamefont {Wibowo}}]{(Dob26b)}%
  \BibitemOpen
  \bibfield  {author} {\bibinfo {author} {\bibfnamefont {J.}~\bibnamefont {Dobaczewski}}, \bibinfo {author} {\bibfnamefont {A.~E.}\ \bibnamefont {Stuchbery}}, \bibinfo {author} {\bibfnamefont {G.}~\bibnamefont {Danneaux}}, \bibinfo {author} {\bibfnamefont {A.}~\bibnamefont {Nagpal}}, \bibinfo {author} {\bibfnamefont {P.~L.}\ \bibnamefont {Sassarini}},\ and\ \bibinfo {author} {\bibfnamefont {H.}~\bibnamefont {Wibowo}},\ }\bibfield  {title} {\bibinfo {title} {Electromagnetic moments of ground and excited states calculated in heavy {odd-$N$} open-shell nuclei},\ }\href {https://doi.org/10.1103/4q99-rv67} {\bibfield  {journal} {\bibinfo  {journal} {Phys. Rev. C}\ }\textbf {\bibinfo {volume} {113}},\ \bibinfo {pages} {024306} (\bibinfo {year} {2026})}\BibitemShut {NoStop}%
\bibitem [{\citenamefont {Dobaczewski}\ \emph {et~al.}(2025)\citenamefont {Dobaczewski}, \citenamefont {Backes}, \citenamefont {de~Groote}, \citenamefont {Restrepo-Giraldo}, \citenamefont {Sun},\ and\ \citenamefont {Wibowo}}]{(Dob25g)}%
  \BibitemOpen
  \bibfield  {author} {\bibinfo {author} {\bibfnamefont {J.}~\bibnamefont {Dobaczewski}}, \bibinfo {author} {\bibfnamefont {B.}~\bibnamefont {Backes}}, \bibinfo {author} {\bibfnamefont {R.}~\bibnamefont {de~Groote}}, \bibinfo {author} {\bibfnamefont {A.}~\bibnamefont {Restrepo-Giraldo}}, \bibinfo {author} {\bibfnamefont {X.}~\bibnamefont {Sun}},\ and\ \bibinfo {author} {\bibfnamefont {H.}~\bibnamefont {Wibowo}},\ }\href {https://arxiv.org/abs/arXiv:2511.04632} {\bibinfo {title} {Electromagnetic and exotic moments in nuclear {DFT}}} (\bibinfo {year} {2025}),\ \bibinfo {note} {submitted to Annual Review of Nuclear and Particle Science},\ \Eprint {https://arxiv.org/abs/2511.04632} {arXiv:2511.04632 [nucl-th]} \BibitemShut {NoStop}%
\bibitem [{\citenamefont {Mi\ifmmode~\acute{s}\else \'{s}\fi{}kiewicz}\ \emph {et~al.}(2025)\citenamefont {Mi\ifmmode~\acute{s}\else \'{s}\fi{}kiewicz}, \citenamefont {Konieczka},\ and\ \citenamefont {Satu\l{}a}}]{nkh2-3kfl}%
  \BibitemOpen
  \bibfield  {author} {\bibinfo {author} {\bibfnamefont {J.}~\bibnamefont {Mi\ifmmode~\acute{s}\else \'{s}\fi{}kiewicz}}, \bibinfo {author} {\bibfnamefont {M.}~\bibnamefont {Konieczka}},\ and\ \bibinfo {author} {\bibfnamefont {W.}~\bibnamefont {Satu\l{}a}},\ }\bibfield  {title} {\bibinfo {title} {Two-neutrino ${0}^{+}\ensuremath{\rightarrow}{0}^{+}$ double-$\ensuremath{\beta}$ decay of $^{48}\mathrm{Ca}$ within the density-functional-theory--based no-core configuration-interaction framework},\ }\href {https://doi.org/10.1103/nkh2-3kfl} {\bibfield  {journal} {\bibinfo  {journal} {Phys. Rev. C}\ }\textbf {\bibinfo {volume} {112}},\ \bibinfo {pages} {055502} (\bibinfo {year} {2025})}\BibitemShut {NoStop}%
\bibitem [{\citenamefont {Dobaczewski}\ \emph {et~al.}(2021)\citenamefont {Dobaczewski}, \citenamefont {B\c{a}czyk}, \citenamefont {Becker}, \citenamefont {Bender}, \citenamefont {Bennaceur}, \citenamefont {Bonnard}, \citenamefont {Gao}, \citenamefont {Idini}, \citenamefont {Konieczka}, \citenamefont {Kortelainen}, \citenamefont {Pr{\'o}chniak}, \citenamefont {Romero}, \citenamefont {Satu{\l}a}, \citenamefont {Shi}, \citenamefont {Yu},\ and\ \citenamefont {Werner}}]{(Dob21f)}%
  \BibitemOpen
  \bibfield  {author} {\bibinfo {author} {\bibfnamefont {J.}~\bibnamefont {Dobaczewski}}, \bibinfo {author} {\bibfnamefont {P.}~\bibnamefont {B\c{a}czyk}}, \bibinfo {author} {\bibfnamefont {P.}~\bibnamefont {Becker}}, \bibinfo {author} {\bibfnamefont {M.}~\bibnamefont {Bender}}, \bibinfo {author} {\bibfnamefont {K.}~\bibnamefont {Bennaceur}}, \bibinfo {author} {\bibfnamefont {J.}~\bibnamefont {Bonnard}}, \bibinfo {author} {\bibfnamefont {Y.}~\bibnamefont {Gao}}, \bibinfo {author} {\bibfnamefont {A.}~\bibnamefont {Idini}}, \bibinfo {author} {\bibfnamefont {M.}~\bibnamefont {Konieczka}}, \bibinfo {author} {\bibfnamefont {M.}~\bibnamefont {Kortelainen}}, \bibinfo {author} {\bibfnamefont {L.}~\bibnamefont {Pr{\'o}chniak}}, \bibinfo {author} {\bibfnamefont {A.~M.}\ \bibnamefont {Romero}}, \bibinfo {author} {\bibfnamefont {W.}~\bibnamefont {Satu{\l}a}}, \bibinfo {author} {\bibfnamefont {Y.}~\bibnamefont {Shi}}, \bibinfo {author} {\bibfnamefont {L.~F.}\ \bibnamefont {Yu}},\ and\ \bibinfo {author} {\bibfnamefont
  {T.~R.}\ \bibnamefont {Werner}},\ }\bibfield  {title} {\bibinfo {title} {Solution of universal nonrelativistic nuclear {DFT} equations in the {Cartesian} deformed harmonic-oscillator basis. {(IX) HFODD} (v3.06h): a new version of the program},\ }\href {https://doi.org/10.1088/1361-6471/ac0a82} {\bibfield  {journal} {\bibinfo  {journal} {J. Phys. G: Nucl. Part. Phys.}\ }\textbf {\bibinfo {volume} {48}},\ \bibinfo {pages} {102001} (\bibinfo {year} {2021})}\BibitemShut {NoStop}%
\bibitem [{\citenamefont {{J. Dobaczewski {\it et al.}}}(2026)}]{(Dob26a)}%
  \BibitemOpen
  \bibfield  {author} {\bibinfo {author} {\bibnamefont {{J. Dobaczewski {\it et al.}}}},\ }\href@noop {} {\bibinfo {title} {Code {\sc hfodd}, version to be published}} (\bibinfo {year} {2026})\BibitemShut {NoStop}%
\bibitem [{\citenamefont {{A. Restrepo-Giraldo {\it et al.}}}(2026)}]{(Res26)}%
  \BibitemOpen
  \bibfield  {author} {\bibinfo {author} {\bibnamefont {{A. Restrepo-Giraldo {\it et al.}}}},\ }\href@noop {} {\bibinfo {title} {Code {\sc hfodd}, the {User Guide}, to be published}} (\bibinfo {year} {2026})\BibitemShut {NoStop}%
\bibitem [{\citenamefont {Bender}\ \emph {et~al.}(2002)\citenamefont {Bender}, \citenamefont {Dobaczewski}, \citenamefont {Engel},\ and\ \citenamefont {Nazarewicz}}]{(Ben02d)}%
  \BibitemOpen
  \bibfield  {author} {\bibinfo {author} {\bibfnamefont {M.}~\bibnamefont {Bender}}, \bibinfo {author} {\bibfnamefont {J.}~\bibnamefont {Dobaczewski}}, \bibinfo {author} {\bibfnamefont {J.}~\bibnamefont {Engel}},\ and\ \bibinfo {author} {\bibfnamefont {W.}~\bibnamefont {Nazarewicz}},\ }\bibfield  {title} {\bibinfo {title} {{Gamow-Teller} strength and the spin-isospin coupling constants of the {Skyrme} energy functional},\ }\href {https://doi.org/10.1103/PhysRevC.65.054322} {\bibfield  {journal} {\bibinfo  {journal} {Phys. Rev. C}\ }\textbf {\bibinfo {volume} {65}},\ \bibinfo {pages} {054322} (\bibinfo {year} {2002})}\BibitemShut {NoStop}%
\bibitem [{\citenamefont {Idini}\ \emph {et~al.}(2017)\citenamefont {Idini}, \citenamefont {Bennaceur},\ and\ \citenamefont {Dobaczewski}}]{(Idi17)}%
  \BibitemOpen
  \bibfield  {author} {\bibinfo {author} {\bibfnamefont {A.}~\bibnamefont {Idini}}, \bibinfo {author} {\bibfnamefont {K.}~\bibnamefont {Bennaceur}},\ and\ \bibinfo {author} {\bibfnamefont {J.}~\bibnamefont {Dobaczewski}},\ }\bibfield  {title} {\bibinfo {title} {Landau parameters for energy density functionals generated by local finite-range pseudopotentials},\ }\href {https://doi.org/10.1088/1361-6471/aa691e} {\bibfield  {journal} {\bibinfo  {journal} {Journal of Physics G: Nuclear and Particle Physics}\ }\textbf {\bibinfo {volume} {44}},\ \bibinfo {pages} {064004} (\bibinfo {year} {2017})}\BibitemShut {NoStop}%
\bibitem [{sup()}]{supp-229Th}%
  \BibitemOpen
  \href@noop {} {}\bibinfo {note} {{See supplemental material at [URL will be inserted by the publisher] for the lists of non-converged and/or singular points, parameters and numerical conditions of the calculation, results presented in numerical form, and plots of electromagnetic moments calculated in this Letter, which cites Ref.~\protect\cite{(Sla51a)} }}\BibitemShut {NoStop}%
\bibitem [{\citenamefont {Brown}(1998)}]{(Bro98c)}%
  \BibitemOpen
  \bibfield  {author} {\bibinfo {author} {\bibfnamefont {B.~A.}\ \bibnamefont {Brown}},\ }\bibfield  {title} {\bibinfo {title} {New {Skyrme} interaction for normal and exotic nuclei},\ }\href {http://link.aps.org/abstract/PRC/v58/p220} {\bibfield  {journal} {\bibinfo  {journal} {Phys. Rev. C}\ }\textbf {\bibinfo {volume} {58}},\ \bibinfo {pages} {220} (\bibinfo {year} {1998})}\BibitemShut {NoStop}%
\bibitem [{\citenamefont {Bartel}\ \emph {et~al.}(1982)\citenamefont {Bartel}, \citenamefont {Quentin}, \citenamefont {Brack}, \citenamefont {Guet},\ and\ \citenamefont {H{\aa}kansson}}]{(Bar82c)}%
  \BibitemOpen
  \bibfield  {author} {\bibinfo {author} {\bibfnamefont {J.}~\bibnamefont {Bartel}}, \bibinfo {author} {\bibfnamefont {P.}~\bibnamefont {Quentin}}, \bibinfo {author} {\bibfnamefont {M.}~\bibnamefont {Brack}}, \bibinfo {author} {\bibfnamefont {C.}~\bibnamefont {Guet}},\ and\ \bibinfo {author} {\bibfnamefont {H.-B.}\ \bibnamefont {H{\aa}kansson}},\ }\bibfield  {title} {\bibinfo {title} {Towards a better parametrisation of \protect{Skyrme-like} effective forces: A critical study of the \protect{SkM} force},\ }\href {https://doi.org/10.1016/0375-9474(82)90403-1} {\bibfield  {journal} {\bibinfo  {journal} {Nuclear Physics A}\ }\textbf {\bibinfo {volume} {386}},\ \bibinfo {pages} {79 } (\bibinfo {year} {1982})}\BibitemShut {NoStop}%
\bibitem [{\citenamefont {Kortelainen}\ \emph {et~al.}(2010)\citenamefont {Kortelainen}, \citenamefont {Lesinski}, \citenamefont {Mor\'e}, \citenamefont {Nazarewicz}, \citenamefont {Sarich}, \citenamefont {Schunck}, \citenamefont {Stoitsov},\ and\ \citenamefont {Wild}}]{(Kor10c)}%
  \BibitemOpen
  \bibfield  {author} {\bibinfo {author} {\bibfnamefont {M.}~\bibnamefont {Kortelainen}}, \bibinfo {author} {\bibfnamefont {T.}~\bibnamefont {Lesinski}}, \bibinfo {author} {\bibfnamefont {J.}~\bibnamefont {Mor\'e}}, \bibinfo {author} {\bibfnamefont {W.}~\bibnamefont {Nazarewicz}}, \bibinfo {author} {\bibfnamefont {J.}~\bibnamefont {Sarich}}, \bibinfo {author} {\bibfnamefont {N.}~\bibnamefont {Schunck}}, \bibinfo {author} {\bibfnamefont {M.~V.}\ \bibnamefont {Stoitsov}},\ and\ \bibinfo {author} {\bibfnamefont {S.}~\bibnamefont {Wild}},\ }\bibfield  {title} {\bibinfo {title} {Nuclear energy density optimization},\ }\href {https://doi.org/10.1103/PhysRevC.82.024313} {\bibfield  {journal} {\bibinfo  {journal} {Phys. Rev. C}\ }\textbf {\bibinfo {volume} {82}},\ \bibinfo {pages} {024313} (\bibinfo {year} {2010})}\BibitemShut {NoStop}%
\bibitem [{\citenamefont {Beiner}\ \emph {et~al.}(1975)\citenamefont {Beiner}, \citenamefont {Flocard}, \citenamefont {Giai},\ and\ \citenamefont {Quentin}}]{(Bei75b)}%
  \BibitemOpen
  \bibfield  {author} {\bibinfo {author} {\bibfnamefont {M.}~\bibnamefont {Beiner}}, \bibinfo {author} {\bibfnamefont {H.}~\bibnamefont {Flocard}}, \bibinfo {author} {\bibfnamefont {N.~V.}\ \bibnamefont {Giai}},\ and\ \bibinfo {author} {\bibfnamefont {P.}~\bibnamefont {Quentin}},\ }\bibfield  {title} {\bibinfo {title} {Nuclear ground-state properties and self-consistent calculations with the {Skyrme} interaction: {(I). Spherical} description},\ }\href {https://doi.org/https://doi.org/10.1016/0375-9474(75)90338-3} {\bibfield  {journal} {\bibinfo  {journal} {Nuclear Physics A}\ }\textbf {\bibinfo {volume} {238}},\ \bibinfo {pages} {29 } (\bibinfo {year} {1975})}\BibitemShut {NoStop}%
\bibitem [{\citenamefont {Reinhard}(1999)}]{(Rei99)}%
  \BibitemOpen
  \bibfield  {author} {\bibinfo {author} {\bibfnamefont {P.-G.}\ \bibnamefont {Reinhard}},\ }\bibfield  {title} {\bibinfo {title} {Skyrme forces and giant resonances in exotic nuclei},\ }\href {https://doi.org/10.1016/S0375-9474(99)00076-7} {\bibfield  {journal} {\bibinfo  {journal} {Nucl. Phys. A}\ }\textbf {\bibinfo {volume} {649}},\ \bibinfo {pages} {305c} (\bibinfo {year} {1999})}\BibitemShut {NoStop}%
\bibitem [{\citenamefont {Chabanat}\ \emph {et~al.}(1998)\citenamefont {Chabanat}, \citenamefont {Bonche}, \citenamefont {Haensel}, \citenamefont {Meyer},\ and\ \citenamefont {Schaeffer}}]{(Cha98a)}%
  \BibitemOpen
  \bibfield  {author} {\bibinfo {author} {\bibfnamefont {E.}~\bibnamefont {Chabanat}}, \bibinfo {author} {\bibfnamefont {P.}~\bibnamefont {Bonche}}, \bibinfo {author} {\bibfnamefont {P.}~\bibnamefont {Haensel}}, \bibinfo {author} {\bibfnamefont {J.}~\bibnamefont {Meyer}},\ and\ \bibinfo {author} {\bibfnamefont {R.}~\bibnamefont {Schaeffer}},\ }\bibfield  {title} {\bibinfo {title} {A \protect{Skyrme} parametrization from subnuclear to neutron star densities \protect{Part II. Nuclei} far from stabilities},\ }\href {https://doi.org/https://doi.org/10.1016/S0375-9474(98)00180-8} {\bibfield  {journal} {\bibinfo  {journal} {Nuclear Physics A}\ }\textbf {\bibinfo {volume} {635}},\ \bibinfo {pages} {231 } (\bibinfo {year} {1998})}\BibitemShut {NoStop}%
\bibitem [{\citenamefont {Kortelainen}\ \emph {et~al.}(2012)\citenamefont {Kortelainen}, \citenamefont {McDonnell}, \citenamefont {Nazarewicz}, \citenamefont {Reinhard}, \citenamefont {Sarich}, \citenamefont {Schunck}, \citenamefont {Stoitsov},\ and\ \citenamefont {Wild}}]{(Kor12b)}%
  \BibitemOpen
  \bibfield  {author} {\bibinfo {author} {\bibfnamefont {M.}~\bibnamefont {Kortelainen}}, \bibinfo {author} {\bibfnamefont {J.}~\bibnamefont {McDonnell}}, \bibinfo {author} {\bibfnamefont {W.}~\bibnamefont {Nazarewicz}}, \bibinfo {author} {\bibfnamefont {P.-G.}\ \bibnamefont {Reinhard}}, \bibinfo {author} {\bibfnamefont {J.}~\bibnamefont {Sarich}}, \bibinfo {author} {\bibfnamefont {N.}~\bibnamefont {Schunck}}, \bibinfo {author} {\bibfnamefont {M.~V.}\ \bibnamefont {Stoitsov}},\ and\ \bibinfo {author} {\bibfnamefont {S.~M.}\ \bibnamefont {Wild}},\ }\bibfield  {title} {\bibinfo {title} {Nuclear energy density optimization: {Large} deformations},\ }\href {https://doi.org/10.1103/PhysRevC.85.024304} {\bibfield  {journal} {\bibinfo  {journal} {Phys. Rev. C}\ }\textbf {\bibinfo {volume} {85}},\ \bibinfo {pages} {024304} (\bibinfo {year} {2012})}\BibitemShut {NoStop}%
\bibitem [{\citenamefont {Dobaczewski}\ \emph {et~al.}(2018)\citenamefont {Dobaczewski}, \citenamefont {Engel}, \citenamefont {Kortelainen},\ and\ \citenamefont {Becker}}]{(Dob18a)}%
  \BibitemOpen
  \bibfield  {author} {\bibinfo {author} {\bibfnamefont {J.}~\bibnamefont {Dobaczewski}}, \bibinfo {author} {\bibfnamefont {J.}~\bibnamefont {Engel}}, \bibinfo {author} {\bibfnamefont {M.}~\bibnamefont {Kortelainen}},\ and\ \bibinfo {author} {\bibfnamefont {P.}~\bibnamefont {Becker}},\ }\bibfield  {title} {\bibinfo {title} {Correlating {Schiff} moments in the light actinides with octupole moments},\ }\href {https://doi.org/10.1103/PhysRevLett.121.232501} {\bibfield  {journal} {\bibinfo  {journal} {Phys. Rev. Lett.}\ }\textbf {\bibinfo {volume} {121}},\ \bibinfo {pages} {232501} (\bibinfo {year} {2018})}\BibitemShut {NoStop}%
\bibitem [{Note2()}]{Note2}%
  \BibitemOpen
  \bibinfo {note} {We use the same convention for $Q^3_0$ as in Ref.~\protect \cite {(Dob18a)}.}\BibitemShut {Stop}%
\bibitem [{\citenamefont {Gaffney}\ \emph {et~al.}(2013)\citenamefont {Gaffney}, \citenamefont {Butler}, \citenamefont {Scheck}, \citenamefont {Hayes}, \citenamefont {Wenander}, \citenamefont {Albers}, \citenamefont {Bastin}, \citenamefont {Bauer}, \citenamefont {Blazhev}, \citenamefont {B\"onig}, \citenamefont {Bree}, \citenamefont {Cederk\"all}, \citenamefont {Chupp}, \citenamefont {Cline}, \citenamefont {Cocolios}, \citenamefont {Davinson}, \citenamefont {De~Witte}, \citenamefont {Diriken}, \citenamefont {Grahn}, \citenamefont {Herzan}, \citenamefont {Huyse}, \citenamefont {Jenkins}, \citenamefont {Joss}, \citenamefont {Kesteloot}, \citenamefont {Konki}, \citenamefont {Kowalczyk}, \citenamefont {Kr\"oll}, \citenamefont {Kwan}, \citenamefont {Lutter}, \citenamefont {Moschner}, \citenamefont {Napiorkowski}, \citenamefont {Pakarinen}, \citenamefont {Pfeiffer}, \citenamefont {Radeck}, \citenamefont {Reiter}, \citenamefont {Reynders}, \citenamefont {Rigby}, \citenamefont {Robledo}, \citenamefont {Rudigier},
  \citenamefont {Sambi}, \citenamefont {Seidlitz}, \citenamefont {Siebeck}, \citenamefont {Stora}, \citenamefont {Thoele}, \citenamefont {Van~Duppen}, \citenamefont {Vermeulen}, \citenamefont {von Schmid}, \citenamefont {Voulot}, \citenamefont {Warr}, \citenamefont {Wimmer}, \citenamefont {Wrzosek-Lipska}, \citenamefont {Wu},\ and\ \citenamefont {Zielinska}}]{(Gaf13)}%
  \BibitemOpen
  \bibfield  {author} {\bibinfo {author} {\bibfnamefont {L.~P.}\ \bibnamefont {Gaffney}}, \bibinfo {author} {\bibfnamefont {P.~A.}\ \bibnamefont {Butler}}, \bibinfo {author} {\bibfnamefont {M.}~\bibnamefont {Scheck}}, \bibinfo {author} {\bibfnamefont {A.~B.}\ \bibnamefont {Hayes}}, \bibinfo {author} {\bibfnamefont {F.}~\bibnamefont {Wenander}}, \bibinfo {author} {\bibfnamefont {M.}~\bibnamefont {Albers}}, \bibinfo {author} {\bibfnamefont {B.}~\bibnamefont {Bastin}}, \bibinfo {author} {\bibfnamefont {C.}~\bibnamefont {Bauer}}, \bibinfo {author} {\bibfnamefont {A.}~\bibnamefont {Blazhev}}, \bibinfo {author} {\bibfnamefont {S.}~\bibnamefont {B\"onig}}, \bibinfo {author} {\bibfnamefont {N.}~\bibnamefont {Bree}}, \bibinfo {author} {\bibfnamefont {J.}~\bibnamefont {Cederk\"all}}, \bibinfo {author} {\bibfnamefont {T.}~\bibnamefont {Chupp}}, \bibinfo {author} {\bibfnamefont {D.}~\bibnamefont {Cline}}, \bibinfo {author} {\bibfnamefont {T.~E.}\ \bibnamefont {Cocolios}}, \bibinfo {author} {\bibfnamefont
  {T.}~\bibnamefont {Davinson}}, \bibinfo {author} {\bibfnamefont {H.}~\bibnamefont {De~Witte}}, \bibinfo {author} {\bibfnamefont {J.}~\bibnamefont {Diriken}}, \bibinfo {author} {\bibfnamefont {T.}~\bibnamefont {Grahn}}, \bibinfo {author} {\bibfnamefont {A.}~\bibnamefont {Herzan}}, \bibinfo {author} {\bibfnamefont {M.}~\bibnamefont {Huyse}}, \bibinfo {author} {\bibfnamefont {D.~G.}\ \bibnamefont {Jenkins}}, \bibinfo {author} {\bibfnamefont {D.~T.}\ \bibnamefont {Joss}}, \bibinfo {author} {\bibfnamefont {N.}~\bibnamefont {Kesteloot}}, \bibinfo {author} {\bibfnamefont {J.}~\bibnamefont {Konki}}, \bibinfo {author} {\bibfnamefont {M.}~\bibnamefont {Kowalczyk}}, \bibinfo {author} {\bibfnamefont {T.}~\bibnamefont {Kr\"oll}}, \bibinfo {author} {\bibfnamefont {E.}~\bibnamefont {Kwan}}, \bibinfo {author} {\bibfnamefont {R.}~\bibnamefont {Lutter}}, \bibinfo {author} {\bibfnamefont {K.}~\bibnamefont {Moschner}}, \bibinfo {author} {\bibfnamefont {P.}~\bibnamefont {Napiorkowski}}, \bibinfo {author} {\bibfnamefont
  {J.}~\bibnamefont {Pakarinen}}, \bibinfo {author} {\bibfnamefont {M.}~\bibnamefont {Pfeiffer}}, \bibinfo {author} {\bibfnamefont {D.}~\bibnamefont {Radeck}}, \bibinfo {author} {\bibfnamefont {P.}~\bibnamefont {Reiter}}, \bibinfo {author} {\bibfnamefont {K.}~\bibnamefont {Reynders}}, \bibinfo {author} {\bibfnamefont {S.~V.}\ \bibnamefont {Rigby}}, \bibinfo {author} {\bibfnamefont {L.~M.}\ \bibnamefont {Robledo}}, \bibinfo {author} {\bibfnamefont {M.}~\bibnamefont {Rudigier}}, \bibinfo {author} {\bibfnamefont {S.}~\bibnamefont {Sambi}}, \bibinfo {author} {\bibfnamefont {M.}~\bibnamefont {Seidlitz}}, \bibinfo {author} {\bibfnamefont {B.}~\bibnamefont {Siebeck}}, \bibinfo {author} {\bibfnamefont {T.}~\bibnamefont {Stora}}, \bibinfo {author} {\bibfnamefont {P.}~\bibnamefont {Thoele}}, \bibinfo {author} {\bibfnamefont {P.}~\bibnamefont {Van~Duppen}}, \bibinfo {author} {\bibfnamefont {M.~J.}\ \bibnamefont {Vermeulen}}, \bibinfo {author} {\bibfnamefont {M.}~\bibnamefont {von Schmid}}, \bibinfo {author}
  {\bibfnamefont {D.}~\bibnamefont {Voulot}}, \bibinfo {author} {\bibfnamefont {N.}~\bibnamefont {Warr}}, \bibinfo {author} {\bibfnamefont {K.}~\bibnamefont {Wimmer}}, \bibinfo {author} {\bibfnamefont {K.}~\bibnamefont {Wrzosek-Lipska}}, \bibinfo {author} {\bibfnamefont {C.~Y.}\ \bibnamefont {Wu}},\ and\ \bibinfo {author} {\bibfnamefont {M.}~\bibnamefont {Zielinska}},\ }\bibfield  {title} {\bibinfo {title} {Studies of pear-shaped nuclei using accelerated radioactive beams},\ }\href {https://doi.org/10.1038/nature12073} {\bibfield  {journal} {\bibinfo  {journal} {Nature}\ }\textbf {\bibinfo {volume} {497}},\ \bibinfo {pages} {199} (\bibinfo {year} {2013})}\BibitemShut {NoStop}%
\bibitem [{\citenamefont {Wollersheim}\ \emph {et~al.}(1993)\citenamefont {Wollersheim}, \citenamefont {Emling}, \citenamefont {Grein}, \citenamefont {Kulessa}, \citenamefont {Simon}, \citenamefont {Fleischmann}, \citenamefont {{de Boer}}, \citenamefont {Hauber}, \citenamefont {Lauterbach}, \citenamefont {Schandera}, \citenamefont {Butler},\ and\ \citenamefont {Czosnyka}}]{WOLLERSHEIM1993261}%
  \BibitemOpen
  \bibfield  {author} {\bibinfo {author} {\bibfnamefont {H.}~\bibnamefont {Wollersheim}}, \bibinfo {author} {\bibfnamefont {H.}~\bibnamefont {Emling}}, \bibinfo {author} {\bibfnamefont {H.}~\bibnamefont {Grein}}, \bibinfo {author} {\bibfnamefont {R.}~\bibnamefont {Kulessa}}, \bibinfo {author} {\bibfnamefont {R.}~\bibnamefont {Simon}}, \bibinfo {author} {\bibfnamefont {C.}~\bibnamefont {Fleischmann}}, \bibinfo {author} {\bibfnamefont {J.}~\bibnamefont {{de Boer}}}, \bibinfo {author} {\bibfnamefont {E.}~\bibnamefont {Hauber}}, \bibinfo {author} {\bibfnamefont {C.}~\bibnamefont {Lauterbach}}, \bibinfo {author} {\bibfnamefont {C.}~\bibnamefont {Schandera}}, \bibinfo {author} {\bibfnamefont {P.}~\bibnamefont {Butler}},\ and\ \bibinfo {author} {\bibfnamefont {T.}~\bibnamefont {Czosnyka}},\ }\bibfield  {title} {\bibinfo {title} {Coulomb excitation of {$^{226}$Ra}},\ }\href {https://doi.org/https://doi.org/10.1016/0375-9474(93)90351-W} {\bibfield  {journal} {\bibinfo  {journal} {Nuclear Physics A}\ }\textbf {\bibinfo
  {volume} {556}},\ \bibinfo {pages} {261} (\bibinfo {year} {1993})}\BibitemShut {NoStop}%
\bibitem [{\citenamefont {McGowan}\ \emph {et~al.}(1974)\citenamefont {McGowan}, \citenamefont {Bemis}, \citenamefont {Milner}, \citenamefont {Ford}, \citenamefont {Robinson},\ and\ \citenamefont {Stelson}}]{PhysRevC.10.1146}%
  \BibitemOpen
  \bibfield  {author} {\bibinfo {author} {\bibfnamefont {F.~K.}\ \bibnamefont {McGowan}}, \bibinfo {author} {\bibfnamefont {C.~E.}\ \bibnamefont {Bemis}}, \bibinfo {author} {\bibfnamefont {W.~T.}\ \bibnamefont {Milner}}, \bibinfo {author} {\bibfnamefont {J.~L.~C.}\ \bibnamefont {Ford}}, \bibinfo {author} {\bibfnamefont {R.~L.}\ \bibnamefont {Robinson}},\ and\ \bibinfo {author} {\bibfnamefont {P.~H.}\ \bibnamefont {Stelson}},\ }\bibfield  {title} {\bibinfo {title} {Coulomb excitation of vibrational-like states in the {even-$A$} actinide nuclei},\ }\href {https://doi.org/10.1103/PhysRevC.10.1146} {\bibfield  {journal} {\bibinfo  {journal} {Phys. Rev. C}\ }\textbf {\bibinfo {volume} {10}},\ \bibinfo {pages} {1146} (\bibinfo {year} {1974})}\BibitemShut {NoStop}%
\bibitem [{\citenamefont {Engel}\ \emph {et~al.}(1975)\citenamefont {Engel}, \citenamefont {Brink}, \citenamefont {Goeke}, \citenamefont {Krieger},\ and\ \citenamefont {Vautherin}}]{(Eng75)}%
  \BibitemOpen
  \bibfield  {author} {\bibinfo {author} {\bibfnamefont {Y.~M.}\ \bibnamefont {Engel}}, \bibinfo {author} {\bibfnamefont {D.~M.}\ \bibnamefont {Brink}}, \bibinfo {author} {\bibfnamefont {K.}~\bibnamefont {Goeke}}, \bibinfo {author} {\bibfnamefont {S.~J.}\ \bibnamefont {Krieger}},\ and\ \bibinfo {author} {\bibfnamefont {D.}~\bibnamefont {Vautherin}},\ }\bibfield  {title} {\bibinfo {title} {{Time-dependent} {Hartree-Fock} theory with {Skyrme's} interaction},\ }\href@noop {} {\bibfield  {journal} {\bibinfo  {journal} {Nucl. Phys. A}\ }\textbf {\bibinfo {volume} {249}},\ \bibinfo {pages} {215} (\bibinfo {year} {1975})}\BibitemShut {NoStop}%
\bibitem [{\citenamefont {Perli\'{n}ska}\ \emph {et~al.}(2004)\citenamefont {Perli\'{n}ska}, \citenamefont {Rohozi\'{n}ski}, \citenamefont {Dobaczewski},\ and\ \citenamefont {Nazarewicz}}]{(Per04c)}%
  \BibitemOpen
  \bibfield  {author} {\bibinfo {author} {\bibfnamefont {E.}~\bibnamefont {Perli\'{n}ska}}, \bibinfo {author} {\bibfnamefont {S.~G.}\ \bibnamefont {Rohozi\'{n}ski}}, \bibinfo {author} {\bibfnamefont {J.}~\bibnamefont {Dobaczewski}},\ and\ \bibinfo {author} {\bibfnamefont {W.}~\bibnamefont {Nazarewicz}},\ }\bibfield  {title} {\bibinfo {title} {Local density approximation for proton-neutron pairing correlations: {Formalism}},\ }\href {https://doi.org/10.1103/PhysRevC.69.014316} {\bibfield  {journal} {\bibinfo  {journal} {Phys. Rev. C}\ }\textbf {\bibinfo {volume} {69}},\ \bibinfo {pages} {014316} (\bibinfo {year} {2004})}\BibitemShut {NoStop}%
\bibitem [{\citenamefont {Schunck}(2019)}]{(Sch19b)}%
  \BibitemOpen
  \bibfield  {author} {\bibinfo {author} {\bibfnamefont {N.}~\bibnamefont {Schunck}},\ }\href {https://doi.org/10.1088/2053-2563/aae0ed} {\emph {\bibinfo {title} {{Energy} density functional methods for atomic nuclei}}},\ IOP Expanding Physics\ (\bibinfo  {publisher} {IOP Publishing},\ \bibinfo {address} {Bristol, UK},\ \bibinfo {year} {2019})\ \bibinfo {note} {oCLC: 1034572493}\BibitemShut {NoStop}%
\bibitem [{Note3()}]{Note3}%
  \BibitemOpen
  \bibinfo {note} {We are bound to keep the traditional notation, denoting by the same symbol $\Omega $ the projections of the angular momentum on the axial symmetry axis and of the magnetic octupole moment.}\BibitemShut {Stop}%
\bibitem [{\citenamefont {Ring}\ and\ \citenamefont {Schuck}(1980)}]{(Rin80b)}%
  \BibitemOpen
  \bibfield  {author} {\bibinfo {author} {\bibfnamefont {P.}~\bibnamefont {Ring}}\ and\ \bibinfo {author} {\bibfnamefont {P.}~\bibnamefont {Schuck}},\ }\href {https://www.springer.com/gp/book/9783540212065} {\emph {\bibinfo {title} {The nuclear many-body problem}}}\ (\bibinfo  {publisher} {Springer-Verlag, Berlin},\ \bibinfo {year} {1980})\BibitemShut {NoStop}%
\bibitem [{\citenamefont {Dobaczewski}\ \emph {et~al.}(2009)\citenamefont {Dobaczewski}, \citenamefont {Satu{\l}a}, \citenamefont {Carlsson}, \citenamefont {Engel}, \citenamefont {Olbratowski}, \citenamefont {Powa{\l}owski}, \citenamefont {Sadziak}, \citenamefont {Sarich}, \citenamefont {Schunck}, \citenamefont {Staszczak}, \citenamefont {Stoitsov}, \citenamefont {Zalewski},\ and\ \citenamefont {Zdu{\'n}czuk}}]{(Dob09g)}%
  \BibitemOpen
  \bibfield  {author} {\bibinfo {author} {\bibfnamefont {J.}~\bibnamefont {Dobaczewski}}, \bibinfo {author} {\bibfnamefont {W.}~\bibnamefont {Satu{\l}a}}, \bibinfo {author} {\bibfnamefont {B.}~\bibnamefont {Carlsson}}, \bibinfo {author} {\bibfnamefont {J.}~\bibnamefont {Engel}}, \bibinfo {author} {\bibfnamefont {P.}~\bibnamefont {Olbratowski}}, \bibinfo {author} {\bibfnamefont {P.}~\bibnamefont {Powa{\l}owski}}, \bibinfo {author} {\bibfnamefont {M.}~\bibnamefont {Sadziak}}, \bibinfo {author} {\bibfnamefont {J.}~\bibnamefont {Sarich}}, \bibinfo {author} {\bibfnamefont {N.}~\bibnamefont {Schunck}}, \bibinfo {author} {\bibfnamefont {A.}~\bibnamefont {Staszczak}}, \bibinfo {author} {\bibfnamefont {M.}~\bibnamefont {Stoitsov}}, \bibinfo {author} {\bibfnamefont {M.}~\bibnamefont {Zalewski}},\ and\ \bibinfo {author} {\bibfnamefont {H.}~\bibnamefont {Zdu{\'n}czuk}},\ }\bibfield  {title} {\bibinfo {title} {Solution of the {Skyrme-Hartree-Fock-Bogolyubov} equations in the {Cartesian} deformed harmonic-oscillator
  basis.: {(VI) {\sc hfodd}} (v2.40h): {A} new version of the program},\ }\href {https://doi.org/10.1016/j.cpc.2009.08.009} {\bibfield  {journal} {\bibinfo  {journal} {Comput. Phys. Commun.}\ }\textbf {\bibinfo {volume} {180}},\ \bibinfo {pages} {2361 } (\bibinfo {year} {2009})}\BibitemShut {NoStop}%
\bibitem [{\citenamefont {Bertsch}\ \emph {et~al.}(2009)\citenamefont {Bertsch}, \citenamefont {Dobaczewski}, \citenamefont {Nazarewicz},\ and\ \citenamefont {Pei}}]{(Ber09d)}%
  \BibitemOpen
  \bibfield  {author} {\bibinfo {author} {\bibfnamefont {G.}~\bibnamefont {Bertsch}}, \bibinfo {author} {\bibfnamefont {J.}~\bibnamefont {Dobaczewski}}, \bibinfo {author} {\bibfnamefont {W.}~\bibnamefont {Nazarewicz}},\ and\ \bibinfo {author} {\bibfnamefont {J.}~\bibnamefont {Pei}},\ }\bibfield  {title} {\bibinfo {title} {{Hartree-Fock-Bogoliubov} theory of polarized {Fermi} systems},\ }\href {https://doi.org/10.1103/PhysRevA.79.043602} {\bibfield  {journal} {\bibinfo  {journal} {Phys. Rev. A}\ }\textbf {\bibinfo {volume} {79}},\ \bibinfo {pages} {043602} (\bibinfo {year} {2009})}\BibitemShut {NoStop}%
\bibitem [{Note4()}]{Note4}%
  \BibitemOpen
  \bibinfo {note} {The dominant Nilsson labels~\protect \cite {(Dob97c)} correspond to the largest component of a given s.p.\ state when it is expanded on the asymptotic Nilsson states $[N_0n_z\Lambda ]\Omega $~\protect \cite {(Rin80b)}.}\BibitemShut {Stop}%
\bibitem [{\citenamefont {Sheikh}\ \emph {et~al.}(2021)\citenamefont {Sheikh}, \citenamefont {Dobaczewski}, \citenamefont {Ring}, \citenamefont {Robledo},\ and\ \citenamefont {Yannouleas}}]{(She21)}%
  \BibitemOpen
  \bibfield  {author} {\bibinfo {author} {\bibfnamefont {J.~A.}\ \bibnamefont {Sheikh}}, \bibinfo {author} {\bibfnamefont {J.}~\bibnamefont {Dobaczewski}}, \bibinfo {author} {\bibfnamefont {P.}~\bibnamefont {Ring}}, \bibinfo {author} {\bibfnamefont {L.~M.}\ \bibnamefont {Robledo}},\ and\ \bibinfo {author} {\bibfnamefont {C.}~\bibnamefont {Yannouleas}},\ }\bibfield  {title} {\bibinfo {title} {Symmetry restoration in mean-field approaches},\ }\href {https://doi.org/10.1088/1361-6471/ac288a} {\bibfield  {journal} {\bibinfo  {journal} {Journal of Physics G: Nuclear and Particle Physics}\ }\textbf {\bibinfo {volume} {48}},\ \bibinfo {pages} {123001} (\bibinfo {year} {2021})}\BibitemShut {NoStop}%
\bibitem [{\citenamefont {Sadoudi}\ \emph {et~al.}(2013{\natexlab{a}})\citenamefont {Sadoudi}, \citenamefont {Bender}, \citenamefont {Bennaceur}, \citenamefont {Davesne}, \citenamefont {Jodon},\ and\ \citenamefont {Duguet}}]{(Sad13)}%
  \BibitemOpen
  \bibfield  {author} {\bibinfo {author} {\bibfnamefont {J.}~\bibnamefont {Sadoudi}}, \bibinfo {author} {\bibfnamefont {M.}~\bibnamefont {Bender}}, \bibinfo {author} {\bibfnamefont {K.}~\bibnamefont {Bennaceur}}, \bibinfo {author} {\bibfnamefont {D.}~\bibnamefont {Davesne}}, \bibinfo {author} {\bibfnamefont {R.}~\bibnamefont {Jodon}},\ and\ \bibinfo {author} {\bibfnamefont {T.}~\bibnamefont {Duguet}},\ }\bibfield  {title} {\bibinfo {title} {{Skyrme} pseudo-potential-based {EDF} parametrization for spuriousity-free {MR} {EDF} calculations},\ }\href {https://doi.org/10.1088/0031-8949/2013/t154/014013} {\bibfield  {journal} {\bibinfo  {journal} {Physica Scripta}\ }\textbf {\bibinfo {volume} {T154}},\ \bibinfo {pages} {014013} (\bibinfo {year} {2013}{\natexlab{a}})}\BibitemShut {NoStop}%
\bibitem [{\citenamefont {Sadoudi}\ \emph {et~al.}(2013{\natexlab{b}})\citenamefont {Sadoudi}, \citenamefont {Duguet}, \citenamefont {Meyer},\ and\ \citenamefont {Bender}}]{(Sad13b)}%
  \BibitemOpen
  \bibfield  {author} {\bibinfo {author} {\bibfnamefont {J.}~\bibnamefont {Sadoudi}}, \bibinfo {author} {\bibfnamefont {T.}~\bibnamefont {Duguet}}, \bibinfo {author} {\bibfnamefont {J.}~\bibnamefont {Meyer}},\ and\ \bibinfo {author} {\bibfnamefont {M.}~\bibnamefont {Bender}},\ }\bibfield  {title} {\bibinfo {title} {{Skyrme} functional from a three-body pseudopotential of second order in gradients: {Formalism} for central terms},\ }\href {https://doi.org/10.1103/PhysRevC.88.064326} {\bibfield  {journal} {\bibinfo  {journal} {Phys. Rev. C}\ }\textbf {\bibinfo {volume} {88}},\ \bibinfo {pages} {064326} (\bibinfo {year} {2013}{\natexlab{b}})}\BibitemShut {NoStop}%
\bibitem [{\citenamefont {Bennaceur}\ \emph {et~al.}(2017)\citenamefont {Bennaceur}, \citenamefont {Idini}, \citenamefont {Dobaczewski}, \citenamefont {Dobaczewski}, \citenamefont {Kortelainen},\ and\ \citenamefont {Raimondi}}]{(Ben17)}%
  \BibitemOpen
  \bibfield  {author} {\bibinfo {author} {\bibfnamefont {K.}~\bibnamefont {Bennaceur}}, \bibinfo {author} {\bibfnamefont {A.}~\bibnamefont {Idini}}, \bibinfo {author} {\bibfnamefont {J.}~\bibnamefont {Dobaczewski}}, \bibinfo {author} {\bibfnamefont {P.}~\bibnamefont {Dobaczewski}}, \bibinfo {author} {\bibfnamefont {M.}~\bibnamefont {Kortelainen}},\ and\ \bibinfo {author} {\bibfnamefont {F.}~\bibnamefont {Raimondi}},\ }\bibfield  {title} {\bibinfo {title} {Nonlocal energy density functionals for pairing and beyond-mean-field calculations},\ }\href {https://doi.org/10.1088/1361-6471/aa5fd7} {\bibfield  {journal} {\bibinfo  {journal} {Journal of Physics G: Nuclear and Particle Physics}\ }\textbf {\bibinfo {volume} {44}},\ \bibinfo {pages} {045106} (\bibinfo {year} {2017})}\BibitemShut {NoStop}%
\bibitem [{rep()}]{rep-229Th}%
  \BibitemOpen
  \href@noop {} {}\bibinfo {note} {{Repository of raw data obtained in this work is available at this URL \url{https://webfiles.york.ac.uk/HFODD/Projects/229Th} }}\BibitemShut {NoStop}%
\bibitem [{\citenamefont {Slater}(1951)}]{(Sla51a)}%
  \BibitemOpen
  \bibfield  {author} {\bibinfo {author} {\bibfnamefont {J.~C.}\ \bibnamefont {Slater}},\ }\bibfield  {title} {\bibinfo {title} {A simplification of the {Hartree-Fock} method},\ }\href {https://doi.org/10.1103/PhysRev.81.385} {\bibfield  {journal} {\bibinfo  {journal} {Phys. Rev.}\ }\textbf {\bibinfo {volume} {81}},\ \bibinfo {pages} {385} (\bibinfo {year} {1951})}\BibitemShut {NoStop}%
\bibitem [{\citenamefont {Dobaczewski}\ and\ \citenamefont {Dudek}(1997)}]{(Dob97c)}%
  \BibitemOpen
  \bibfield  {author} {\bibinfo {author} {\bibfnamefont {J.}~\bibnamefont {Dobaczewski}}\ and\ \bibinfo {author} {\bibfnamefont {J.}~\bibnamefont {Dudek}},\ }\bibfield  {title} {\bibinfo {title} {Solution of the {Skyrme-Hartree-Fock} equations in the {Cartesian} deformed harmonic oscillator basis {II. The program {\sc hfodd}}},\ }\href {https://doi.org/https://doi.org/10.1016/S0010-4655(97)00005-2} {\bibfield  {journal} {\bibinfo  {journal} {Comput. Phys. Comm.}\ }\textbf {\bibinfo {volume} {102}},\ \bibinfo {pages} {183 } (\bibinfo {year} {1997})}\BibitemShut {NoStop}%
\end{thebibliography}
%\bibliographystyle{unsrt}

\clearpage

% How to produce a separate stand-alone LaTeX file containing the main text only.
% -------------------------------------------------------------------
% 1. Remove all lines from "How to produce a separate" above to the line before "\end{document}" of this file.
% 2. Remove the line "\clearpage" above.
% 3. Comment the line "\bibliography" above
% 4. Add line \label{LastBibItem} at the end of file  "output.bbl".
% 5. Copy the file "output.bbl" below the line "\bibliography" above.
% 6. ATTENTION! Inclusion of "output.bbl" in the main file does not work well with footnotes!
% -------------------------------------------------------------------------------
% How to produce a separate LaTeX file containing the Supplemental material only.
% -------------------------------------------------------------------------------
% 1. Remove all lines between and including "\date{today}" and "How to produce a separate" above.
% 2. Remove lines et the end:
%    \clearpage
%    \begin{widetext}
%    \newpage
%    \end{widetext}
% 3. Include the lines like:
%    \setcounter{figure}{4}
% 4. Copy all lines "\newlabel" from the "output.aux" file to after the lines added in 4.

\title{Supplemental Material for: \protect\mytitle}

\date{\today}

\begin{abstract}
\protect\myabstract
\end{abstract}

\maketitle

{\it Convergence and singularities}---Our calculations are limited by divergences and singularities arising from the self-interaction and self-pairing properties of the currently available Skyrme functionals~\cite{(She21)}, which often render solutions unreliable. We were forced to discard such calculations at the mean-field or post-mean-field stages. For the parity-breaking TE and TE+TO calculations, Table~\ref{table_mixing} summarizes the properties of the six configurations per functional studied in this Letter.
\newcommand{\br}[1]{\color{red}{{\bf #1}}}

\begin{widetext}

\begin{table}[h!]
%\centering

\vspace*{-2mm}
\caption
%
%\noindent
{Divergences and singularities encountered in the calculations reported in this Letter. Numbers in parentheses, (1) and (0), denote converged and non-converged single-reference solutions, respectively, obtained for the Nilsson labels of the tag states $(N_0n_z\Lambda)$. Numbers without parentheses, 1 and 0, denote non-singular and singular Hamiltonian matrix elements $H^{I^+\Omega}_{nn'}$. Boldface entries denote single reference solutions corresponding to the self-consistent Nilsson labels of [633]5/2 or [631]3/2.\label{table_mixing}}

\vspace*{2mm}
\begin{tabular}{|l|clll|clll|clll|clll|}
           \cline{2-17}
\multicolumn{1}{c|}{}           & \multicolumn{4}{c|}{$\Omega=5/2$ (TE)   } & \multicolumn{4}{c|}{$\Omega=3/2$ (TE)   }
                               & \multicolumn{4}{|c|}{$\Omega=5/2$ (TE+TO)} & \multicolumn{4}{c|}{$\Omega=3/2$ (TE+TO)} \\
\hline
Skyrme     & $(N_0n_z\Lambda)$
           & \rotatebox[origin=c]{90}{~(633)~}
           &  \rotatebox[origin=c]{90}{~(752)~}
           &  \rotatebox[origin=c]{90}{~(622)~}
           & $(N_0n_z\Lambda)$
           & \rotatebox[origin=c]{90}{~(642)~}
           &  \rotatebox[origin=c]{90}{~(741)~}
           &  \rotatebox[origin=c]{90}{~(631)~}
           & $(N_0n_z\Lambda)$
           & \rotatebox[origin=c]{90}{~(633)~}
           &  \rotatebox[origin=c]{90}{~(752)~}
           &  \rotatebox[origin=c]{90}{~(622)~}
           & $(N_0n_z\Lambda)$
           & \rotatebox[origin=c]{90}{~(642)~}
           &  \rotatebox[origin=c]{90}{~(741)~}
           &  \rotatebox[origin=c]{90}{~(631)~} \\
\hline
                   &      (633)  & \br{1(1)} &     1      &     1     &    (642)   &     1(1) &     0    &     0    &      (633)  & \br{0(0)} &     0     &     0     &     (642)   &     0(1)  &     0     &     0      \\
SkX$_{\text{c}}$   &      (752)  &     1     &     1(1)  &     1     &     (741)   &     0     &     0(1)  &     0     &      (752)  &     0     &     1(1)  &     1     &     (741)   &     0     &     0(1)  &     0      \\
                   &      (622)  &     1     &     1     &     1(1)  &     (631)   &     0     &     0     &  \br{0(1)}&      (622)  &     0     &     1     &     1(1)  &     (631)   &     0     &     0     &\br{0(1)}   \\

\hline
                   &      (633)  & \br{0(1)} &     0     &     0     &     (642)   &     1(1)  &     0     &     1     &      (633)  & \br{1(1)} &     0     &     0     &     (642)   & \br{1(1)}  &     0     &     1      \\
SkM*               &      (752)  &     0     &     1(1)  &     0     &     (741)   &     0     &     0(0)  &     0     &      (752)  &     0     &     0(0)  &     0     &     (741)   &     0     &     0(0)   &     0      \\
                   &      (622)  &     0     &     0     &     0(1)  &     (631)   &     1     &     0     & \br{1(1)} &      (622)  &     0     &     0     &     0(1)  &     (631)   &     1     &     0     &     1(1)   \\

\hline
                   &      (633)  & \br{0(0)} &     0     &     0     &     (642)   &     1(1)  &     0     &     0     &      (633)  &     0(0)  &     0     &     0     &     (642)   &     0(1)  &     0     &     0      \\
UNEDF1             &      (752)  &     0     &     1(1)  &     0     &     (741)   &     0     &     0(1)  &     0     &      (752)  &     0     &     1(1)  &     0     &     (741)   &     0     &     0(1)  &     0      \\
                   &      (622)  &     0     &     0     &     0(1)  &     (631)   &     0     &     0     & \br{0(1)} &      (622)  &     0     &     0     &     0(1)  &     (631)   &     0     &     0     & \br{1(1)}   \\

\hline
                   &      (633)  & \br{0(0)} &     0     &     0     &     (642)   &     1(1)  &     0     &     1     &      (633)  & \br{1(1)} &     1     &     1     &     (642)   &     0(0)  &     0     &     0      \\
SIII               &      (752)  &     0     &     0(1)  &     0     &     (741)   &     0     &     0(1)  &     0     &      (752)  &     1     &     1(1)  &     1     &     (741)   &     0     &     1(1)  &     1      \\
                   &      (622)  &     0     &     0     &     1(1)  &     (631)   &     1     &     0     & \br{1(1)} &      (622)  &     1     &     1     &     1(1)  &     (631)   &     0     &     1     & \br{1(1)}   \\
\hline
                   &      (633)  &\br{1(1)}  &     0     &     0     &     (642)   &     1(1)  &     1     &     1     &      (633)  &  \br{0(1)} &     0     &     0     &     (642)   &    0(0)  &     0     &     0      \\
SkO$^\prime$       &      (752)  &     0     &     0(0)  &     0     &     (741)   &     1     & 1(1)      &     1     &      (752)  &     0     &     0(0)  &     0     &     (741)   &     0     &     0(1)  &     0      \\
                   &      (622)  &     0     &     0     &     0(1)  &     (631)   &     1     &     1     &  \br{1(1)}&      (622)  &     0     &     0     &     0(1)  &     (631)   &     0     &     0     & \br{0(1)}   \\

\hline
                   &      (633)  &\br{1(1) } &     0     &     1     &     (642)   &     1(1)  &     0     &     0     &      (633)  & \br{1(1)} &     0     &     0     &     (642)   &     0(1)  &     0     &     0      \\
SLy4               &      (752)  &     0     &     0(1)  &     0     &     (741)   &     0     &     0(1)  &     0     &      (752)  &     0     &     0(1)  &     0     &     (741)   &     0     &     0(1)  &     0      \\
                   &      (622)  &     1     &     0     &     1(1)  &     (631)   &     0     &     0     & \br{0(0)} &      (622)  &     0     &     0     &     0(1)  &     (631)   &     0     &     0     & \br{1(1)}  \\

\hline
                   &      (633)  & \br{1(1)} &     0     &     1     &     (642)   &     1(1)  &     0     &     1     &      (633)  &\br{0(1)}  &     0     &     0     &     (642)   &     1(1)  &     1     &     0      \\
UNEDF0             &      (752)  &     0     &     0(1)  &     0     &     (741)   &     0     & 0(1)      &     0     &      (752)  &     0     &     0(1)  &     0     &     (741)   &     1     & \br{1(1)} &     0      \\
                   &      (622)  &     1     &     0     &     1(1)  &     (631)   &     1     &     0     & \br{1(1)} &      (622)  &     0     &     0     &     1(1)  &     (631)   &     0     &     0     &     0(0)   \\

\hline
\end{tabular}
\end{table}

\end{widetext}

{\it Parameters and numerical conditions}---We performed the calculations using a 3D harmonic oscillator basis with $N_0\leq$16 shells for the harmonic oscillator frequencies~\cite{(Dob97c)}, $\hbar\omega_0=1.2\times41\,A^{-1/3}=8.1012$\,{MeV}, and the oscillator length $b= \sqrt{{\hbar}/{m\omega_0}}= 2.2625862$\,fm, identical in three Cartesian directions.

\begin{table}[b!]
\caption{The Landau parameters $g_0'$, strengths of the proton and neutron volume pairing interactions, $V_p$ and $V_n$ (in MeV), and intrinsic octupole moments, $Q^3_0$ (in 1000\,fm$^3$), of $^{226}$Ra and $^{230}$Th used or calculated in this Letter.}
\label{table:parameters}
\vspace*{2mm}
\centering
\begin{tabular}{l c c c c c c}
\hline
Skyrme & $g_0'$ & $V_{0,n}$ & $V_{0,p}$  & $Q^3_0(^{226}$Ra) &  $Q^3_0(^{230}$Th)  \\
\hline
SkX$_{\text{c}}$   & 1.2 & 139.02 & 173.63 & 0.6654 & 0.6446\\
SkM*               & 1.2 & 181.47 & 216.25 & 1.0763 & 0.9667\\
UNEDF1             & 1.7 & 145.35 & 169.79 & 0.8194 & 1.0797\\
SIII               & 1.2 & 181.14 & 220.19 & 0.8742 & 2.1131\\
SkO$^\prime$       & 1.0 & 163.82 & 184.34 & 0.9482 & 1.2839\\
SLy4               & 1.3 & 207.76 & 231.89 & 1.0475 & 1.0088\\
UNEDF0             & 1.2 & 130.61 & 158.39 & 0.8887 & 1.2077\\
\hline
\end{tabular}
\end{table}
We used the standard time-even-sector parameters of the Skyrme functionals, as given in the original publications~\cite{(Bro98c),(Bar82c),(Kor10c),(Bei75b),(Rei99),(Cha98a),(Kor12b)}.
In the time-odd sector, for the UNEDF1, SKO$'$, and SLy4 functionals, the Landau isovector parameters $g_0'$ were taken from Ref.~\cite{(Sas22c)}, and for the remaining functionals, they were set to the recommended value of 1.2~\cite{(Ben02d)}. For all functionals, the Landau isoscalar parameters were fixed at 0.4~\cite{(Ben02d)}. The pairing strengths $V_{0,n}$ and $V_{0,p}$ were determined from the mass staggering of neighboring isotopes of $^{229}$Th and $^{227}$Ac and taken from Ref.~\cite{Athanasakis-Kaklamanakis2025}. All these parameters are listed in Table~\ref{table:parameters}.

In the convention of Ref.~\cite{(Per04c)}, the coupling constants of the Skyrme functionals used in this Letter are displayed in Table~\protect\ref{table_Skyrme}. Values of $C_t^s$ where obtained from the Landau parameters $g_0$ and $g'_0$ with $C^{s}_t[\rho_0]=C^{s}_t[0]$, where $\rho_0$ is the nuclear matter saturation density. The Coulomb exchange terms were evaluated within the Slater approximation~\cite{(Sla51a)}.

{\it Results in numerical form}---In Table~\protect\ref{table:numerical1}, we show the unrounded numerical values of the B(M1; $3/2^+_1\rightarrow 5/2^+_1)$ transition probabilities (in $\mu_N^2$), spectroscopic magnetic dipole moments $\mu$ (in $\mu_N$), spectroscopic electric quadrupole moments $Q$ (in barn), and spectroscopic magnetic octupole moments $\Omega$ (in $\mu_N$\,b) calculated for the mixed, not mixed, and no-octupole $5/2^+$ and $3/2^+$ states. The upper, middle, and lower groups of rows correspond to the left, center, and right panels of Fig.~\protect\ref{Summary}. Values and uncertainties obtained from the regression analysis relative to the $^{226}$Ra and $^{230}$Th data are also shown. Values without uncertainties correspond to a two-point regression. The last row lists the experimental data~\cite{PhysRevLett.132.182501,(Bee25),Yamaguchi2024}.
%\newline

\begin{figure}
\centering\includegraphics[width=0.44\textwidth]{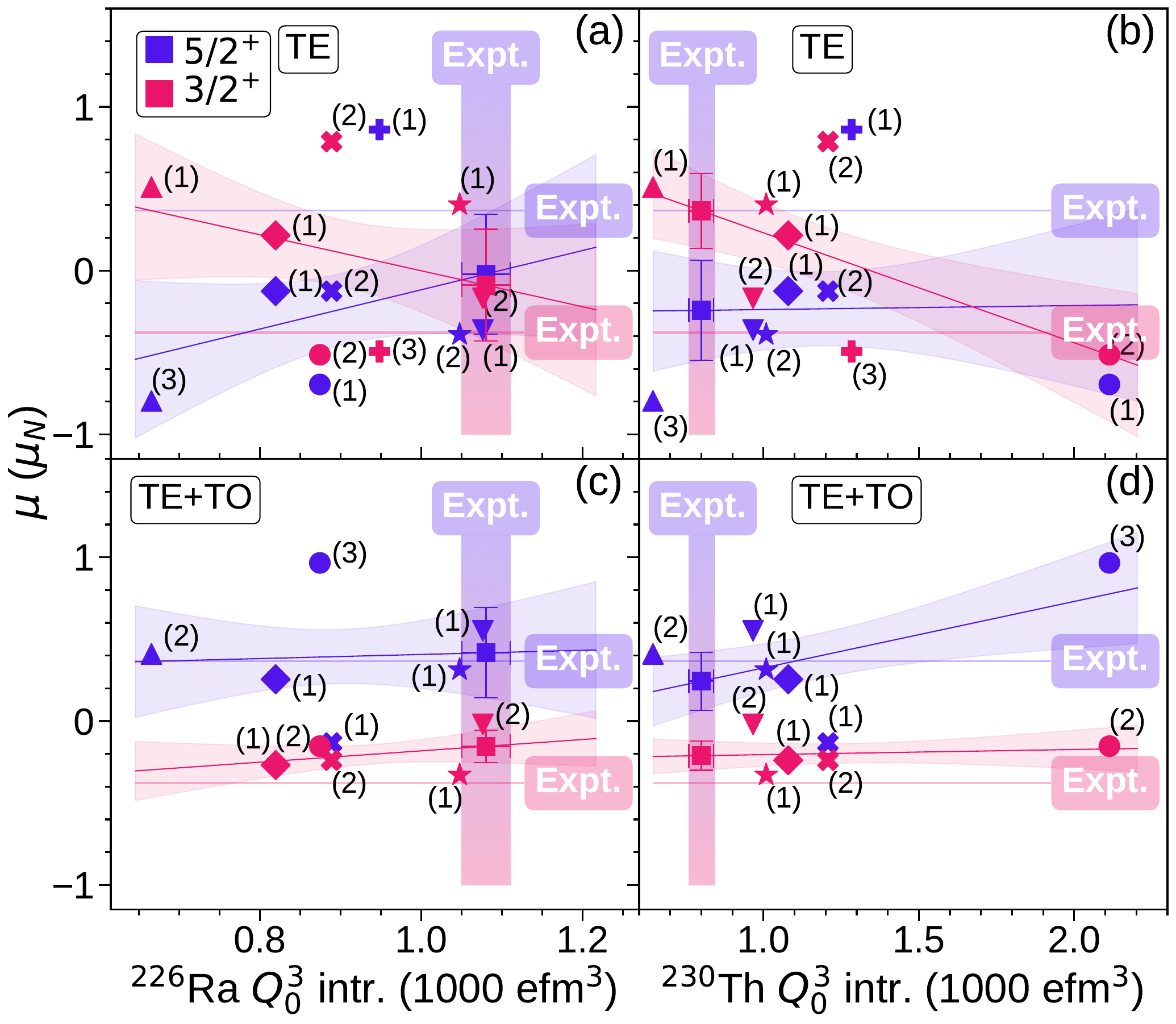}
\caption{Same as Fig.~\protect\ref{Corr1a} but for magnetic dipole moments.
\label{CorrMM1}}
%\end{figure}

\vspace*{3mm}
%\begin{figure}
\centering\includegraphics[width=0.44\textwidth]{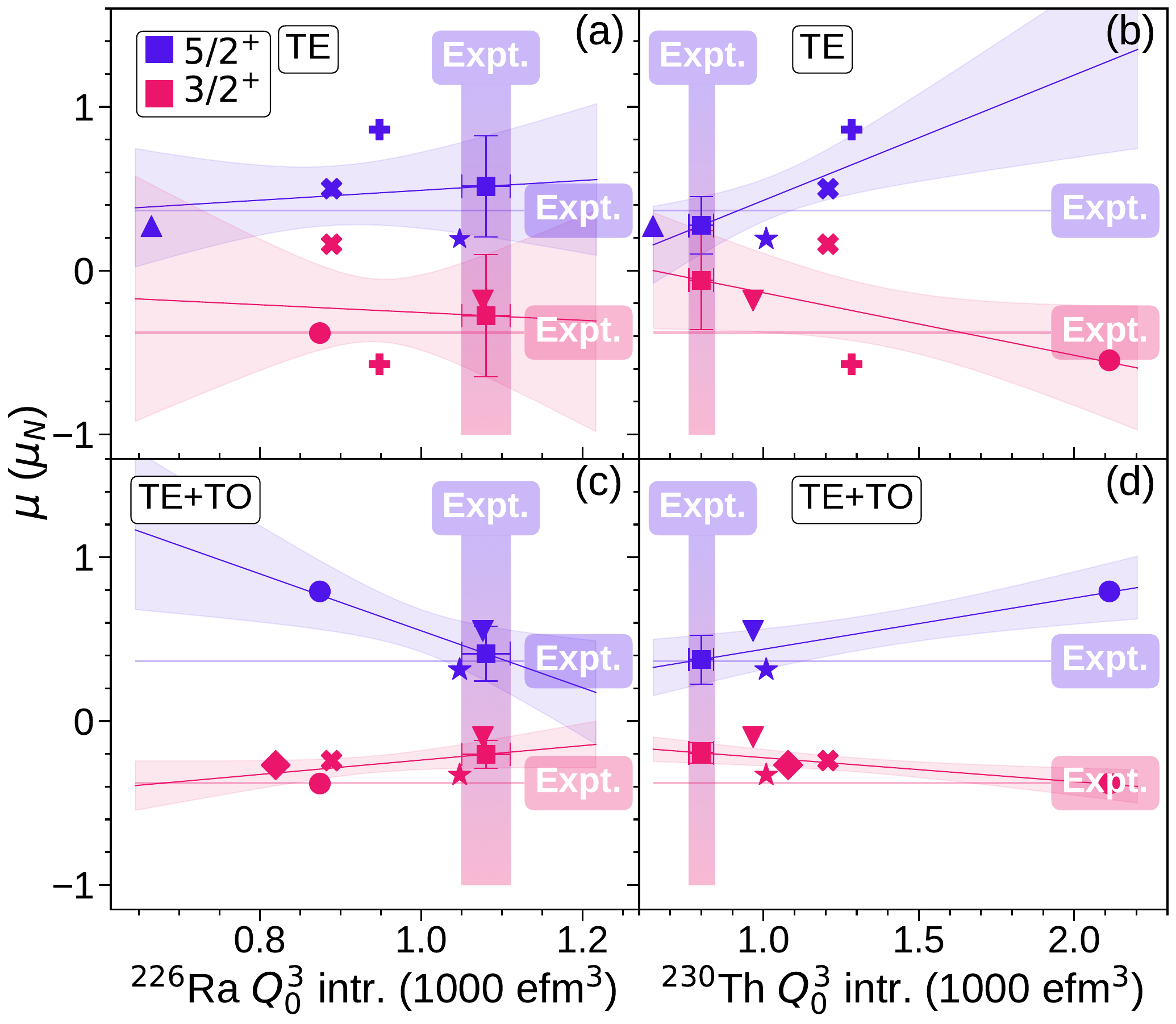}
\caption{Same as Fig.~\protect\ref{Corr2a} but for magnetic dipole moments.
\label{CorrM1}}
%\end{figure}

\vspace*{3mm}
%\begin{figure}
\centering\includegraphics[width=0.44\textwidth]{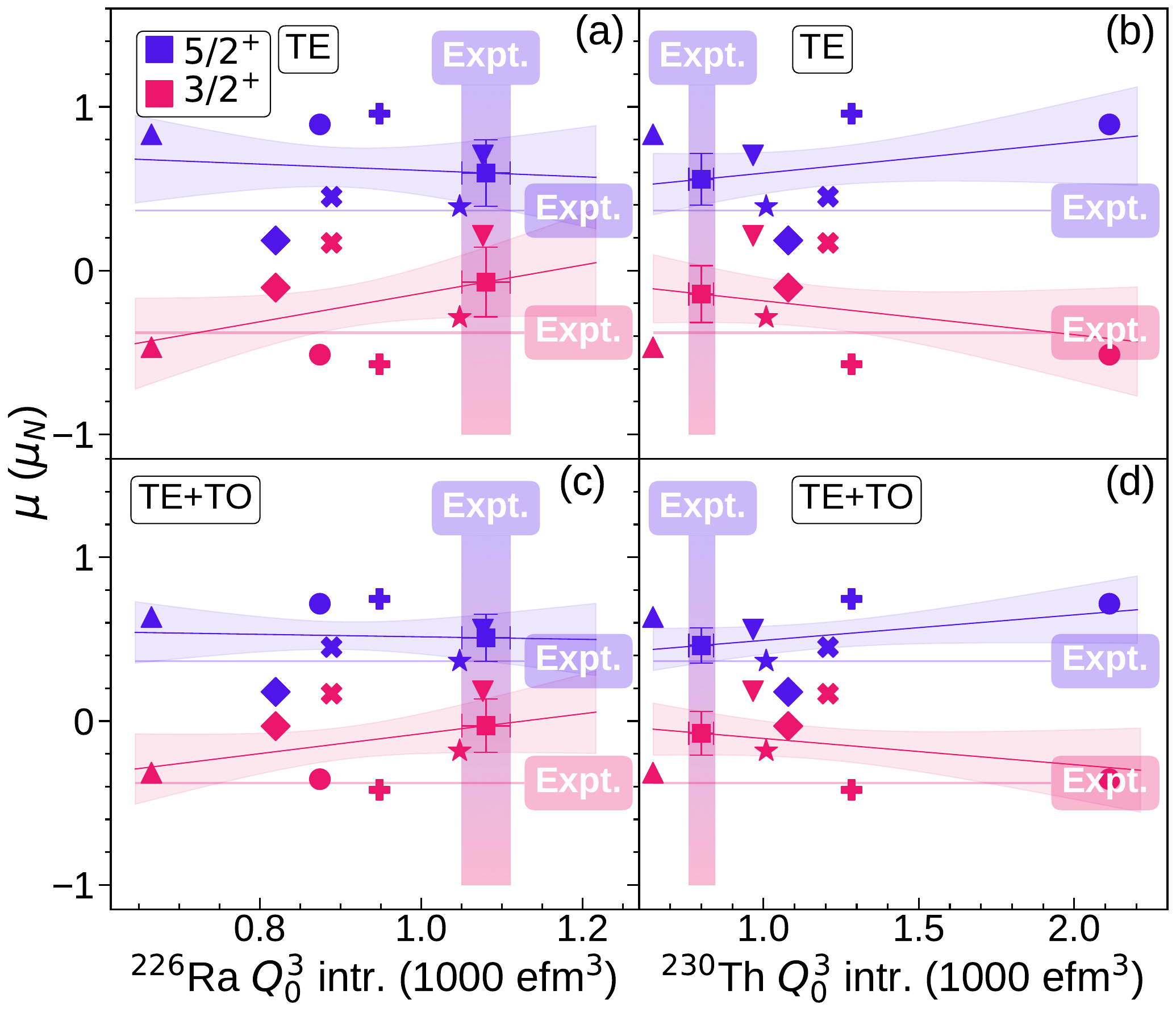}
\caption{Same as Fig.~\protect\ref{Corr3a} but for magnetic dipole moments.
\label{CorrMM2}}
\end{figure}

\begin{figure}
\centering\includegraphics[width=0.44\textwidth]{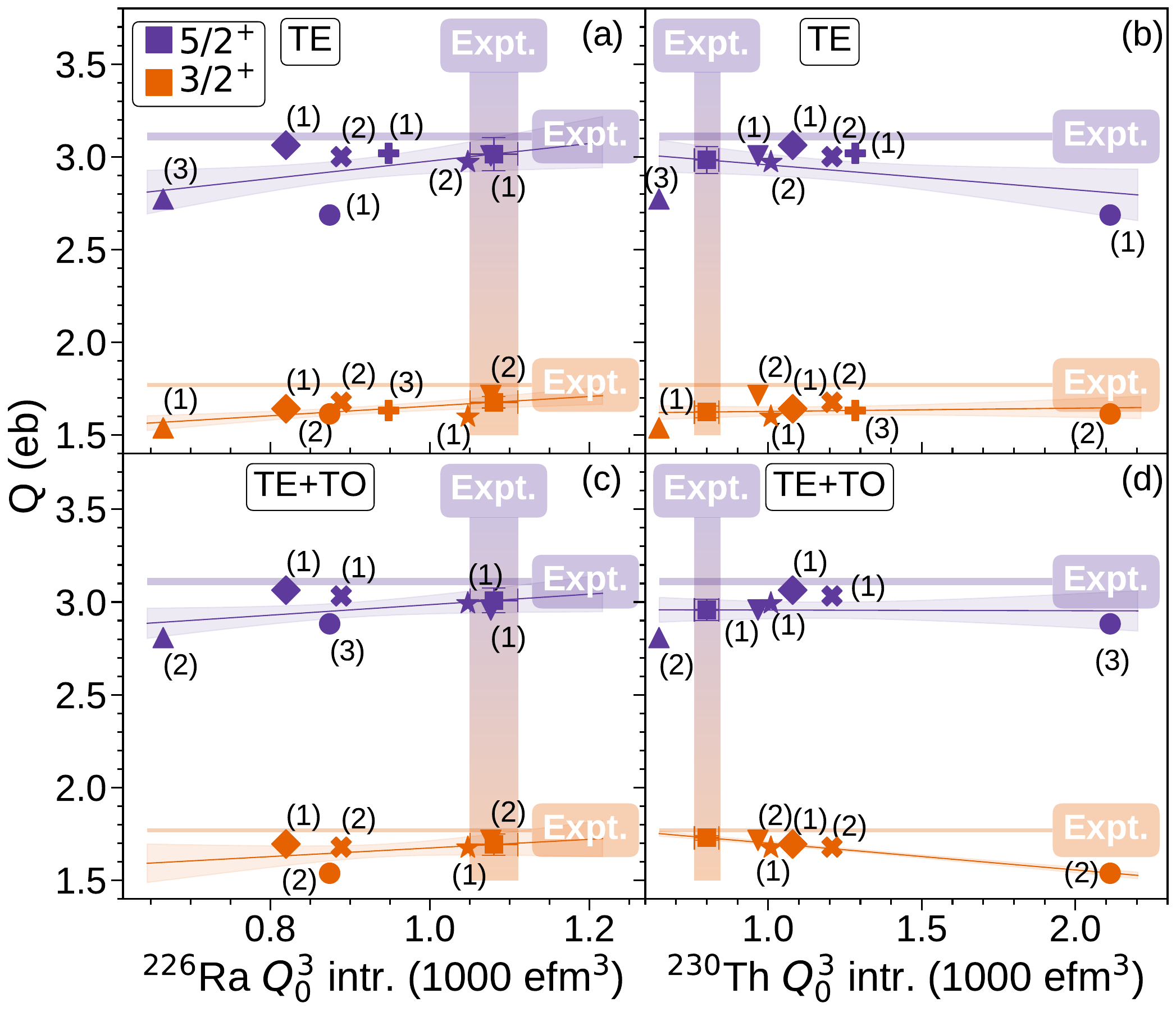}
\caption{Same as Fig.~\protect\ref{Corr1a} but for electric quadrupole moments.
\label{CorrQQ1}}
%\end{figure}

\vspace*{1mm}
%\begin{figure}
\centering\includegraphics[width=0.44\textwidth]{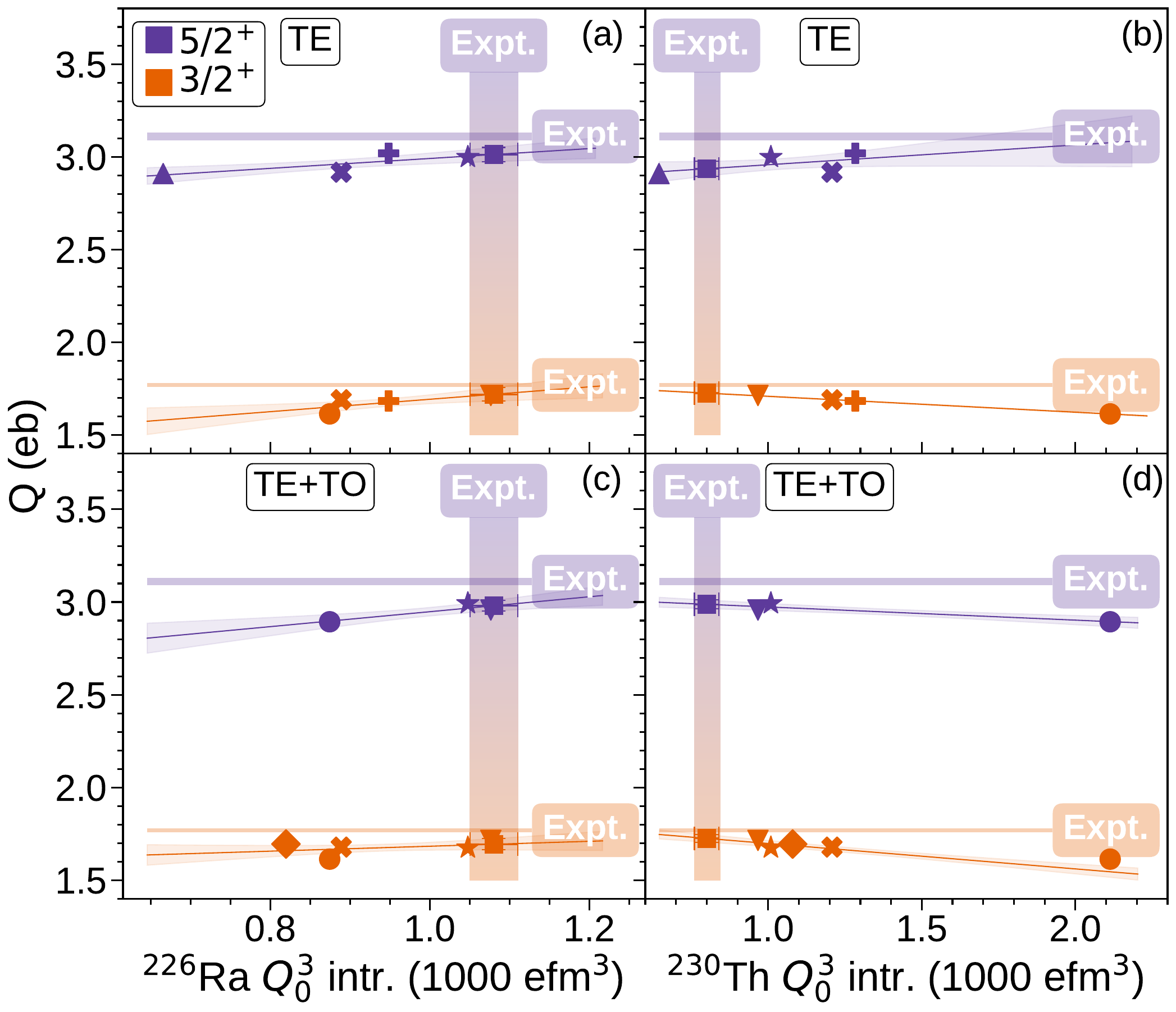}
\caption{Same as Fig.~\protect\ref{Corr2a} but for electric quadrupole moments.
\label{CorrQ1}}
%\end{figure}

\vspace*{1mm}
%\begin{figure}
\centering\includegraphics[width=0.44\textwidth]{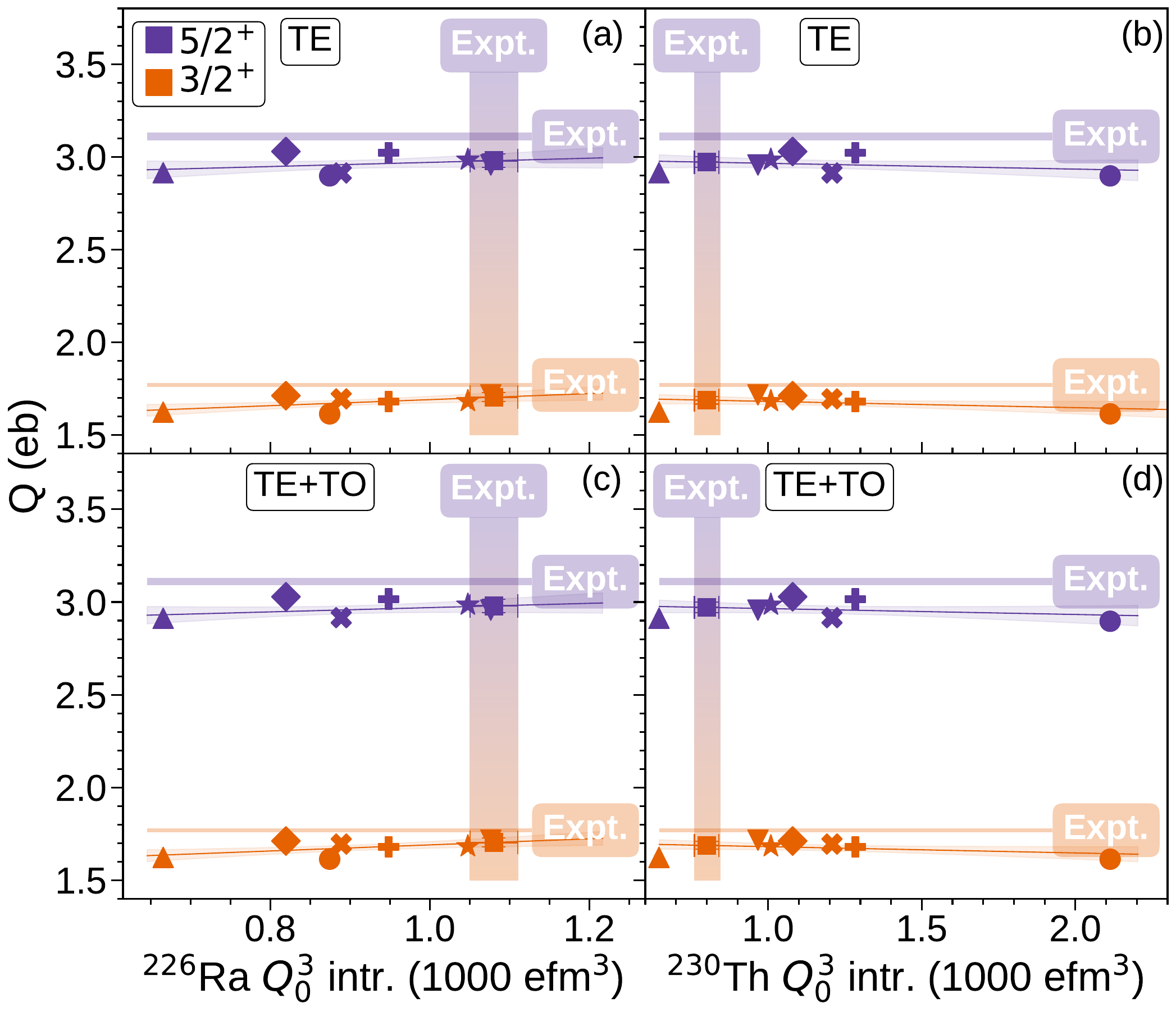}
\caption{Same as Fig.~\protect\ref{Corr3a} but for electric quadrupole moments.
\label{CorrQ2}}
\end{figure}

\begin{figure}
\centering\includegraphics[width=0.44\textwidth]{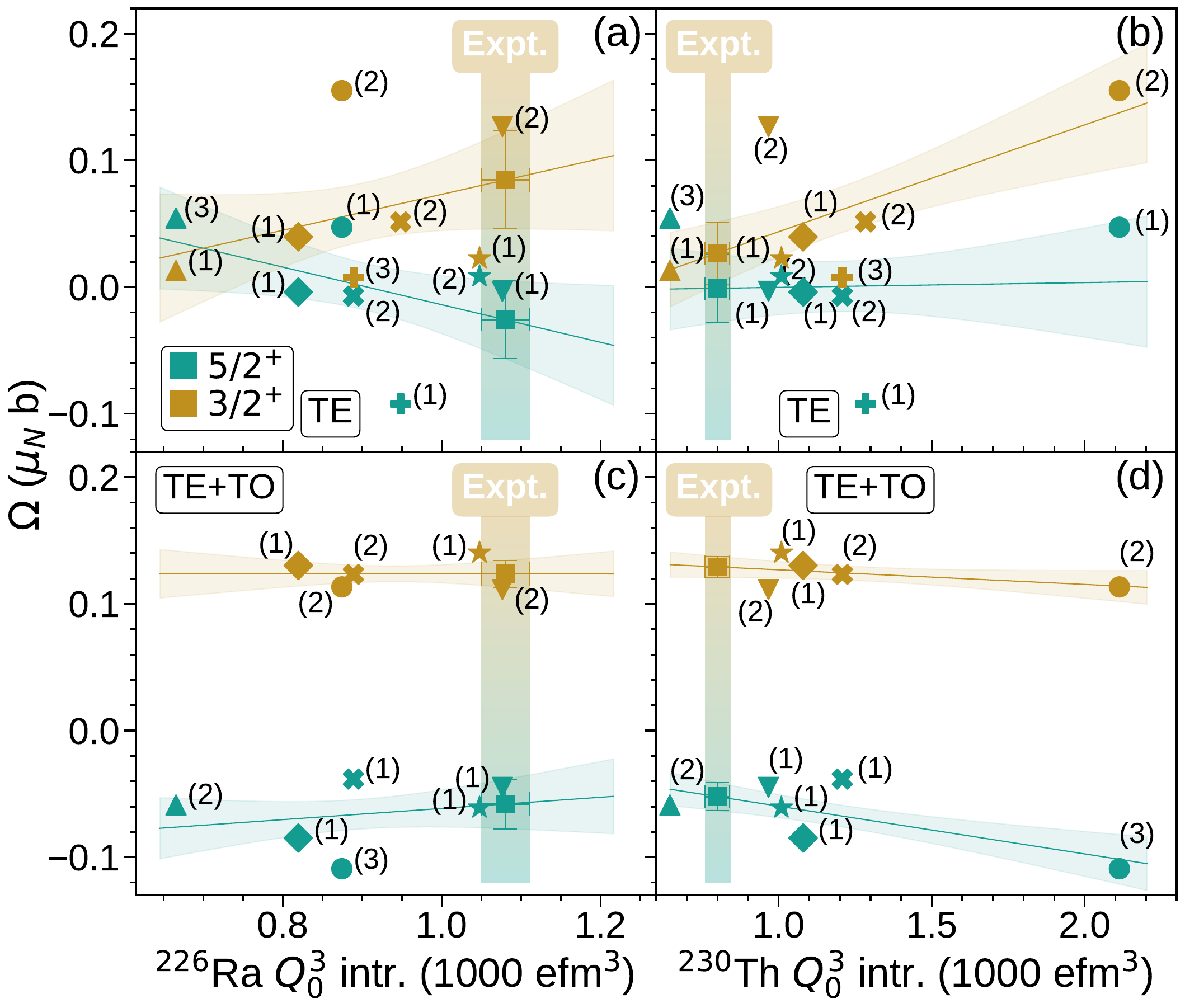}
\caption{Same as Fig.~\protect\ref{Corr1a} but for magnetic octupole moments.
\label{CorrOO1}}
%\end{figure}

\vspace*{3mm}
%\begin{figure}
\centering\includegraphics[width=0.44\textwidth]{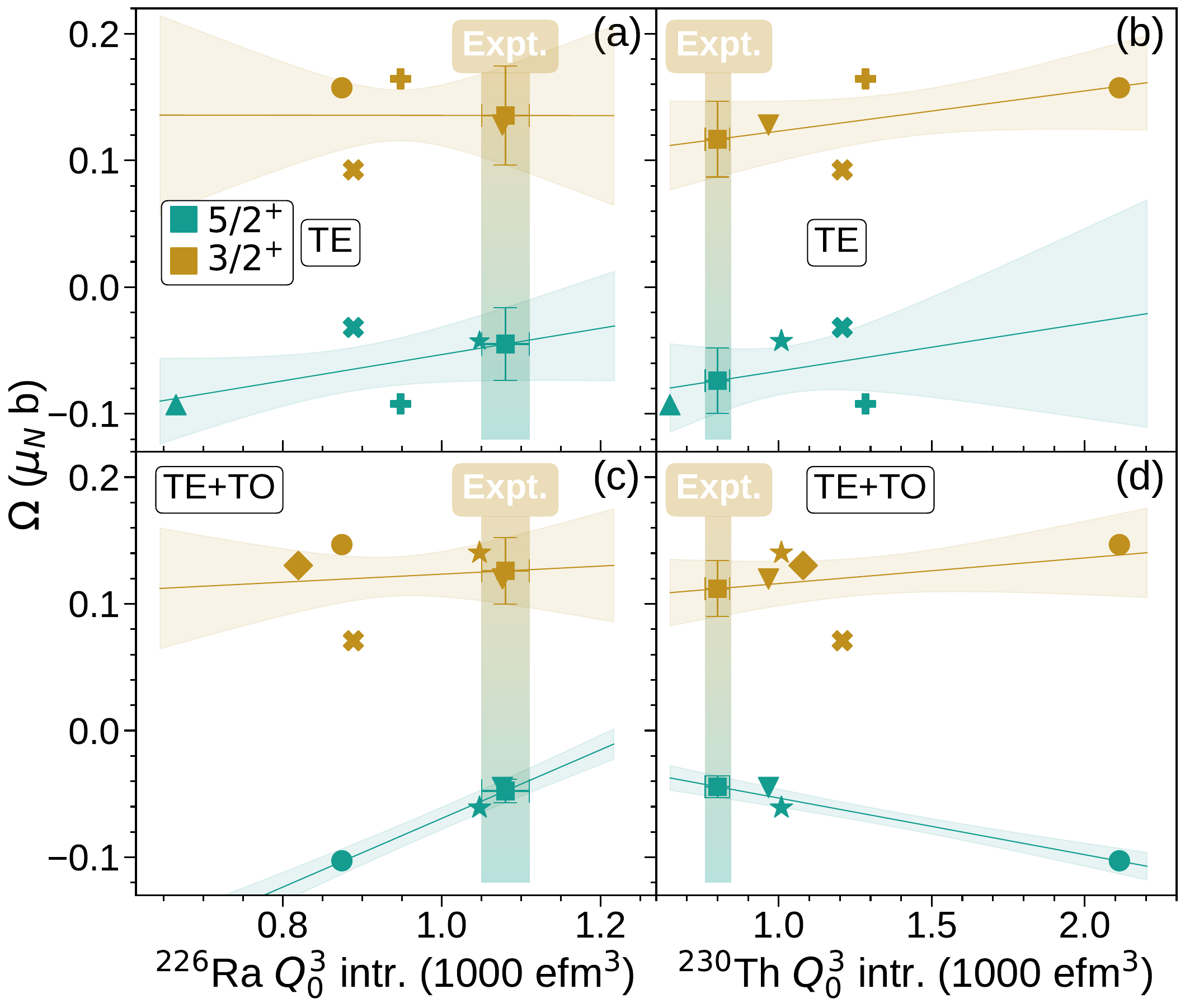}
\caption{Same as Fig.~\protect\ref{Corr2a} but for magnetic octupole moments.
\label{CorrO1}}
%\end{figure}

\vspace*{3mm}
%\begin{figure}
\centering\includegraphics[width=0.44\textwidth]{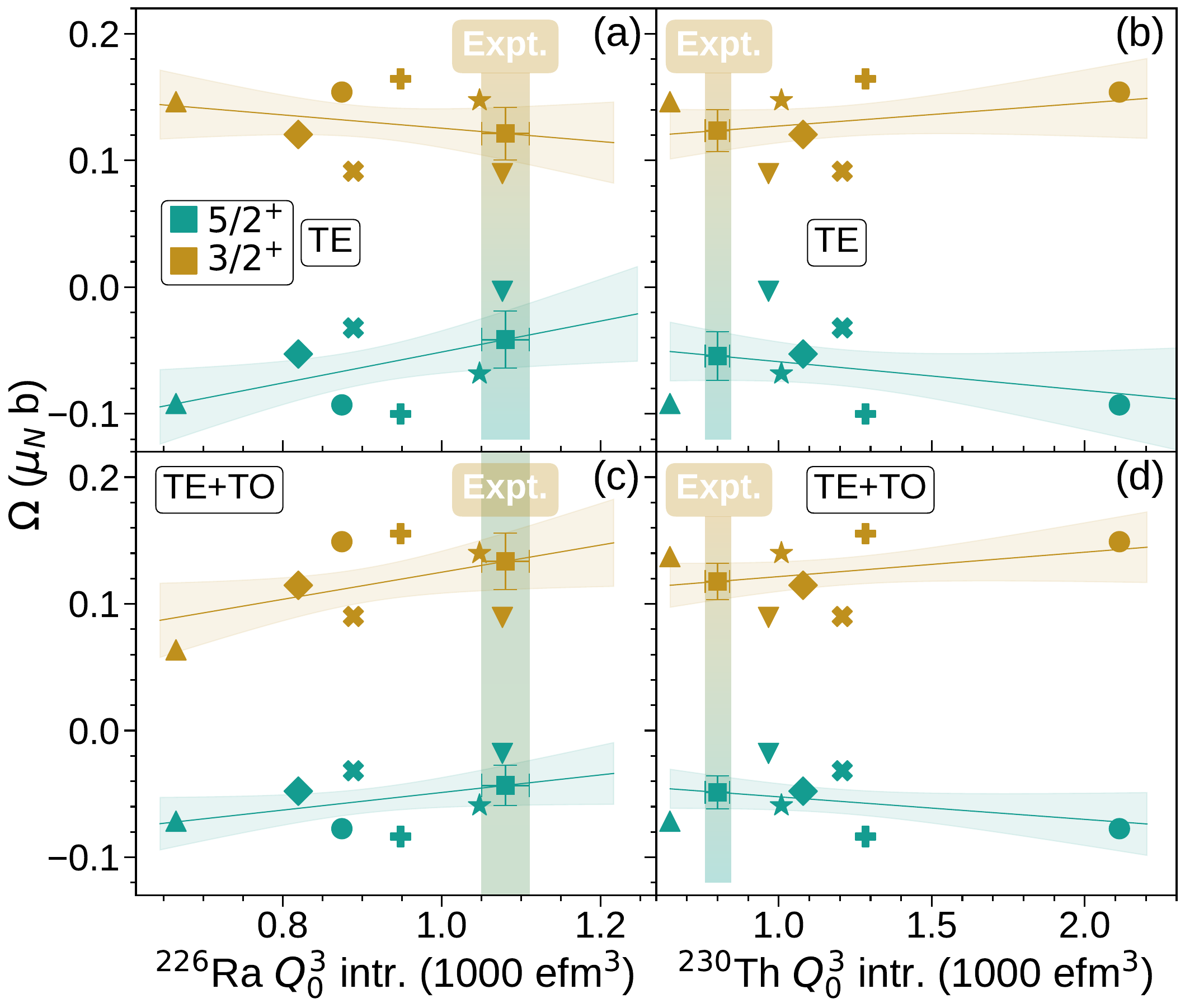}
\caption{Same as Fig.~\protect\ref{Corr3a} but for magnetic octupole moments.
\label{CorrO2}}
\end{figure}

{\it Plots of electromagnetic moments}---Figures~\ref{CorrMM1}--\ref{CorrO2} show the results of the regression analysis for the magnetic dipole moments (Figs.~\ref{CorrMM1}--\ref{CorrMM2}), electric quadrupole moments (Figs.~\ref{CorrQQ1}--\ref{CorrQ2}), and magnetic octupole moments (Figs.~\ref{CorrOO1}--\ref{CorrO2}) of the 3/2$^+$ isomeric and 5/2$^+$ ground states, and for the TE and TE+TO variants of calculations. Figures~\ref{CorrMM1} ,\ref{CorrQQ1}, and~\ref{CorrOO1}, analogous to Fig.~\ref{Corr1a}, show the results for parity breaking and mixing of configurations, where the numbers in parentheses represent the number of configurations mixed to obtain the results. Figures~\ref{CorrM1}, \ref{CorrQ1}, and~\ref{CorrO1}, analogous to Fig.~\ref{Corr2a}, show the results for parity breaking with no mixing, while Figs.~\ref{CorrMM2}, \ref{CorrQ2}, and~\ref{CorrO2}, analogous to Fig.~\ref{Corr3a}, display the result for the parity-conserving variant with no mixing.

\begin{widetext}

\begin{table}[h!]
\centering
\caption
{{Unrounded coupling constants of the Skyrme functionals used in this Letter.}\label{table_Skyrme}}
\vspace*{2mm}
%\begin{tabular}{l d d d d d d d }
\begin{tabular}{|l|ddddddd|}
\cline{2-8}
   \multicolumn{1}{r|}{}
 & \multicolumn{1}{r}{SkX$_\text{c}$\protect\cite{(Bro98c)}}
 & \multicolumn{1}{r}{SkM*\protect\cite{(Bar82c)}}
 & \multicolumn{1}{r}{UNEDF1\protect\cite{(Kor10c)}}
 & \multicolumn{1}{r}{SIII\protect\cite{(Bei75b)}}
 & \multicolumn{1}{r}{SkO'\protect\cite{(Rei99)}}
 & \multicolumn{1}{r}{SLy4\protect\cite{(Cha98a)}}
 & \multicolumn{1}{r|}{UNEDF0\protect\cite{(Kor12b)}} \\
\hline
$C^{\rho}_0$         & -539.250000 & -991.875000 & -779.373009 & -423.281250 & -787.282125 & -933.341250 & -706.382929 \\
$C^{\rho}_1$         & 283.286000  & 390.137500  & 287.722131  &  268.078125 & 246.942585  & 830.051485  & 240.049522  \\
$C^{s}_0$            & 44.138239   & 38.905862   & 30.554818   & 41.584341   & 219.237417  & 44.243392   & 34.079267   \\
$C^{s}_1$            & 135.000706  & 116.717587  & 129.857975  & 124.753022  & 101.970111  & 143.791024  & 102.237800  \\
$C^{\tau}_0$         & -0.821625   & 34.687500   & -0.989914   & 44.375000   & 15.000072   & 57.129375   & -12.917242  \\
$C^{\tau}_1$         & -44.706725  & -34.062500  & -33.632096  & -30.625000  & -4.156240   & 24.656385   & -45.189417  \\
$C^{T}_0$            & -7.389900   & 0.000000    & 0.000000    & 0.000000    & -104.093563 & 0.000000    & 0.000000    \\
$C^{T}_1$            & -23.625000  & 0.000000    & 0.000000    & 0.000000    & -9.171875   & 0.000000    & 0.000000    \\
$C^{j}_0$         & 0.821625   & -34.687500   & 0.989914   & -44.375000   & -15.000072   & -57.129375   & 12.917242  \\
$C^{j}_1$         & 44.706725  & 34.062500  & 33.632096  & 30.625000  & 4.156240   & -24.656385   & 45.189417  \\
$C^{J}_0$         & 7.389900   & 0.000000    & 0.000000    & 0.000000    & 104.093563 & 0.000000    & 0.000000    \\
$C^{J}_1$         & 23.625000  & 0.000000    & 0.000000    & 0.000000    & 9.171875   & 0.000000    & 0.000000    \\
$C^{\Delta\rho}_0$   & -46.011656  & -68.203125  & -45.135131  & -62.968750  & -52.787044  & -76.996406  & -55.260600  \\
$C^{\Delta\rho}_1$   & 22.750481   & 17.109375   & -145.382168 & 17.031250   & -32.162034  & 15.657086   & -55.622600  \\
$C^{\Delta s}_t$, $C^{\nabla s}_t$     & 0.000000    & 0.000000    & 0.000000    &  0.000000   & 0.000000    & 0.000000    & 0.000000    \\
% $C^{\Delta s}_1$     & 0.000000    & 0.000000    & 0.000000    &  0.000000   & 0.000000    & 0.000000    & 0.000000    \\
$C^{\nabla J}_0=C^{\nabla j}_0$     & -72.850000  & -97.500000  & -74.026333  & -90.000000  & -102.450600 & -92.250000  & -79.530800  \\
$C^{\nabla J}_1=C^{\nabla j}_1$     & 0.000000    & -32.500000  & -35.658261  & -30.000000  & 41.444400   & -30.750000  & 45.630200   \\
% $C^{\nabla s}_0$     & 0.000000 & 0.000000 & 0.000000 & 0.000000 & 0.000000 & 0.000000 & 0.000000 \\
% $C^{\nabla s}_1$     & 0.000000 & 0.000000 & 0.000000 & 0.000000 & 0.000000 & 0.000000 & 0.000000 \\
\hline
\end{tabular}
\end{table}
%\end{widetext}

%\begin{widetext}

\renewcommand{\arraystretch}{1.2}
\begin{table}[h!]
%\noindent
%\centering
\caption{Unrounded numerical values of results determined in this Letter, see text. \label{table:numerical1}}
\vspace*{2mm}
\rotatebox{90}{
\begin{tabular}{|@{~~}l@{~~}|l|ll|ll|ll|ll|ll|ll|ll|}
\cline{3-16}
\multicolumn{1}{c}{}&\multicolumn{1}{c|}{}
                     & \multicolumn{2}{c}{B(M1)}
                     & \multicolumn{2}{|c}{$\mu$($5/2^+$)}
                     & \multicolumn{2}{|c}{$Q$($5/2^+$)}
                     & \multicolumn{2}{|c}{$\Omega$($5/2^+$)}
                     & \multicolumn{2}{|c}{$\mu$($3/2^+$)}
                     & \multicolumn{2}{|c}{$Q$($3/2^+$)}
                     & \multicolumn{2}{|c|}{$\Omega$($3/2^+$)}  \\
\cline{2-16}
\multicolumn{1}{c|}{}&Skyrme
                     & \multicolumn{1}{c}{TE}  & \multicolumn{1}{c}{TE+TO}
                     & \multicolumn{1}{|c}{TE} & \multicolumn{1}{c}{TE+TO}
                     & \multicolumn{1}{|c}{TE} & \multicolumn{1}{c}{TE+TO}
                     & \multicolumn{1}{|c}{TE} & \multicolumn{1}{c}{TE+TO}
                     & \multicolumn{1}{|c}{TE} & \multicolumn{1}{c}{TE+TO}
                     & \multicolumn{1}{|c}{TE} & \multicolumn{1}{c}{TE+TO}
                     & \multicolumn{1}{|c}{TE} & \multicolumn{1}{c|}{TE+TO}  \\
\hline
\multirow{9}{*}{\rotatebox[origin=c]{270}{MIXED}}
%                            BM1TE           BM1TE+TO   m5/2TE       m5/2TE+TO    Q5/2TE    Q5/2TE+TO    O5/2TE       O5/2TE+TO     m3/2TE       m3/2TE+TO    Q3/2TE    Q3/2TE+TO  O3/2TE    O3/2TE+TO
& SkX$_{\text{c}}$         & \ph0.01026    & ~~---    & $-$0.79535 & \ph0.41030 & 2.7726  & 2.8080     & \ph0.05458  & $-$0.58669   & \ph0.51042 & \ph~~---   & 1.5391  & ~~---    & 0.01306  & ~~---     \\
& SkM*                     & \ph0.04508    & 0.07176  & $-$0.36250 & \ph0.55101 & 3.0054  & 2.9568     & $-$0.00334  & $-$0.44858   & $-$0.16870 & $-$0.01990 & 1.7135  & 1.7154   & 0.12664  & 0.11122    \\
& UNEDF1                   & \ph0.05547    & 0.04833  & $-$0.12505 & \ph0.25525 & 3.0628  & 3.0642     & $-$0.00708  & $-$0.84788   & \ph0.21499 & $-$0.26823 & 1.6418  & 1.6953   & 0.03966  & 0.13031    \\
& SIII                     & \ph0.32901    & 0.06880  & $-$0.69481 & \ph0.96563 & 2.6864  & 2.8825     & \ph0.04724  & $-$0.10904   & $-$0.51371 & $-$0.15283 & 1.6143  & 1.5379   & 0.15510  & 0.11340    \\
& SkO$^\prime$             & \ph0.07900    & ~~---    & \ph0.86083 & \ph~~---   & 3.0198  & ~~---      & $-$0.09226  & \ph~~---    & $-$0.49344 & \ph~~---   & 1.6320  & ~~---     & 0.05142  & ~~---     \\
& SLy4                     & \ph0.03430    & 0.00456  & $-$0.38997 & \ph0.31410 & 2.9714  & 2.9926     & \ph0.00848  & $-$0.60853   & \ph0.40281 & $-$0.32877 & 1.5991  & 1.6756   & 0.02272  & 0.14034    \\
& UNEDF0                   & \ph0.00377    & 0.00588  & \ph0.69722 & $-$0.13349 & 3.0010  & 3.0328     & $-$0.00390  & $-$0.38276   & \ph0.78717 & $-$0.23921 & 1.6769  & 1.6790   & 0.00763  & 0.12326    \\
\cline{2-16}
& $^{226}$Ra reg.          & \ph0.08(8)    & 0.04(3)  & \ph0.0(4)  & \ph0.4(3)  & 3.01(9) & 3.01(7)    & $-$0.00(3) & $-$0.06(2)  & $-$0.1(3)  & $-$0.15(10)& 1.68(3) & 1.69(6)  & 0.08(4) & 0.128(11)   \\
& $^{230}$Th reg.          & $-$0.01(3)    & 0.03(2)  & $-$0.2(3)  & \ph0.2(2)  & 2.98(7) & 2.96(6)    & $-$0.03(3) & $-$0.052(11)  & \ph0.4(2)  & $-$0.21(9) & 1.62(3) & 1.729(12)& 0.03(2) & 0.129(8)   \\
\hline

\multirow{9}{*}{\rotatebox[origin=c]{270}{NOT MIXED}}
%                            BM1TE           BM1TE+TO   m5/2TE       m5/2TE+TO    Q5/2TE    Q5/2TE+TO    O5/2TE       O5/2TE+TO     m3/2TE       m3/2TE+TO    Q3/2TE    Q3/2TE+TO  O3/2TE    O3/2TE+TO
& SkX$_{\text{c}}$        & \ph~~---       & ~~---    & \ph0.27118 & \ph~~---   & 2.9095  & ~~---      & $-$0.09283  & \ph~~---    & \ph~~---   & \ph~~---   & ~~---   & ~~---    & ~~---   & ~~---     \\
& SkM*                    & \ph~~---       & 0.07380  & \ph~~---   & \ph0.55101 & ~~---   & 2.9568     & \ph~~---   & $-$0.04485   & $-$0.18195 & $-$0.09812 & 1.7133  & 1.7149   & 0.12800  & 0.11965    \\
& UNEDF1                  & \ph~~---       & ~~---    & \ph~~---   & \ph~~---   & ~~---   & ~~---      & \ph~~---   & \ph~~---    & \ph~~---   & $-$0.26823 & ~~---   & 1.6953   & ~~---   & 0.13031    \\
& SIII                    & \ph~~---       & 0.02179  & \ph~~---   & \ph0.79171 & ~~---   & 2.8937     & \ph~~---   & $-$0.10276   & $-$0.54818 & $-$0.38116 & 1.6141  & 1.6141   & 0.15733  & 0.14679    \\
& SkO$^\prime$            & \ph0.06019     & ~~---    & \ph0.86083 & \ph~~---   & 3.0198  & ~~---      & $-$0.09226  & \ph~~---    & $-$0.57155 & \ph~~---   & 1.6835  & ~~---    & 0.16448  & ~~---     \\
& SLy4                    & \ph~~---       & 0.00456  & \ph0.19361 & \ph0.31410 & 2.9984  & 2.9926     & $-$0.04260  & $-$0.06085   & \ph~~---   & $-$0.32877 & ~~---   & 1.6756   & ~~---   & 0.14034    \\
& UNEDF0                  & \ph0.00145     & ~~---    & \ph0.50012 & \ph~~---   & 2.9168  & ~~---      & $-$0.03184  & \ph~~---    & \ph0.16136 & $-$0.24113 & 1.6904  & 1.6789   & 0.09251  & 0.07091    \\
\cline{2-16}
& $^{226}$Ra reg.         & \ph0.19        & 0.04(4)  & \ph0.5(3)  & \ph0.4(2)   & 3.01(4)& 2.98(3)    & $-$0.04(3) & $-$0.048(9) & $-$0.3(4)  & $-$0.20(10)& 1.72(4) & 1.69(3)  & 0.13(4) & 0.13(3)   \\
& $^{230}$Th reg.         & $-$0.31        & 0.04(4)  & \ph0.3(2)  & \ph0.38(15) & 2.94(4)& 2.99(2)    & $-$0.07(3) & $-$0.044(9) & $-$0.1(3)  & $-$0.19(6) & 1.73(3) & 1.73(2)  & 0.12(3) & 0.11(2)   \\
\hline
\multirow{9}{*}{\rotatebox[origin=c]{270}{NO OCTUPOLE}}
%                            BM1TE           BM1TE+TO   m5/2TE       m5/2TE+TO    Q5/2TE    Q5/2TE+TO    O5/2TE       O5/2TE+TO     m3/2TE       m3/2TE+TO    Q3/2TE    Q3/2TE+TO  O3/2TE    O3/2TE+TO
& SkX$_{\text{c}}$        & \ph0.07804     & 0.09297  & \ph0.83200 & \ph0.63589 & 2.9140  & 2.9127     & $-$0.09187  & $-$0.07147   & $-$0.46792 & $-$0.31471 & 1.6240  & 1.6234   & 0.14641  & 0.13731    \\
& SkM*                    & \ph0.20008     & 0.19810  & \ph0.70337 & \ph0.55873 & 2.9561  & 2.9565     & $-$0.00326  & $-$0.01802   & \ph0.21260 & \ph0.18240 & 1.7166  & 1.7165   & 0.08959  & 0.08932    \\
& UNEDF1                  & \ph0.33397     & ~~---    & \ph0.18417 & \ph0.17797 & 3.0287  & 3.0285     & $-$0.05277  & $-$0.04785   & $-$0.10385 & $-$0.03062 & 1.7125  & 1.7118   & 0.12045  & 0.11468    \\
& SIII                    & \ph0.06343     & 0.06659  & \ph0.89248 & \ph0.71634 & 2.8973  & 2.8964     & $-$0.09293  & $-$0.07743   & $-$0.51329 & $-$0.35483 & 1.6143  & 1.6137   & 0.15396  & 0.14909    \\
& SkO$^\prime$            & \ph0.05298     & 0.08053  & \ph0.95830 & \ph0.74518 & 3.0210  & 3.0165     & $-$0.10022  & $-$0.08364   & $-$0.57189 & $-$0.41960 & 1.6814  & 1.6809   & 0.16450  & 0.15533    \\
& SLy4                    & \ph0.20014     & 0.20001  & \ph0.39101 & \ph0.36633 & 2.9840  & 2.9852     & $-$0.06824  & $-$0.05910   & $-$0.28603 & $-$0.18165 & 1.6829  & 1.6829   & 0.14739  & 0.14003    \\
& UNEDF0                  & \ph0.26571     & 0.25370  & \ph0.50392 & \ph0.45116 & 2.9129  & 2.9156     & $-$0.03216  & $-$0.03173   & \ph0.16926 & \ph0.16695 & 1.6952  & 1.6952   & 0.09145  & 0.09007    \\
\cline{2-16}
& $^{226}$Ra reg.         & \ph0.19(8)     & 0.19(5)  & \ph0.6(2)  & \ph0.51(14)& 2.98(4) & 2.98(4)    & $-$0.04(2) & $-$0.04(2)  & $-$0.1(2)  & $-$0.0(2)  & 1.70(2) & 1.70(2)  & 0.12(2) & 0.12(2)   \\
& $^{230}$Th reg.         & \ph0.20(6)     & 0.17(4)  & \ph0.6(2)  & \ph0.46(11)& 2.97(3) & 2.97(3)    & $-$0.05(2) & $-$0.049(13)& $-$0.1(2)  & $-$0.13(7) & 1.69(2) & 1.69(2)  & 0.12(2) & 0.116(15) \\
\hline
\multicolumn{1}{c|}{}& \multicolumn{1}{l|}{Experiment}
                     & \multicolumn{2}{c}{0.0388(12)}
                     & \multicolumn{2}{|c}{0.366(6)}
                     & \multicolumn{2}{|c}{3.11(2)}
                     & \multicolumn{2}{|c}{~~---  }
                     & \multicolumn{2}{|c}{$-$0.378(8)}
                     & \multicolumn{2}{|c}{1.77(1)}
                     & \multicolumn{2}{|c|}{~~---  }  \\
\cline{2-16}
\end{tabular}}
%\vspace*{3mm}
%\noindent
%\begin{minipage}{\textwidth}{\hspace*{1.5cm}Table V. See text.
%}
%\end{minipage}
\end{table}

\end{widetext}

%\vspace*{3mm}

\clearpage

\clearpage

\begin{widetext}

\newpage

\end{widetext}

\end{document}